\documentclass{JHEP3}
\usepackage{amsfonts,amsmath,bbm,cite}
\usepackage{graphicx}
\usepackage{xy}
\xyoption{all}

\newcommand{\fref}[1]{Figure~\ref{#1}}
\newcommand{\cref}[1]{Chapter~\ref{#1}}
\newcommand{\beq}{\begin{equation}}
\newcommand{\eeq}{\end{equation}}
\newcommand{\ba}{\begin{array}}
\newcommand{\ea}{\end{array}}
\newcommand{\bcenter}{\begin{center}}
\newcommand{\ecenter}{\end{center}}

\def\IC{\mathbb{C}}

\def\IGa{\relax\hbox{${\rm I}\kern-.18em\Gamma$}}

\def\IP{\mathbb{P}}
\def\IR{\mathbb{R}}

\def\IZ{\mathbb{Z}}

\def\smiley{\hbox{\large$\bigcirc$\hspace{-0.80em}\raise.2ex
\hbox{$\cdot\cdot$}\kern-.61em\lower.2ex\hbox{\scriptsize$\smile$}}\ }
\def\frowny{\hbox{\large$\bigcirc$\hspace{-0.80em}\raise.2ex
\hbox{$\cdot\cdot$}\kern-.635em\lower.2ex\hbox{\scriptsize$\frown$}}\ }

\makeatletter
\let\hangafter\@hangfrom
\makeatother

\newtheorem{thm}{Theorem}[subsection]
\newcommand{\btheorem}{\begin{thm}}
\newcommand{\etheorem}{\end{thm}}
\newtheorem{lem}[thm]{Lemma}
\newcommand{\blemma}{\begin{lem}}
\newcommand{\elemma}{\end{lem}}
\newtheorem{dfn}[thm]{Definition}
\newcommand{\bdefn}{\begin{dfn}}
\newcommand{\edefn}{\end{dfn}}
\newtheorem{cor}[thm]{Corollary}
\newcommand{\bcor}{\begin{cor}}
\newcommand{\ecor}{\end{cor}}
\def\bproof{\begin{proof}} 
\def\eproof{\end{proof}}

\newcommand{\be}{\begin{equation}}
\newcommand{\ee}{\end{equation}}
\newcommand{\bea}{\begin{eqnarray}}
\newcommand{\eea}{\end{eqnarray}}
\newcommand{\bean}{\begin{eqnarray*}}
\newcommand{\eean}{\end{eqnarray*}}
\newcommand{\bc}{\begin{center}}
\newcommand{\ec}{\end{center}}
\newcommand{\comment}[1]{}

\def\kahler{K\"{a}hler\,}
\DeclareFontFamily{U}{rsf}{} \DeclareFontShape{U}{rsf}{m}{n}{  <5> <6> rsfs5 <7> <8> <9> rsfs7 <10-> rsfs10}{}
\DeclareMathAlphabet\Scr{U}{rsf}{m}{n} \DeclareMathAlphabet\mathbi{U}{cmr}{bx}{it}

\def\ses#1#2#3{\xymatrix@1{0 \ar[r] & #1 \ar[r] & #2 \ar[r] & #3 \ar[r] & 0}}

\def\IR{\mathbb{R}}
\def\IC{\mathbb{C}}
\def\IZ{\mathbb{Z}}
\def\IP{\mathbb{P}}

\def\be{\begin{equation}}
\def\ee{\end{equation}}

\def\kahler{K\"{a}hler }

\newcommand{\mtwo}[4]{
\left( \begin{array}{cc}
 #1 & #2 \\
 #3 & #4   \end{array}  \right)}

\def\CN{{\mathcal{N}}}
\newcommand{\ie}{{\it i.e.\ }}
\newcommand{\eg}{{\it e.g.\ }}

\preprint{MIT-CTP-3960}

\title{Semi-Flatland}

\author{David Vegh and John McGreevy
\\
~\\
Center for Theoretical Physics,
Massachusetts Institute of Technology,\\
Cambridge, MA 02139, USA.
\\~\\
 \email{dvegh {\it at} mit.edu}
}

\abstract{We study perturbative compactifications of Type II string theory that rely on a
fibration structure of the extra dimensions {\it \` a la} SYZ. Non-geometric spaces are obtained
by using T-dualities as monodromies. These vacua generically preserve $\mathcal{N}=1$
supersymmetry in four dimensions, and are U-dual to M-theory on $G_2$ manifolds. Several
examples are discussed, some of which admit an asymmetric orbifold description. The massless
spectrum is matched to that of the dual M-theory compactification on a Joyce manifold when a
comparison is possible. We explore the possibility of twisted reductions where left-moving
spacetime fermion number Wilson lines are turned on in the fiber. We also give an explanation
from this semiflat viewpoint for the Hanany-Witten brane-creation effect and for the equivalence
of the Type IIA orientifold on $T^5/\IZ_2$ and Type IIB on $S^1\times K3$. }

\begin{document}


\section{Introduction}


A great deal of progress has been made in the study of string compactification using the
ten-dimensional supergravity approximation
(for a review, see \cite{Douglas:2006es}). However, it has become clear that certain interesting
physical features of our world are difficult (if not impossible) to realize when this
description is valid. Examples which come to mind include a period of slow-roll inflation
\cite{Hertzberg:2007wc, Dimopoulos:2005ac, Grimm:2007hs}, certain models of dynamical
supersymmetry breaking \cite{Florea:2006si}, chiral matter with stabilized moduli
\cite{Blumenhagen:2007sm} and parametrically-small perturbatively-stabilized extra dimensions
\cite{Douglas:2006es}. This strongly motivates attempts to find descriptions of
moduli-stabilized string vacua which transcend the simple geometric description.

One approach to vacua outside the domain of validity of 10d supergravity is to rely only on the
4d gravity description, as in \eg \cite{Shelton:2005cf, Silverstein:2007ac}. This can be
combined with insight into the microscopic ingredients to give a description of much more
generic candidate string vacua. A drawback of this approach is that it is difficult to control
systematically the interactions between the ingredients. Another promising direction is
heterotic constructions, which do not require RR flux and hence are more amenable to a
worldsheet treatment \cite{Adams:2006kb, Adams:2007vp}. However, stabilization of the dilaton in
these constructions requires non-perturbative physics.

A third technique, which is at an earlier state of development, was implemented in
\cite{Hellerman:2002ax}, and was inspired by \cite{Greene:1989ya, Vafa:1996xn}. The idea is to
build a compactification out of locally ten-dimensional geometric descriptions, glued together
by transition functions which include large gauge transformations, such as stringy dualities.
This technique is uniquely adapted to construct examples with no global geometric description.
In this paper, we build on the work of \cite{Hellerman:2002ax} to give 4d ${\cal N} = 1$
examples.

With S. Hellerman and B. Williams \cite{Hellerman:2002ax}, one of us constructed early examples
of vacua involving such `non-geometric fluxes'. These examples were constructed by compactifying
string theory on a flat $n$-torus, and allowing the moduli of this torus to vary over some base
manifold. The description of these spaces where the torus fiber is flat is called the {\it
semi-flat approximation} \cite{Strominger:1996it}. Allowing the torus to degenerate at real
codimension two on the base reduces the construction of interesting spaces to a Riemann-Hilbert
problem; the relevant data is in the monodromy of the torus around the degenerations
\cite{Greene:1989ya}. Generalizing this monodromy group to include not just modular
transformations of the torus, but more general discrete gauge symmetries of string theory
(generally known as string dualities) allows the construction of vacua of string theory which
have no global geometric description \cite{Hellerman:2002ax}. The examples studied in detail in
\cite{Hellerman:2002ax} had two-torus fibers, which allowed the use of complex geometry.

A natural explanation of mirror symmetry is provided by the conjecture \cite{Strominger:1996it}
that any CY has a description as a three-torus $(T^3)$ fibration, over a 3-manifold base. In the
large complex structure limit, the locus in the base where the torus degenerates is a trivalent
graph; the data of the CY is encoded in the monodromies experienced by the fibers in
circumnavigating this graph. Further, the edges of the graph carry energy and create a deficit
angle -- in this description a compact CY is a self-gravitating cosmic string network whose
back-reaction compactifies the space around itself. In this paper, our goal is to use this
description of ordinary CY manifolds to construct non-geometric vacua, again by enlarging the
monodromy group. We find a number of interesting new examples of non-geometric vacua with 4d
${\cal N}=1$ supersymmetry. In a limit, they have an exact CFT description as asymmetric
orbifolds, and hence can be considered `blowups' thereof. We study the spectrum, particularly
the massless scalars, and develop some insight into how these vacua fit into the web of known
constructions.

We emphasize at the outset two limitations of our analysis. First, the examples constructed so
far are special cases which have arbitrarily-weakly-coupled perturbative descriptions and
(therefore) unfixed moduli. Our goal is to use them to develop the semiflat techniques in a
controllable context. Generalizations with nonzero RR fluxes are naturally incorporated by
further enlarging the monodromy group to include large RR gauge transformations, as in F-theory
\cite{Vafa:1996xn}. There one can hope that all moduli will be lifted. This is the next step
once we have reliable tools for understanding such vacua using the fibration description.

The second limitation is that we have not yet learned to describe configurations where the base
of the $T^3$-fibration is not flat away from the degeneration locus.
The examples of SYZ fibrations we construct
(analogous to F-theory at constant coupling \cite{Dasgupta:1996ij}) all involve
composite degenerations which we do not know how to resolve. The set of rules we find for
fitting these composite degenerations into compact examples will be a useful guide to the more
difficult general case.

A number of intriguing observations arise in the course of our analysis. One can ``geometrize''
these non-geometric compactifications by realizing the action of the T-duality group as a
geometric action on a $T^4$ fiber. The semi-flat metric on the fiber contains the original
metric and the Hodge dual of the B-field. Hence, we are led to study seven-manifolds ${\cal
X}_7$ which are $T^4$ fibrations over a 3d base. They can be embedded into flat $T^4$
compactifications of M-theory down to seven dimensions where the reduced theory has an $SL(5)$
U-symmetry. U-duality then suggests that ${\cal X}_7$ may be a $G_2$ manifold since the
non-geometric Type IIA configuration can be rotated into a purely geometric solution of maximal
supergravity in seven dimensions. Whether or not these solutions can in general be lifted to
eleven dimensions is a question for further investigation. In this paper, we study explicit
examples of $G_2$ (and Calabi-Yau) manifolds and show that they do provide perturbative
non-geometric solutions to Type IIA in ten dimensions through this correspondence. The spectrum
of these spaces can be computed by noticing that they admit an asymmetric orbifold description,
and it matches that computed from M-theory when a comparison is possible.

The paper is organized as follows. In the next section we review the semiflat approximation to
geometric compactification in various dimensions. We describe in detail the semiflat
decomposition of an orbifold limit of a Calabi-Yau threefold; this will be used as a starting
point for nongeometric generalizations in section four. In section three we describe the
effective field theory for type II strings on a flat $T^3$. We show that the special class of
field configurations which participate in $T^3$-fibrations with {\it perturbative} monodromies
can alternatively be described in terms of geometric $T^4$-fibrations. We explain the U-duality
map which relates these constructions to M-theory on $T^4$-fibered $G_2$-manifolds. In sections
four and five we put this information together to construct nongeometric compactifications. In
section six we consider generalizations where the fiber theory involves discrete Wilson lines.
Hidden after the conclusions are many appendices. Appendix A gives more detail of the reduction
on $T^3$. The purpose of Appendices B--D is to build confidence in and intuition about the
semiflat approximation: Appendix B is a check on the relationship between the semiflat
approximation and the exact solution which it approximates; Appendix C is a derivation of the
Hanany-Witten brane-creation effect using the semiflat limit; Appendix D derives a known duality
using the semiflat description. In Appendix E we record asymmetric orbifold descriptions of the
nongeometric constructions of section four. In Appendices F through H, we study in detail the
massless spectra of many of our constructions, and compare to the spectra of M-theory on the
corresponding $G_2$-manifolds when we can. Appendix I contains templates to help the reader to
build these models at home.

\newpage
\section{Semi-flat limit}

Since we want to construct non-geometric spaces by means of T-duality, we exhibit the spaces as
torus fibrations. We need isometries in the fiber directions in which the dualities act. Hence,
we wish to study manifolds in a {\it semi-flat limit} where the fields do not depend on the
fiber coordinates. This is the realm of the SYZ conjecture \cite{Strominger:1996it}. Mirror
symmetry of Calabi-Yau manifolds implies that they have a special Lagrangian $T^n$ fibration.
Branes can be wrapped on the fibers in a supersymmetric way and their moduli space is the mirror
Calabi-Yau. At tree level, this moduli space is a semi-flat fibration, \ie the metric has a
$U(1)^n$ isometry along the fiber. However, there are world-sheet instanton corrections to this
tree-level metric. Such corrections are suppressed (away from singular fibers) in the {\it large
volume limit}. The mirror Calabi-Yau is then in the {\it large complex structure limit}. In this
limit the metric is semi-flat and mirror symmetry boils down to T-duality along the fiber
directions\footnote{It is best to think of the fiber as being very small compared to the size of
the base. It is thought that in the large complex structure limit, the total space of the CY
collapses to a metric space homeomorphic to $S^n$ which is the base of the fibration (see \eg
\cite{GrossWilson}). }.


As a warm-up, we will now briefly review the one-{\it complex}-dimensional case of a torus, and
the two-dimensional case of stringy cosmic strings \cite{Greene:1989ya}. These sections may be
skipped by experts. In Section \ref{threedim}, we construct a fibration for a three-dimensional
orbifold that will in later sections be modified to a non-geometric compactification.

\subsection{One dimension}
\label{onedimsec}

The simplest example is the flat two-torus. Its complex structure is given by modding out the
complex plane by a lattice generated by 1 and $\tau = \tau_1 + i \tau_2 \in \IC$ (with
$\tau_2>0$). The \kahler structure is $\rho = b +iV/2$ where $b=\int_{T^2} B$ and $V$ the area of
the torus (again, $V>0$).

\begin{figure}[ht]
\begin{center}
  \includegraphics[totalheight=6cm,angle=0,origin=c]{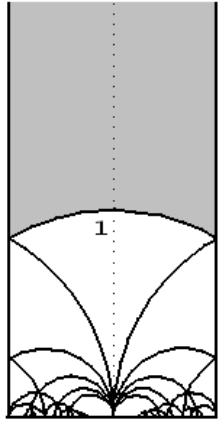}
  \caption{A possible fundamental domain (gray area) for the action of the $SL(2,\IZ)$ modular group on the upper half-plane.
  The upper-half plane parametrizes the possible values of $\tau$ (or $\rho$): the moduli of a two-torus.
  The gray domain can be folded into an $S^2$ with three special points (the two orbifold points: $\tau_{\IZ_6} = e^{2\pi i /6}$ and $\tau_{\IZ_4} =
  i$, and the decompactification point: $\tau\rightarrow i \infty$).}
  \label{fundom}
\end{center}
\end{figure}

There is an $SL(2,\IZ)_\tau$ group acting on the complex modulus $\tau$. This is a redundancy in
defining the lattice. The group action is generated by $\tau \mapsto \tau + 1$ and $\tau \mapsto
-1/\tau$. Another $SL(2,\IZ)_\rho$ group acts on $\rho$. This is generated by the shift in the
B-field $b \mapsto b+1$ and a double T-duality combined with a $90^\circ$ rotation that is $\rho
\mapsto -1/\rho$. The fundamental domain for the moduli is shown in \fref{fundom}.

The torus can naturally be regarded as a semi-flat circle fibration over a circle. For special
Lagrangian fibers, we choose the real slices in the complex plane. In the $\tau_2 \rightarrow
\infty$ large complex structure limit, these fibers are small compared to the base $S^1$ which
is along the imaginary axis.

Mirror symmetry exchanges the complex structure $\tau$ with the \kahler structure $\rho$. This
boils down to T-duality along the fiber direction according to the Buscher rules
\cite{Buscher:1987sk, Buscher:1987qj}. It maps the large complex structure into large \kahler
structure that is $\rho_2 = V \rightarrow \infty$.


\subsection{Two dimensions}
\label{twodimsec}



In order to construct semi-flat fibrations in two dimensions, let us consider the dynamics
first. Type~IIA on a flat two-torus can be described by the effective action in Einstein frame
\be
  S=\int d^8 x \sqrt{g} \left( R + \frac{\partial_\mu \tau \partial^\mu \bar\tau}{\tau_2^2} +
  \frac{\partial_\mu \rho \partial^\mu \bar\rho}{\rho_2^2} \right)
  \label{acti}
\ee
where $\tau$ is the complex structure of the torus, and $\rho=b+iV/2$ is the \kahler modulus as
described earlier.
The action is invariant under the
$SL(2,\IZ)_\tau \times SL(2,\IZ)_\rho$ perturbative duality group,
which acts on $\tau$ and $\rho$ by fractional linear transformations.

Variation with respect to $\tau$ gives
\be\label{einsteineqn}
  \partial \bar\partial \tau + \frac{2\partial\tau\bar\partial\tau}{\bar\tau-\tau}=0 ;
\ee
and $\rho$ obeys the same equation. Stringy cosmic string solutions to the EOM can be obtained
by choosing a complex coordinate $z$ on two of the remaining eight dimensions, and taking
$\tau(z)$ a holomorphic section of an $SL(2, \IZ)$ bundle. Such solutions are not modified by
considering the following ansatz for the metric around the string\footnote{ By an appropriate
coordinate transformation of the base coordinate, this metric can be recast into a symmetric $g
\oplus g$ form (see \cite{Strominger:1996it,Loftin:2004qu}).}
\be
  ds^2=ds_\textrm{Mink}^2+e^{\psi(z,\bar z)} dz d\bar z + ds_\textrm{fiber}^2
  \label{flateq1}
\ee
where
\be
  ds_\textrm{fiber}^2 = \frac{1}{\tau_2} \left( \begin{array}{cc}
  1 & \tau_1  \\
  \tau_1 & \ |\tau|^2
  \end{array}
  \right)
\ee

The Einstein equation is the Poisson equation,
\be
  \partial \bar\partial \psi =   \partial \bar\partial \textrm{log} \, \tau_2
  \label{flateq2}
\ee
Far away from the strings, the metric of the base goes like \cite{Greene:1989ya}
\be
  ds^2_{2D} \sim | z^{-N/12} dz |^2
\ee
where $N$ is the number of strings. This can be coordinate transformed by $\tilde z =
z^{1-N/12}$ to a flat metric with $2\pi N/12$ deficit angle.

\vskip 0.5cm \noindent {\bf Solutions and orbifold points.} One could in principle write down
solutions by means of the $j$-function,
\be
  j(\tau) = \eta(\tau)^{-24} (\theta^8_1(\tau)+\theta^8_2(\tau)+\theta^8_3(\tau))^3
\ee
which maps the $\tau_{\IZ_6} = e^{2\pi i /6}$ and $\tau_{\IZ_4} = i$ orbifold points to 0 and 1,
respectively. The $\tau_2 \rightarrow \infty$ degeneration point gets mapped to $j\rightarrow
\infty$. A simple solution would then be
\be
  j(\tau) = \frac{1}{z-z_0} + j_0
\ee

At infinity, the shape of the fiber is constant, \ie $\tau_\infty = j^{-1}(j_0)$  and thus this
non-compact solution may be glued to any other solution with constant $\tau$ at infinity.
However, since $\tau$ covers the entire fundamental domain once, there will be two points in the
base where $\tau(z) = \tau_{\IZ_6}$ or $\tau_{\IZ_4}$. Over these points, the fiber is an
orbifold of the two-torus. These singular points cannot be resolved in a Ricci-flat way and we
can't use this solution for superstrings.

There is, however, a six-string solution which evades this problem \cite{Greene:1989ya}. It is
possible to collect six strings together in a way that $\tau$ approaches a constant value at
infinity. $\tau$ can be given implicitly by \eg
\be
  y^2 = x (x-1)(x-2)(x-z)
\ee
There are no orbifold points now because $\tau$ can be written as a holomorphic function over
the base. The above equation describes three double degenerations, that is, three strings of
tension twice the basic unit. In the limit when the strings are on top of one another, we obtain
what is known (according to the Kodaira classification) as a $D_4$ singularity with deficit
angle $180^\circ$.

 The monodromy of the fiber around this singularity
is described by
\be
  \mathcal{M}_{D_4} = \left( \begin{array}{rr}
  -1 & 0  \\
  0 & \ -1
  \end{array}
  \right)
\ee
acting on $\binom{\omega_1}{\omega_2}$with $ \tau \equiv {\omega_1\over \omega_2} $. This
monodromy decomposes into that of six elementary strings which are mutually
non-local\footnote{For explicit monodromies for the six strings, see \cite{Gaberdiel:1997ud}.}.

This can be generalized to more than six strings using the Weierstrass equation
\be\label{weierstrasseqn}
  y^2 = x^3 + f(z)x + g(z)
\ee
The modular parameter of the torus is determined by
\be
  j(\tau(z)) = \frac{4 f^3}{4f^3+27g^2}
\ee
Whenever the numerator vanishes, $\tau=\tau_{\IZ_6}$ and we are at an orbifold point. We see
however that it is a triple root of $f^3$ and no orbifolding of the fiber is necessary. The same
applies for the $\IZ_4$ points. The strings are located where $\tau_2 \rightarrow \infty$ that
is where the {\it modular discriminant} $\Delta\equiv 4f^3+27g^2$ vanishes.
Note that the monodromy of the fibers around a smooth point is automatically the identity
in such a construction.


\vskip 0.5cm \noindent {\bf Kodaira classification.} Degenerations of elliptic fibrations have
been classified according to
their monodromy
by Kodaira.
For convenience, we summarize the result
in the following table
\cite{Bershadsky:1996nh}:

\vskip 0.5cm \vskip 0cm

\begin{center}
\begin{tabular}{|c | c | c | c | c | }
\hline  {\bf ord(f)} &  {\bf ord(g)}  &  {\bf ord($\Delta$)} &  {\bf monodromy}  &  {\bf
singularity}
\\ \hline
  $\ge 0$ & $\ge 0$ & 0 & {\footnotesize $\mtwo{1}{0}{0}{1}$} & none \\ \hline
  $0 $ & $ 0$ & $n$ & {\footnotesize $\mtwo{1}{n}{0}{1}$} & $A_{n-1}$ \\ \hline
  $\ge 1 $ & $1$ & $2$ & {\footnotesize $\mtwo{1}{1}{-1}{0}$} & none \\ \hline
  $1$ & $\ge 2$ & $3$ & {\footnotesize $\mtwo{0}{1}{-1}{0}$} & $A_{1}$ \\ \hline
  $\ge 2$ & $2$ & $4$ & {\footnotesize $\mtwo{0}{1}{-1}{-1}$} & $A_{2}$ \\ \hline
  $2$ & $\ge 3$ & $n+6$ & {\footnotesize $\mtwo{-1}{-n}{0}{-1}$} & $D_{n+4}$ \\ \hline
  $\ge 2$ & $3$ & $n+6$ & {\footnotesize $\mtwo{-1}{-n}{0}{-1}$} & $D_{n+4}$ \\ \hline
  $\ge 3$ & $4$ & $8$ & {\footnotesize $\mtwo{-1}{-1}{1}{0}$} & $E_{6}$ \\ \hline
  $3$ & $\ge 5$ & $9$ & {\footnotesize $\mtwo{0}{-1}{1}{0}$} & $E_{7}$ \\ \hline
  $\ge 4$ & $5$ & $10$ & {\footnotesize $\mtwo{0}{-1}{1}{1}$} & $E_{8}$ \\ \hline
\end{tabular}
\end{center}

\clearpage

\begin{figure}[ht]
\begin{center}
  \includegraphics[totalheight=3.0cm,angle=0,origin=c]{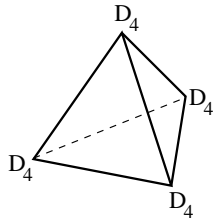}
  \caption{Flat $S^2$ base constructed from four triangles: base of $K3$ in the $\IZ_2$ orbifold limit.}
  \label{tetrah}
\end{center}
\end{figure}

\vskip 0.5cm \noindent {\bf Constructing K3.} One can construct a compact example
where the fiber experiences 24 $A_0$ degenerations.
In the Weierstrass description (\ref{weierstrasseqn}),
this means that $f$ has degree 8, $g$ has degree 12, and $\Delta$ has degree 24.
This is the semi-flat description of a $K3$ manifold.
In a certain limit where we group the strings
into four composite $D_4$ singularities, the base is flat and the total space becomes
$T^4/\IZ_2$. The base can be obtained by gluing four flat triangles as seen in \fref{tetrah}. At
each $D_4$ degeneration, the base has $180^\circ$ deficit angle which adds up to $4\pi$ and
closes the space into a flat sphere with the curvature concentrated at four points.

\begin{figure}[ht]
\begin{center}
  \includegraphics[totalheight=3.5cm,angle=0,origin=c]{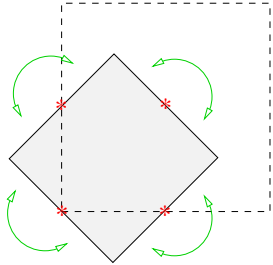}
  \caption{Base of the $T^4/\IZ_2$ orbifold. The $\IZ_2$ action inverts the base coordinates and has four fixed points denoted by red stars.
  They have $180^\circ$ deficit angle. As the arrows show, one has to fold the diagram and this gives an $S^2$. }
  \label{t4fund}
\end{center}
\end{figure}



\vskip -0.5cm

As we have seen, in two dimensions the Weierstrass equation solves the problem of orbifold
points. In higher dimensions, we don't have this tool but we can still try to glue patches of
spaces in order to get compact solutions. Gluing is especially easy if the base is flat.
However, generically this is not the case. Having a look at the Einstein equation (\ref{einsteineqn}), we see that a
flat base can be obtained if $\tau(z)$ is constant. This happens in the case of $D_4$ and $E_n$
singularities.
Our discussion in this paper will (unfortunately) be restricted to these singularities.

The cosmic string metric is singular in the above semi-flat description. It must be slightly
modified in order to get a smooth Calabi-Yau metric for the total space. This will be discussed
in Appendix \ref{sfvs}.

\newpage
\subsection{Three dimensions}

\label{threedim}

In two dimensions, the only smooth compact Calabi-Yau is the $K3$ surface. In three dimensions,
there are many different spaces and therefore the situation is much more complicated. The SYZ
conjecture \cite{Strominger:1996it} says that every Calabi-Yau threefold which has a geometric
mirror, is a special Lagrangian $T^3$ fibration with possibly degenerate fibers at some points.
For the generic case, the base is an $S^3$. Without the special Lagrangian condition, the
conjecture has been well understood in the context of topological mirror symmetry
\cite{Gross:1999hc, Tomasiello:2005bp}. There, the degeneration loci form a (real) codimension
two subset in the base. A graph $\Gamma$ is formed by edges and trivalent vertices. The fiber
suffers from monodromy around the edges. This is specified by a homomorphism
\be
  M: \ \pi_1(S^3 \setminus \Gamma) \longrightarrow SL(3,\IZ)
\ee
There are two types of vertices which contribute $\pm 1$ to the Euler character of the total
space\footnote{These positive and negative vertices are also called type (1,2) / type (2,1)
\cite{Gross:1999hc} or type III / type II \cite{Ruan1} vertices by different authors. For an
existence proof of metric on the vertex, see \cite{Loftin:2004qu}.}. At the vertices, the
topological junction condition relates the monodromies of the edges.

One of the most studied non-trivial Calabi-Yau spaces is the quintic in $\IP^4$. However, even
the topological description of this example is fairly complicated \cite{Gross:1999hc}. The
topological construction contains $250+50$ vertices and $450$ edges in the $S^3$ base.


Constructing not only topological, but {\it special Lagrangian} SYZ fibrations is a much harder
task. In fact, it is expected that away from the semi-flat limit, the real codimension two
singular loci in the base get promoted to codimension one singularities, \ie surfaces in three
dimensions. These were termed ribbon graphs \cite{Joyce:2000ru} and their description remains
elusive.

\vskip 0.5cm \noindent {\bf A compact orbifold example.} In the following, we will describe the
singular $T^6 / \IZ_2 \times \IZ_2$ orbifold in the SYZ fibration picture. One starts with $T^6$
that is a product of three tori with complex coordinates $z_i$. Without discrete torsion, the
orbifold action is generated by the geometric transformations,
\bea
  \alpha: (z_1, z_2, z_3) \mapsto  (-z_1, -z_2, z_3) \\
  \beta: (z_1, z_2, z_3) \mapsto  (-z_1, z_2, -z_3)
\eea
These transformations have unit determinant and thus the resulting space may be resolved into a
smooth Calabi-Yau manifold.

In order to obtain a fibration structure, we need to specify the base and the fibers. For the
base coordinates, we choose $x_i \equiv \textrm{Re}(z_i)$ and for the fibers $y_i \equiv
\textrm{Im}(z_i)$. Under the orbifold action, fibers are transformed into fibers and they don't
mix with the base\footnote{It is much harder in the general case to find a fibration that
commutes with the group action.}.

\begin{figure}[ht]
\begin{center}
  \includegraphics[totalheight=4.5cm,angle=0,origin=c]{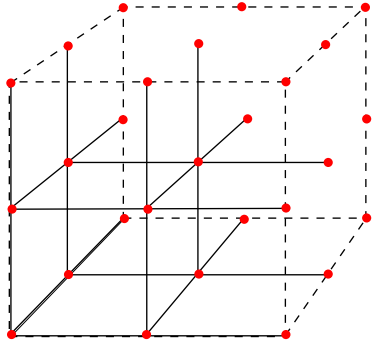}
  \caption{Singularities in the base of $T^6 / \IZ_2 \times \IZ_2$. The big dashed cube is the original $T^3$ base.
  The orbifold group generates the singular lines as depicted in the figure. The red dots show the intersection points of these edges.}
  \label{singut6z2z2}
\end{center}
\end{figure}

\vskip 0.5cm \noindent {\bf Degeneration loci in the base.} The base originally is a $T^3$. What
happens after orbifolding? If we fix, for instance, the $x_3$ coordinate, then the orbifold
action locally reduces to $\alpha$ (since the other two non-trivial group elements change
$x_3$). This means that we simply obtain four fixed points in this slice of the base. This is
exactly analogous to the $T^4/\IZ_2$ example. The fixed points correspond to $D_4$ singularities
with a deficit angle of $180^\circ$. As we change $x_3$, we obtain four parallel edges in the
base. By keeping instead $x_1$ or $x_2$ fixed, we get perpendicular lines corresponding to
conjugate $D_4$s whose monodromies act on another $T^2$ in the $T^3$ fiber. Altogether, we get
$3\times 4$ lines of degeneration as depicted in \fref{singut6z2z2}. These edges meet at
(half-)integer points in the $T^3$ base.

\begin{figure}[ht]
\begin{center}
  \includegraphics[totalheight=6.5cm,angle=0,origin=c]{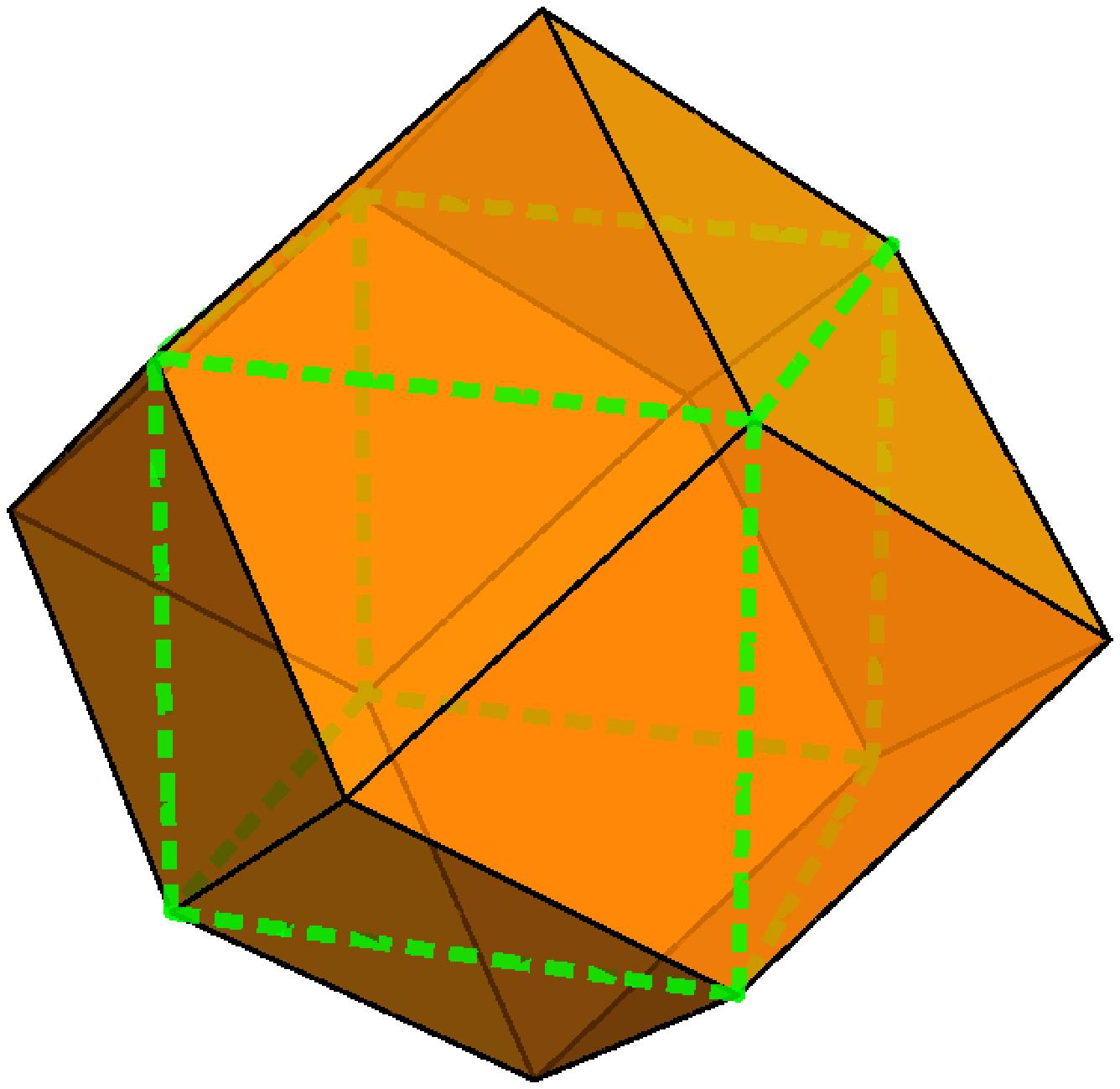} \qquad
  \includegraphics[totalheight=6.5cm,origin=c]{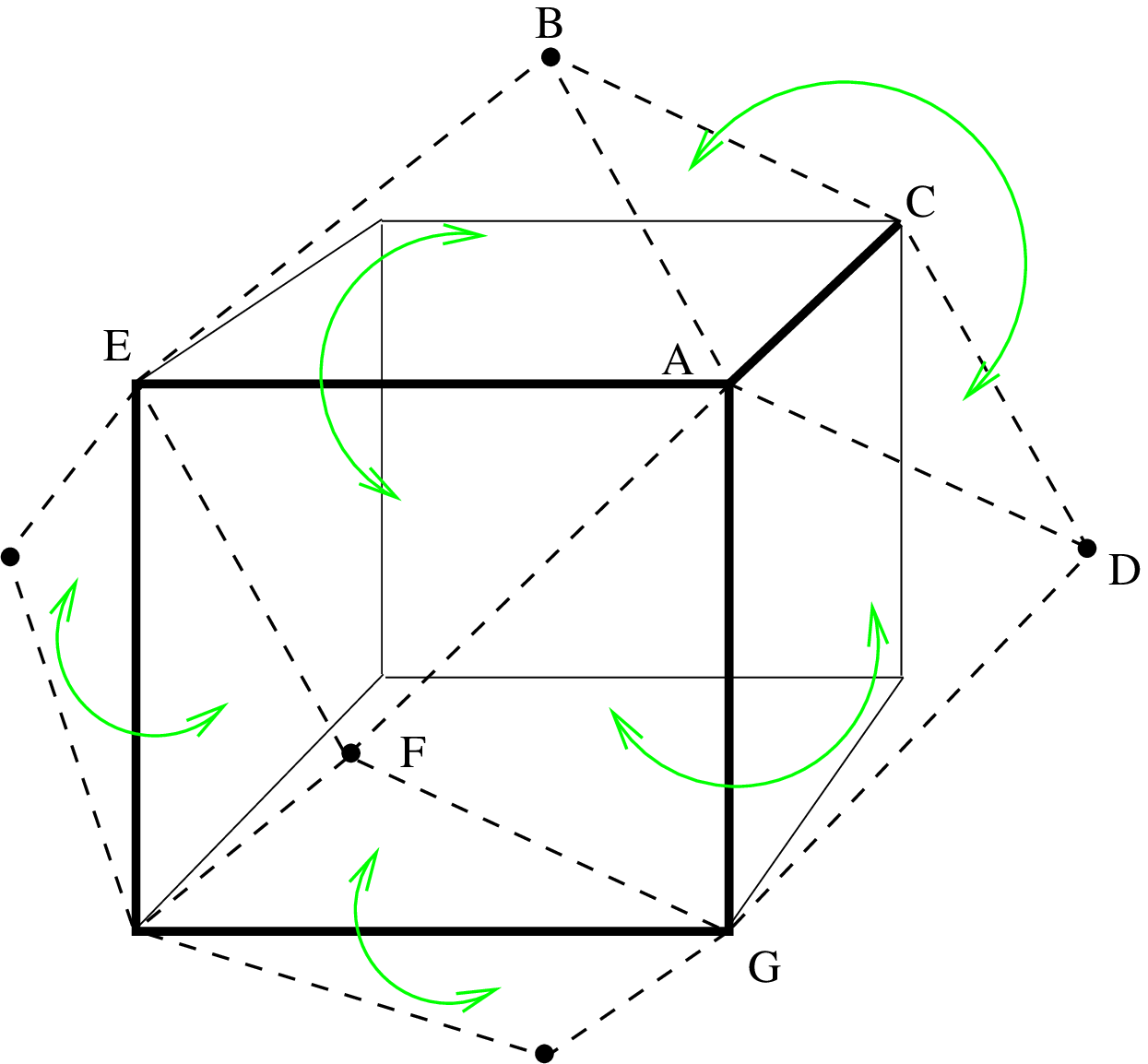}
  \caption{(i) Rhombic dodecahedron: fundamental domain for the base of $T^6/\IZ_2\times\IZ_2$. Six pyramids are glued on top of the faces of a cube.
  Neighboring pyramid triangles give rhombi since the vertices are coplanar (\eg $ABCD$).  (ii) The $S^3$ base can be constructed by
  identifying triangles as shown by the arrows. After gluing, the deficit angle around cube edges is $180^\circ$ which is appropriate for a
  $D_4$ singularity. The dihedral angles of the dashed lines are $120^\circ$ and since three of
them are glued together, there is no deficit angle. The tips of the
  pyramids get identified and the space finally becomes an $S^3$.}
  \label{cube_fold}
\end{center}
\end{figure}

Some parts of the base have been identified by the orbifold group. We can take this into account
by a folding procedure which we have already seen for $T^4 / \IZ_2$. The degeneration loci are
the edges of a cube. The volume of this cube is $\frac{1}{8}$ of the volume of the original
$T^3$. The base can be obtained by gluing six pyramids on top of the faces (see
\fref{cube_fold}). The top vertices of these pyramids are the reflection of the center of the
cube on the faces and thus the total volume is twice that of the cube. This polyhedron is a
Catalan solid\footnote{Catalan solids are duals to Archimedean solids which are convex polyhedra
composed of two or more types of regular polygons meeting in identical vertices. The dual of the
rhombic dodecahedron is the cuboctahedron.}: the {\it rhombic dodecahedron}. (Note that one can
also construct the same base by gluing two separate cubes together along their faces.)

In order to have a compact space, we finally glue the faces of the pyramids to neighboring faces
(see the right-hand side of \fref{cube_fold}). This is analogous to the case of $T^4 / \IZ_2$
where triangles were glued along their edges (\fref{t4fund}).

\newpage

\begin{figure}[ht]
\begin{center}
  \includegraphics[totalheight=4cm,angle=0,origin=c]{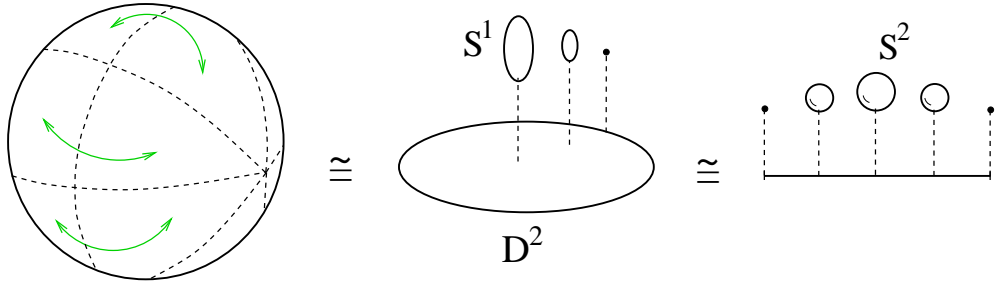}
  \caption{The base of $T^6/\IZ_2\times\IZ_2$ is homeomorphic to a three-ball with an $S^2$ boundary which has to be folded as shown in the figure.}
  \label{cube_s3}
\end{center}
\end{figure}

\vskip 0.5cm \noindent {\bf The topology of the base.} The base is an $S^3$ which can be seen as
follows\footnote{We thank A. Tomasiello for help in proving this.}. First fold the three rombi
$ABCD$, $AFGD$ and $ABEF$, and the corresponding three on the other side of the fundamental
domain. Then, we are still left with six rhombi that we need to fold. It is not hard to see that
the problem is topologically the same as having a $B^3$ ball with boundary $S^2$. Twelve
triangles cover the $S^2$ and we need to glue them together as depicted in \fref{cube_s3}. This
operation is the same as taking the $S^2$ and identifying its points by an $x\mapsto -x$ flip.
This on the other hand, exhibits the space as an $S^1$ fibration over $D^2$. The fiber vanishes
at the boundary of the disk. This is further equivalent to an $S^2$ fibration over an interval
where the fiber vanishes at both endpoints. This space is simply an $S^3$. The degeneration loci
are on the $S^2$ equator of this $S^3$ base and form the edges of the cube.

\begin{figure}[ht]
\begin{center}
  \includegraphics[totalheight=3.5cm,angle=0,origin=c]{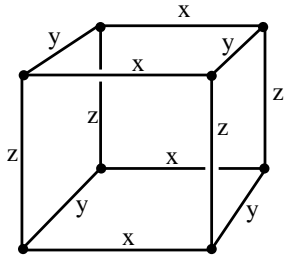}
  \caption{Monodromies for the edges.}
  \label{cube_mono}
\end{center}
\end{figure}

\vskip 0.5cm \noindent {\bf Edges and vertices.} The monodromies of the edges are shown in
\fref{cube_mono}. The letters on the degeneration edges denote the following $SL(3)$
monodromies:
\be
  x =
  \left( \begin{array}{rrr}
  1 & 0 & 0  \\
  0 & -1 & 0  \\
  0 & 0 & -1
  \end{array}
  \right)
  \quad
  y =
  \left( \begin{array}{rrr}
  -1 & 0 & 0  \\
  0 & \ 1 & 0  \\
  0 & 0 & -1
  \end{array}
  \right)\quad
  z =
  \left( \begin{array}{rrr}
  -1 & 0 & 0  \\
  0 & -1 & 0  \\
  0 & 0 & \ 1
  \end{array}
  \right)
\ee

This orbifold example contained $D_4$ strings. These are composite edges made out of six
``mutually non-local'' elementary edges. The edges have $180^\circ$ deficit angle around them
which is $6\times \frac{\pi}{6}$ where $\frac{\pi}{6}$ is the deficit angle of the elementary
string.

\comment{ This can be easily seen as the dihedral angles of the cube are $90^\circ$ and we glued
two cubes together ($180^\circ = 360^\circ - 2\times 90^\circ$). There are eight trivalent
(composite) vertices. Using the fact that the solid angle of a vertex can be obtained from the
neighboring dihedral angles as
\be
  \theta = \alpha + \beta + \gamma - \pi
\ee
or just simply noticing that the cube extends into an octant from any of its vertices, we arrive
at the conclusion that the {\it solid deficit angle} of a vertex is $3\pi$. }

Note that the base is flat. This made it possible to easily glue the fundamental cell to itself
yielding a compact space. Since the edges around any vertex meet in a symmetric way, the
cancellation of forces is automatic.

There are other spaces that one can describe using $D_4$ edges and the above mentioned composite
vertices. Some examples are presented in Section \ref{examplesec}. The strategy is to make a
compact space by gluing polyhedra like the above described cubes, then make sure that the
dihedral deficit angles are appropriate for the $D_4$ singularity.

\newpage
\subsection{Flat vertices}

Codimension two degeneration loci meet at vertices in the base. In the generic case, these are
trivalent vertices of elementary strings. Such strings have $30^\circ$ deficit angle around them
measured at infinity. This creates a solid deficit angle around the vertex.

In some cases when composite singularities meet, the base is flat and the vertex is easier to
understand. In particular, the total deficit angle arises already in the vicinity of the
strings. An example was given in Section \ref{threedim} where composite vertices arise from the
``collision'' of three $D_4$ singularities (see \fref{cube_fold}). The singular edges have a
deficit angle $\pi$. The vertex can be constructed by taking an octant of three dimensional
space and gluing another octant to it along the boundary walls. The curvature is then
concentrated in the axes. The solid angle can be computed as twice the solid angle of an octant.
This gives $\pi$ (or a deficit solid angle of $3\pi$).

\begin{figure}[ht]
\begin{center}
  \includegraphics[totalheight=4cm,angle=0,origin=c]{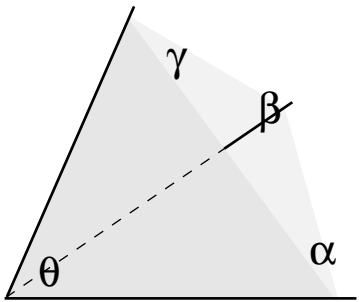}
  \caption{The solid angle at the apex is determined by the dihedral angles between the planes.}
  \label{conepic}
\end{center}
\end{figure}

In the general (flat) case, a composite vertex may be described by gluing two identical cones
(the analogs of octants). Such a cone is shown in \fref{conepic}. Note that the solid angle
spanned by three vectors is given by the formula
\be
  \theta = \alpha + \beta + \gamma - \pi
\ee
where $\alpha$, $\beta$ and $\gamma$ are the dihedral angles at the edges. This can be used to
compute the solid angle around a composite vertex.

\begin{figure}[ht]
\begin{center}
  \includegraphics[totalheight=3cm,angle=0,origin=c]{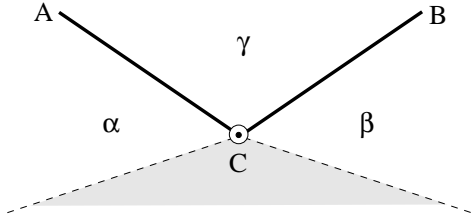}
  \caption{Flat vertex. $A$, $B$ and $C$ are singular edges. $C$ is pointing towards the reader. The dashed lines must be glued together
  to account for the deficit angle around $C$.}
  \label{balpic}
\end{center}
\end{figure}

The singular edges have a tension which is proportional to the deficit angle around them. This
leads to the problem of force balance. In \fref{balpic}, a flat vertex is shown. The two solid
lines ($A$ and $B$) are degeneration loci. The third edge ($C$) is pointing towards the reader
as indicated by the arrow head. The deficit angle around $C$ is shown by the shaded area. In the
weak tension limit (where we rescale the deficit angles by a small number), one condition for
force balance is that these edges are in a plane. (Otherwise, energy could be decreased by
moving the vertex.) This can be generalized for almost flat spaces by ensuring that $\alpha +
\beta = \gamma$. This is automatic when we construct the neighborhood of a vertex by gluing two
identical cones\footnote{In the weak tension limit, the two identical cones almost fill two
half-spaces. The slopes of the edges are dictated by the tensions as in \cite{Sen:1997xi}. We
leave the proof to the interested reader.}.

\begin{figure}[ht]
\begin{center}
  \includegraphics[totalheight=3.0cm,angle=0,origin=c]{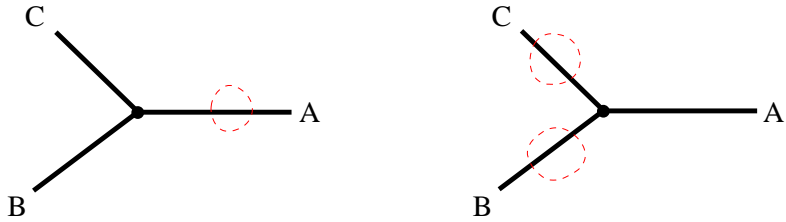}
  \caption{Junction condition for monodromies. The red loop around $A$ can be smoothly deformed into two loops around $B$ and $C$.}
  \label{jconpic}
\end{center}
\end{figure}

Another problem to be solved is related to the fiber monodromies. These can be described by
matrices $A$, $B$ and $C$ (see \fref{jconpic}). The loop around one of the edges (say $A$) can
be smoothly deformed into the union of the other two ($B, C$). This gives the monodromy
condition\footnote{Since monodromy matrices do not generically commute, it is important to keep
track of the branch cut planes.} $ABC=1$.

Some composite strings can be easier described than elementary ones because the base metric can
be flat around them. Such singularities are $D_4$, $E_6$, $E_7$ and $E_8$ with deficit angles
$\pi$, $4\pi/3$, $3\pi/2$ and $5\pi/3$, respectively \cite{Greene:1989ya}. Vertices where
composite lines meet can also be easily found by studying flat $\IC^3$ orbifolds. Here we list
some of the vertices that will later arise in the examples.

\vskip 0.5cm \vskip 0cm

\begin{table}[htdp]
\begin{center}
\begin{tabular}{|c | c | c | c |}
\hline  {\bf orbifold group} &  {\bf colliding singularities} &  {\bf solid angle}  \\
\hline
  $\IZ_2 \times \IZ_2$  &  $D_4 - D_4  - D_4$  & $\pi$ \\
  $\IZ_2 \times \IZ_4$  &  $D_4 - D_4  - E_7$  & $\pi/2$ \\
  $\Delta_{12}$  &  $D_4 - E_6  - E_6$  & $\pi/3$ \\
  $\Delta_{24}$  &  $D_4 - E_6  - E_7$  & $\pi/6$ \\
\hline
\end{tabular}
\end{center}
\caption{Examples for composite vertices.}
\end{table}

\newpage

We have already seen the  $\IZ_2 \times \IZ_2$ vertex in Section \ref{threedim}. If the vertex
is located at the origin, then the strings are stretched along the coordinate axes,
\be
  D^{(1)}_4 : (1,0,0) \qquad D^{(2)}_4 : (0,1,0) \qquad D^{(3)}_4 : (0,0,1)
\ee
The second example is generated by
\bean
  \alpha : (z_1, z_2, z_3) & \mapsto & (-z_3, z_2, z_1) \\
  \beta : (z_1, z_2, z_3) & \mapsto & (z_1, -z_2, - z_3)
\eean
It contains different colliding singularities. Their directions are given by
\be
  D^{(1)}_4 : (1,0,0) \qquad D^{(2)}_4 : (1,0,1) \qquad E_7 : (0,1,0)
\ee
The $\Delta_{12}$ group has $(\IZ_2)^2$ and $\IZ_3$ subgroups. It is generated by
\bean
  \alpha: \, (z_1, z_2, z_3) &\mapsto & (z_2, z_3, z_1) \\
  \beta: \, (z_1, z_2, z_3) &\mapsto & (-z_1, -z_2, z_3)
\eean
The strings directions are
\be
  D_4 : (1,0,0) \qquad E^{(1)}_6 : (1,1,1) \qquad E^{(2)}_6 : (1,1,-1)
\ee
The last example is generated by combining $\IZ_3$ and $\IZ_4$ generators,
\bean
  \alpha : (z_1, z_2, z_3) & \mapsto & (z_2, z_3, z_1) \\
  \beta : (z_1, z_2, z_3) & \mapsto & (-z_2, z_1, z_3)
\eean
which generate the $\Delta_{24}$ group. The direction of the strings are the following,
\be
  D_4 : (1,1,0) \qquad E_6 : (1,1,1) \qquad E_7 : (1,0,0)
\ee

This is not an exhaustive list; a thorough study based on the finite subgroups of $SU(3)$
\cite{Fairbairn} would be interesting.

\newpage
\section{Stringy monodromies}

In this section, we wish to extend the discussion by including the full perturbative duality
group of type II string theory on $T^3$ in the possible set of monodromies. We will find that
this duality group can be interpreted as the geometric duality group of an auxiliary $T^4$. The
extra circle is to be distinguished from the M-theory circle but it is related to it by a
U-duality transformation.

For simplicity, the Ramond-Ramond field strengths will be turned off. This allows us to use
perturbative dualities only. However, in moduli stabilization these fields play an important
role. In fact, in the Appendices \ref{hwapp} and \ref{t5app}, we use U-duality
\cite{Hull:1994ys} monodromies which act on RR-fields in order to describe two familiar
phenomena.

From the worldsheet point of view, string compactifications are expected to be typically
non-geometric, since the 2d CFT does not necessarily have a geometric target space. Even though
we construct our examples directly based on intuition from supergravity, they will have a
worldsheet description as modular invariant asymmetric orbifolds.

For other related works on non-geometric spaces, see \cite{Kumar:1996zx, Hellerman:2002ax,
Kaloper:1999yr, Hitchin:2004ut, Gualtieri:2003dx, Grana:2004bg, Kachru:2002sk, Gurrieri:2002wz,
Grana:2006kf, Lawrence:2006ma, Dabholkar:2005ve, Hull:2004in, Flournoy:2004vn, Hull:2006va,
Shelton:2005cf, Shelton:2006fd, Becker:2006ks} and references therein.

In the following, we study the perturbative duality group in Type IIA string theory compactified
on a flat three-torus. We gain intuition by studying the reduced 7d Lagrangian of the
supergravity approximation. Finally we discuss how U-duality relates non-geometric
compactifications to $G_2$ manifolds in M-theory which will be fruitful when constructing
examples in the next section.

\subsection{Reduction to seven dimensions}

\noindent {\bf Action and symmetries.} Let us consider the bosonic sector of (massless) 10d Type
IIA supergravity,
\be
  S_\textrm{IIA} = S_\textrm{NS}+S_\textrm{R}+S_\textrm{CS}
\ee
where
\be
  S_\textrm{NS} = \frac{1}{2\kappa^2_{10}} \int d^{10}x \sqrt{-g}  \, e^{-2\phi} (R+4\partial_\mu
  \phi \partial^\mu \phi -\frac 1 2 |H_3|^2)
\ee
\be
  S_\textrm{R} = -\frac{1}{4\kappa^2_{10}} \int d^{10}x \sqrt{-g}  \, (|F_2|^2+|\tilde F_4|^2)
\ee
and the Chern-Simons term is
\be
  S_\textrm{CS} = -\frac{1}{4\kappa^2_{10}} \int  B \wedge F_4 \wedge F_4
\ee
with $\tilde F_4 = dA_3 -A_1 \wedge dB$ and $\kappa^2_{10}=\kappa^2_{11}/2\pi R$.

First we set the RR fields to zero\footnote{This can be done consistently since the
$(-1)^{F_L}$ symmetry forbids a tadpole for any RR field.}. This truncates the theory to the NS
part which is identical to the IIB $S_\textrm{NS}$ action. Compactifying Type IIA on a flat
$T^3$ yields the perturbative T-duality group $SO(3,3,\IZ)$ which acts on the coset
$SO(3,3,\IR)/SO(3)^2$.

\newpage
The equivalences of Lie algebras
\be
  \mathfrak{so}(3,3) \cong \mathfrak{sl}(4)
\ee
\be
  \mathfrak{so}(3) \oplus \mathfrak{so}(3) \cong \mathfrak{su}(2) \oplus \mathfrak{su}(2) \cong \mathfrak{so}(4)
\ee
enable us to realize the T-duality group as an $SL(4,\IZ)$ action on $SL(4,\IR)/SO(4)$. This
latter space is simply the moduli space of a flat $T^4$ with constant volume. Therefore,
we can think of the T-duality group as the mapping class group of an auxiliary four-torus of
unit volume. What is the metric on this $T^4$
in terms of the data of the $T^3$? To answer this question, we have to study the
Lagrangian.


\vskip 0.5cm \noindent {\bf Reduction to seven dimensions.} One obtains the following terms
after reduction on $T^3$ \cite{Maharana:1992my} (see Appendix \ref{redapp} for more details and
notation)
\be
  S=\int dx \sqrt{-g} e^{-\phi} \mathcal{L}
\ee
with $\mathcal{L}=\mathcal{L}_1+\mathcal{L}_2+\mathcal{L}_3+\mathcal{L}_4$ and
\bea
  \mathcal{L}_1 &=& R + \partial_\mu \phi \partial^\mu \phi \\
  \mathcal{L}_2 &=& \frac{1}{4}(\partial_\mu G_{\alpha\beta} \partial^\mu G^{\alpha\beta} -
  G^{\alpha\beta}G^{\gamma\delta} \partial_\mu B_{\alpha\gamma} \partial^\mu B_{\beta\delta}) \\
   \mathcal{L}_3 &=& -\frac 1 4 g^{\mu\rho}g^{\nu\lambda} (G_{\alpha\beta} F^{(1)\alpha}_{\mu\nu}
  F^{(1)\beta}_{\rho\lambda} + G^{\alpha\beta} H_{\mu\nu\alpha} H_{\rho\lambda\beta}) \\
  \mathcal{L}_4 &=& -\frac{1}{12}H_{\mu\nu\rho} H^{\mu\nu\rho}
\eea
The relation of these fields and the ten dimensional fields are presented in Appendix
\ref{redapp}. In order to see the $SO(d,d,\IZ)$ symmetry, one introduces the symmetric positive
definite $2d\times 2d$ matrix
\be
  M=\left( \begin{array}{cc}
 G^{-1}  & G^{-1}B \\
 BG^{-1} & \ \ G-BG^{-1}B
  \end{array}
  \right) \in SO(3,3)
\ee
The kinetic terms $ \mathcal{L}_2$ can be written as the $\sigma-$model Lagrangian
\be
   \mathcal{L}_2 = \frac 1 8 \textrm{Tr}(\partial_\mu M^{-1} \partial^\mu M)
\ee
The other terms in the Lagrangian are also invariant under $SO(3,3)$.

\vskip 0.5cm \noindent {\bf The SL(4) duality symmetry and ``N-theory''.} Let us now put the
bosonic action in a manifestly $SL(4)$ invariant form (see \cite{Brace:1998xz}). Rewrite
$\mathcal{L}_2$ as
\be
  \mathcal{L}_2= \frac 1 8 \textrm{Tr}(\partial_\mu M^{-1} \partial^\mu M) = \frac 1 4 \textrm{Tr}(\partial_\mu N^{-1} \partial^\mu
  N), 
\ee
where we introduced the symmetric $SL(4)$ matrix\footnote{This matrix parametrizes the eight
complex structure moduli, and one \kahler modulus of $T^4$. }
\be
  N_{4\times 4}=(\textrm{det} \, G)^{-1/2} \left( \begin{array}{cc}
 G  & G\vec b \\
 \vec b^T G & \ \ \textrm{det} \, G + \vec b^T G  \vec b
  \end{array}
  \right)
  \label{nmetric}
\ee
\be
  B_{ij}=\epsilon_{ijk} b_k \qquad     b_i=\frac 1 2 \epsilon_{ijk} B_{jk}.
\ee
The equality of the Lagrangians can be checked by lengthy algebraic manipulations (or a computer
algebra software). We included the Hodge-dualized B-field in the metric as a Kaluza-Klein
vector. The inverse of $N$ is
\be
  N^{-1}=(\textrm{det} \, G)^{-1/2} \left( \begin{array}{cc}
 (\textrm{det} \, G)G^{-1} + \vec b^T b  & \ \ -\vec b \\
 -\vec b^T  & \ \ 1
  \end{array}
  \right).
  \label{nmetrici}
\ee
Keeping $N$ symmetric, the Lagrangian is invariant under the global transformation,
\be
  N(x) \mapsto U^T N(x) U, \ \ \textrm{with} \ U\in SL(4).
\ee

A useful device for interpreting $N$ is the following. Note that we would get the exact same
bosonic terms of $\mathcal{L}_{1}$ and $\mathcal{L}_{2}$, if we were to reduce an eleven
dimensional classical theory to seven dimensions. This theory is given by the Einstein-Hilbert
action plus a scalar, the ``11d dilaton''\footnote{We will denote the extra dimension by
$x^{10}$. This is not to be confused with the M-theory circle denoted by $x^{11}$.}
\be
  S=\int d^{11}x \sqrt{-\tilde g} \,  e^{-\phi}  (R(\tilde g) + \partial_\mu \phi\partial^\mu \phi)
  \label{ntheory}
\ee
This Lagrangian contains no B-field. The description in terms of (\ref{ntheory}) is only useful
when $\mathcal{L}_3$ and $\mathcal{L}_4$ vanish. This means that $ F^{(1)\alpha}_{\mu\nu} =
H_{\mu\nu\alpha} = H_{\mu\nu\rho} = 0$. Since the size of $T^4$ is constant, its dimensions are
not treated on the same footing as the three geometric fiber dimensions. It is similar to the
situation in F-theory \cite{Vafa:1996xn}, where the area of the $T^2$ is fixed and the \kahler
modulus of the torus is not a dynamical parameter.

We have seen that the matrix $N$ can be interpreted as a semi-flat metric on a $T^4$ torus
fiber. Part of this torus is the original $T^3$ fiber and the overall volume is set to one. The
T-duality group $SO(3,3,\IZ)$ acts on $T^4$ in a geometric way. This means that we can hope to
study non-geometric compactifications by studying purely geometric ones in higher dimension.

\subsection{The perturbative duality group}

In the previous section, we have transformed the coset space $SO(3,3)/SO(3)^2$ into
$SL(4)/SO(4)$ via Eq. (\ref{nmetric}). We also would like to see how the discrete T-duality
group $SO(3,3,\IZ)$ maps to $SL(4,\IZ)$. We will denote the $SO(3,3)$ matrices by $\mathcal{Q}$,
and the $SL(4)$ matrices by $\mathcal{W}$.

\begin{table}[htdp]
\begin{center}
\begin{tabular}{|c | c | c | c |}
\hline  {\bf SO(3,3)} &  {\bf SL(4)} &  {\bf dim} &  {\bf examples} \\ \hline
  \textrm{spinor} & \textrm{fundamental} & 4 & \textrm{RR fields} \\
  \textrm{fundamental} & \textrm{antisym. tensor} & 6 & \textrm{momenta \& winding} \\
\hline
\end{tabular}
\end{center}
\caption{The two basic representations of the duality group.}
\end{table}

\vskip 0.5cm \noindent {\bf Generators of SO(3,3,Z).} It was shown in \cite{Schwarz:1998qj} that
the following $SO(3,3,\IZ)$ elements generate the whole group
\be
\mathcal{Q}_1(n) = \left( \begin{array}{c|c}
 \mathbbm{1}_{3\times 3}  & n    \\ \hline
 0 &  \ \mathbbm{1}_{3\times 3}
  \end{array}
  \right)
  \qquad
\mathcal{Q}_2(R) =   \left( \begin{array}{c|c}
 R \ & 0    \\ \hline
 0 \ & \  (R^{-1})^T
  \end{array}
  \right)
  \qquad
\mathcal{Q}_3 = \left( \begin{array}{ccc|ccc}
 0 &   &   & \ 1 &   &      \\
   & 0 &   &   & 1 &      \\
   &   & 1 \ &   &   & 0    \\
 \hline
 1 &   &   & \ 0 &   &      \\
   & 1 &   &   & 0 &      \\
   &   & 0 \ &   &   & 1
  \end{array}
  \right)
\ee
where $n^T = -n$, $\textrm{det}\, R = \pm 1$. The first two matrices correspond to a change of
basis of the compactification lattice. The last matrix is T-duality along the $x^7-x^8$
coordinates. Instead of using $\mathcal{Q}_3$ directly, we combine double T-duality with a
$90^\circ$ rotation. This gives the $SO(3,3)$ matrix
\be
 \mathcal{  \widetilde Q}_3 =
  \left( \begin{array}{ccc|ccc}
  &   &   &   & -1  &      \\
  &   &   & \ 1 &   &      \\
  &   & 1  &   &   &      \\ \hline
  & 1  &   &   &   &      \\
-1  &   &   &   &   &      \\
  &   &   &   &   & 1
  \end{array}
  \right)
\ee

\vskip 0.5cm \noindent {\bf Generators of SL(4,Z).}  In the Appendix of \cite{Brace:1998ku}, it
was shown that the above matrices have an integral $4\times 4$ spinor representation and in fact
generate the entire $SL(4,\IZ)$. We now list the spinor representations corresponding to these
generators\footnote{Note that \cite{Brace:1998ku} uses a different basis for the spinors.}.

\begin{itemize}

\item
$\mathcal{ Q}_1(n)$ is mapped to matrices
\be
 \mathcal{  W}_1 (n) =
 \left( \begin{array}{rr|rc}
  1 &     &    & \ n_{23}  \\
    & 1 \ &    & \ n_{31}  \\ \hline
    &     & \ 1 & \ n_{12}  \\
    &     &    &  1
  \end{array}
  \right)
\ee
These are the generators corresponding to ``T'' transformations of various $SL(2)$
subgroups.

\item
$\mathcal{\widetilde Q}_3$ is mapped to
\be
 \mathcal{  \widetilde W}_3 =
 \left( \begin{array}{rr|rr}
  1 &     &     &    \\
   &  1 \ &   &    \\ \hline
   &   &   & -1  \\
   &   & \ 1 &
  \end{array}
  \right)
\ee
This corresponds to a modular ``S'' transformation. Note that $(\mathcal{  \widetilde W}_3)^2
\ne \mathbbm{1}$.

\item
When $\textrm{det} \, R = + 1$, the matrix $\mathcal{Q}_2(R)$ is mapped to the $SL(4,\IZ)$
matrix
\be
\mathcal{W}_2(R) = \left( \begin{array}{cc}
 R \ & 0    \\
 0 \ & 1
  \end{array}
  \right)
\ee
For symmetric $R$ matrices it coincides with the prescription of Eq. (\ref{nmetric}).

\item
The $\textrm{det} \, R = -1$ case is more subtle. Even though Type IIA string theory is parity
invariant, in the microscopic description reflecting an odd number of coordinates does not give
a symmetry by itself. Since this transformation flips the spinor representations $16
\leftrightarrow 16'$, it must be accompanied by an internal symmetry $\Omega$ which changes the
orientation of the world-sheet and thus exchanges the left-moving and right-moving spinors.

$SO(3,3)$ has maximal subgroup $S(O(3)\times O(3))$ and hence has two connected components
\cite{Gualtieri:2003dx}. Inversion of an odd number of coordinates is not in the identity
component. $SL(4,\IZ)$ is the double cover of the connected component of $SO(3,3,\IZ)$ only. We
must allow for $\textrm{det} \, \mathcal{W} = \pm 1$ to obtain $Spin(3,3, \IZ)$, the double
cover of the full $SO(3,3,\IZ)$. Then, the reflections of the $x^7$, $x^8$ or $x^9$ coordinates
have the following representations\footnote{The only non-trivial element in the center of
$SL(4)$ is $-\mathbbm{1}$. This sign may be attached to all the group elements not in the
identity component, giving an automorphism of $Spin(3,3, \IZ)$.}
\be
  \mathcal{W}_{I_7}=\textrm{diag}(-1,1,1,1) \qquad   \mathcal{W}_{I_8}=\textrm{diag}(1,-1,1,1) \qquad    \mathcal{W}_{I_9}=\textrm{diag}(1,1,-1,1)
\ee

Upon restriction to $GL(3) \subset SO(3,3)$, the $Spin(3,3)$ group is a trivial covering.

Ramond-Ramond fields transform in the spinor representation of the T-duality group\footnote{As
discussed in \cite{Brace:1998xz}, the fields that have simple transformation properties are
$C^{(3)} = A^{(3)} + A^{(1)} \wedge B$.}. Therefore they form fundamental $SL(4)$ multiplets,
for instance $ (C_7, C_8, C_9, C_{789})$. We can check the above representation for the
coordinate reflections. Reflection of say $x^7$ combined with a flip of the three-form field
gives
\be
  (C_7, C_8, C_9, C_{789}) \mapsto    (-C_7, C_8, C_9, C_{789})
\ee
which is precisely the action of $\mathcal{W}_{I_7}$.

\end{itemize}

\subsection{Embedding $SL(2)^2$ in $SL(4)$}

In order to get some intuition for the $SL(4)$ duality group that we discussed in the previous
section, we first look at the simpler case of $T^2$ compactifications. In this section we
describe how the T-duality group of $T^2$ compactifications can be embedded into the bigger
$SL(4)$ group.

In eight dimensions, the duality group is $SL(2)_\tau\times SL(2)_\rho$ with the first factor
acting on the $\tau$ complex structure of the torus and the second factor acting on
$\rho=b+iV/2$ where $b=\int_{T^2} B$ and $V$ is the volume of $T^2$. If we consider a two
dimensional base with complex coordinate $z$, then the equations of motion are satisfied if
$\tau(z)$ and $\rho(z)$ are holomorphic sections of $SL(2,\IZ)$ bundles. Monodromies of $\tau$
around branch points points describe the geometric degenerations of the fibration. Monodromies
of $\rho$, however, correspond to T-dualities and to the semi-flat description of NS5-branes. In
particular, if there is a monodromy $\rho \mapsto \rho+1$ around a degeneration point in the
base, then it implies $b\mapsto b+1$ which describes a unit magnetic charge for the B-field, \ie
an NS5-brane. The $\rho \mapsto -1/\rho$ monodromy on the other hand is a double T-duality along
the $T^2$ combined with a $90^\circ$ rotation.

 Let us denote the two-torus coordinates by $x^{7,8}$. In order to embed this $SL(2)\times SL(2)$ duality
group into the $SL(4)$ of $T^3$ compactifications, we need to further compactify on a
``spectator'' circle of size $L$. We denote its coordinate by $x^9$. The metric on $T^3$
($x^9-x^{10}-x^{11}$) is now
\be
G_{3\times 3}=\left( \begin{array}{cc|c}
 g_{11}  & g_{12} & \    \\
 g_{21}  & g_{22} & \    \\
\hline & & \ L^2
  \end{array}
  \right)
\ee
Then, one can construct the $4\times 4$ metric on $T^4$ by the prescription of (\ref{nmetric})
which gives
\be
  N=(\textrm{det} \,  g)^{-1/2}\left( \begin{array}{c|cc}
  \frac{1}{L} g_{2\times 2} & & \\
 \hline
      & \ L & Lb \\
      & \ Lb & L(\textrm{det} \,  g + b^2) \\
  \end{array}
  \right) \equiv
  \left( \begin{array}{cc}
 \frac{1}{L}\mathcal{T}_{2\times 2} & \   \\
  \  & L \mathcal{R}_{2\times 2}
  \end{array}
  \right)
\ee
with
\be
  \mathcal{T}=\frac{1}{\tau_2}\left( \begin{array}{cc}
 1 & \tau_1   \\
 \tau_1 & |\tau|^2
  \end{array}
  \right)
  \qquad
  \mathcal{R}=\frac{1}{\rho_2}\left( \begin{array}{cc}
 1 & \rho_1   \\
 \rho_1 & |\rho|^2
  \end{array}
  \right)
\ee
The $\sigma$-model Lagrangian
\be
 \textrm{Tr}(\partial_\mu N^{-1} \partial^\mu
  N)= -2\left( \frac{\partial_\mu \tau\partial^\mu \bar\tau}{\tau_2^2} + \frac{\partial_\mu \rho\partial^\mu \bar\rho}{\rho_2^2}\right)
\ee
indeed gives the familiar kinetic terms for the torus moduli (in seven dimensions).

We have seen how the metric and the B-field parametrize the relevant subset of the
$SL(4,\IR)/SO(4)$ coset space. The generators of the $SL(2,\IZ)\times SL(2,\IZ)$ duality group
are also mapped to elements in $SL(4,\IZ)$. We now verify that these images in fact give the
transformations that we expect.

\begin{itemize}

\item {\bf Geometric transformations}

These are simply generated by
\be
 T=\left( \begin{array}{cc}
 1 & 1   \\
 0 & 1
  \end{array}
  \right) \oplus
  \mathbbm{1}_{2\times 2}
  \quad \textrm{and} \quad
 S=\left( \begin{array}{cc}
 0 & -1   \\
 1 & 0
  \end{array}
  \right) \oplus
  \mathbbm{1}_{2\times 2}
\ee

They act on $g_{2\times 2}$ by conjugation with the non-trivial $SL(2)$ part as expected. The
determinant of $g$ stays the same. The first one is a Dehn-twist and the second one is a
$90^\circ$ rotation.

\item {\bf Non-geometric transformations}

The generators
\be
  T'=\mathbbm{1}_{2\times 2} \oplus \left( \begin{array}{cc}
 1 & 1   \\
 0 & 1
  \end{array}
  \right)
  \quad \textrm{and} \quad
 S'=\mathbbm{1}_{2\times 2} \oplus \left( \begin{array}{cc}
 0 & -1   \\
 1 & 0
  \end{array}
  \right)
\ee
correspond respectively to the shift of the B-field and to a double T-duality on $x^{7,8}$
combined with a $90^\circ$ rotation. The latter one has the $SL(4)$ monodromy
\be
  M=\left( \begin{array}{rr|rr}
 1 & \ 0 & \ 0 & 0  \\
 0 & 1  & 0 & 0  \\ \hline
 0 & 0 & 0 & -1  \\
 0 & 0 & 1 & 0
  \end{array}
  \right)
\ee

This is basically an exchange of the $x^{9}-x^{10}$ coordinates and it transforms the
$\mathcal{R}_{2\times 2}$ submatrix of $N$ into its inverse
\be
\mathcal{R}^{-1}=(\textrm{det} \,  g)^{-1/2}\left( \begin{array}{cc}
 \textrm{det}  \, g + b^2 & -b   \\
 -b & 1
  \end{array}
  \right)
\ee
After this double T-duality, the (geometric) metric on $T^3$  becomes
\be
G_{3\times 3} \mapsto  \widetilde G_{3\times 3}=\left( \begin{array}{c|r}
 \frac{1}{\textrm{det} g + b^2}\ g_{2\times 2}  &  0   \\
\hline  0 &  \ L^2
  \end{array}
  \right)
\ee
The B-field transforms as
\be
  b \mapsto \tilde b = -\frac{b}{\textrm{det} \, g + b^2}
\ee
The metric $g$ on $T^2$ changes, in particular if $b=0$, then the volume gets inverted. 
Since we exchanged the $x^9-x^{10}$ coordinates, one might have expected that this affects the
metric on $x^9$. However, we see that it remains the same as it should since it was only a
spectator circle.

\newpage

\item {\bf Left-moving spacetime fermion number: $(-1)^{F_L}$}

This is a {\it global} transformation which inverts the sign of the Ramond-Ramond fields. It
acts trivially on the vector representation of $SO(3,3)$ (which is the antisymmetric tensor of
$SL(4)$). It will be important since T-duality squares to $(-1)^{F_L}$.  In
\cite{Hellerman:2002ax}, its representation was determined,
\be
 \mathcal{M}_{(-1)^{F_L}}=\left( \begin{array}{cc}
 -1 & 0   \\
 0 & -1
  \end{array}
  \right) \oplus
   \left( \begin{array}{cc}
 -1 & 0   \\
 0 & -1
  \end{array}
  \right) \in SL(2)\times SL(2)
\ee
that is a $D_4$ monodromy combined with a $D'_4$ (\ie a conjugate $D_4$). This statement can be
proven as follows. Let us define complex coordinates
\be
  z_L = x^7_L + i x^8_L
\ee
\be
  z_R = x^7_R + i x^8_R
\ee
where $x_L$ and $x_R$ are the left- and right-moving components of the bosonic coordinates. We
denote a transformation
\be
  (z_L, z_R) \mapsto  (e^{\theta_L} z_L, \ e^{\theta_R} z_R)
\ee
by $\theta=(\theta_L, \theta_R)$. Then,
\be
  \theta_{D_4} = (-\pi, -\pi)
\ee
as it is a reflection of the bosonic coordinates. Moreover, we can use $D'_4 = S^2$ where $S$ is
a double T-duality with a $90^\circ$ rotation. We have
\be
  \theta_{S} = \underbrace{(-\pi,0)}_\textrm{double T-duality} +
  \underbrace{(\frac{\pi}{2},\frac{\pi}{2})}_{\textrm{$90^\circ$ rotation}} =  (-\frac{\pi}{2},\frac{\pi}{2})
\ee
from which we obtain
\be
  \theta_{D'_4} = 2\times \theta_{S} =  (-\pi, \pi)
\ee
Finally,
\be
 \theta_{D_4+D'_4} =   \theta_{D_4} + \theta_{D'_4} = (-2\pi,0)
\ee
which acts trivially on the bosons. However, it inverts the sign of the spinors from left movers
which is precisely the action of $(-1)^{F_L}$. Finally, it can be embedded into $SL(4)$ simply
as
\be
   \mathcal{M}_{(-1)^{F_L}} = \textrm{diag}(-1,-1,-1,-1)
\ee


\end{itemize}

\subsection{U-duality and $G_2$ manifolds}
\label{UdualityandGtwo}

We have seen that upon compactifying Type IIA on $T^3$, a $T^4$ torus emerges. We will be
eventually interested in compactifications to four dimensions. For vacua without fluxes and
T-dualities, the total space of the $T^3$ fibration is a Calabi-Yau threefold. What can we say
about the total space of the $T^4$ fibration?

Note that there is an analogous (more general) story in M-theory. Reducing eleven dimensional
supergravity on a flat $T^4$ yields a Lagrangian that is symmetric under the $SL(5,\IR)$
U-duality group \cite{Hull:1994ys, Cremmer:1979up, Sezgin:1982gi, Cremmer:1997ct}. By
Hodge-dualizing the three-form $A_{IJK}=:\epsilon_{IJKL}X^L$ ($I,J,K,L=7,8,9,11$), one can
define a $5\times 5$ matrix\footnote{
 The relation to F-theory \cite{Vafa:1996xn} can roughly be understood as follows. In
the lower right corner of the $5\times 5$ metric there is a $2\times 2$ submatrix (with
coordinates $x^{11,10}$). In the ten dimensional language, this matrix contains the dilaton and
the three-form $X^{11} \sim C^{(3)}$ which is ``mirror'' to the $C^{(0)}$ axion in Type IIB.
Roughly speaking, (conjugate) S-duality acts on this $T^2 \subset T^5$. }
\be
  G^{-1}=\left( \begin{array}{c|c}
 \omega g^{IJ} + \frac{1}{\omega} X^I X^J & \ -\frac{1}{\omega}X^I   \\
 \hline
 -\frac{1}{\omega}X^I & \frac{1}{\omega}
  \end{array}
  \right)
\ee
which contains the geometric metric $g$ on $T^4$ as well. We denote the dimensions\footnote{Note
that $x^{10}$ and $x^{11}$ are switched. This is because we want to denote the extra M-theory
dimension by $x^{11}$. We stick to this notation throughout the paper.} by $x^7$, $x^8$, $x^9$,
$x^{11}$, $x^{10}$, respectively. The bosonic kinetic terms can be written as a manifestly
$SL(5)$ invariant $\sigma$-model in terms of this metric \cite{Cremmer:1997ct}.

We can embed the $4\times 4$ unit determinant matrix $N^{-1}$ (see Eq. \ref{nmetrici}) into the
$5\times 5$ unit-determinant matrix $G^{-1}$ as follows
\be
  G^{-1}=\left( \begin{array}{cc|c}
 \delta g^{ij} + \frac{1}{\delta} b^i b^j & \ 0 \ &  -\frac{1}{\delta}b^i   \\
 0 & 1 & 0 \\
 \hline
 -\frac{1}{\delta}b^i & 0 & \frac{1}{\delta}
  \end{array}
  \right)
\ee
with $\delta\equiv(\textrm{det} \ g_{ij})^{1/2}$. By setting $\omega :=\delta$, we arrive at the
previous form of the metric. If we now perform a U-duality corresponding to the $x^{10}-x^{11}$
flip, then the solution is transformed into pure geometry in the 11d picture,
\be
  G^{-1}=\left( \begin{array}{cc|c}
 \delta g^{ij} + \frac{1}{\delta} b^i b^j & \ -\frac{1}{\delta}b^i \ &     \\
 -\frac{1}{\delta}b^i &  \frac{1}{\delta} &   \\
 \hline
 & & \ 1
  \end{array}
  \right) \equiv
  \left( \begin{array}{c|c}
g^{IJ}_\textrm{new} \ &     \\
 \hline
 & \ 1
  \end{array}
  \right)
\ee
In 10d Type IIA language, this flip roughly corresponds to the exchange of the Ramond-Ramond
one-form and the Hodge-dual of the B-field in the fiber directions.

In order to preserve minimal supersymmetry in four dimensions, one compactifies M-theory on a
$G_2$ manifold. Semi-flat limits of $G_2$ manifolds are expected to exist by an SYZ-like
argument \cite{Gukov:2002jv}. Then, by the above U-duality in seven dimensions, a solution is
obtained which is non-geometric from a 10d point of view as shown in this diagram
\begin{displaymath}
\xymatrix{ \textrm{\framebox{\ \ M-theory on semi-flat $G_2$ \ }}
\ar@{=>}[d]_{\textrm{reduction}}^{\textrm{on flat $T^4$}} &  \textrm{\framebox{\ \ Type IIA on ``non-geometric space'' \ }}  \\
\textrm{\framebox{\ \ 7d theory on $S^3$ (the base of $G_2$) \ }}
\ar@{<=>}[r]^{\textrm{\qquad\quad \ $SL(5)$ }}_{\textrm{\qquad\quad \ U-duality }} &
\textrm{\framebox{\ \  dual 7d theory on $S^3$ \ }} \ar@{~>}[u]_{\textrm{oxidation?}} }
\end{displaymath}

``Oxidation'' seems obscure in this context since we only have the 7d spacetime equations of
motion. However, for the special case of $D_4$ singularities, we will be able to ``lift the
solutions'' to 10d: they turn out to be asymmetric orbifolds, similar to some examples in
\cite{Hellerman:2002ax}.

\section{Compactifications with ${\bf D_4}$ singularities}
\label{examplesec}

In the previous sections, we studied the semi-flat limit of various geometries which had a
fibration structure. This corner of the moduli space is a natural playground for T-duality since
isometries appear along the fiber directions. Almost everywhere the space locally looks like
$\IR^n \times T^n$ and the duality group can simply be studied by a torus reduction of the
supergravity Lagrangian. The idea is then to glue patches of the base manifold by also including
the T-duality group in the transition functions. Since the duality group is discrete, such
deformations are ``topological'' and {\it a priori} cannot be achieved continuously. From the
10d point of view, the total space becomes non-geometric in general. In seven dimensions, the
$SO(3,3,\IZ)$ group can be realized as the mapping class group of a $T^4$ of unit volume. This
geometrizes the non-geometric space by going one real dimension higher. Considering such
compactifications to four dimensions which preserve $\CN=1$ supersymmetry, U-duality suggests
that the total space of the geometrized internal non-geometric space is a $G_2$ manifold.

In this section, we use these ideas to build non-geometric compactifications. We deform
geometric orbifold spaces by hand and also study particular examples of $G_2$ manifolds. These
examples will only contain (conjugate) $D_4$ singularities. This allows for a constant arbitrary
shape for the fiber and the base is also locally flat. Even though the examples are singular and
supergravity breaks down at the orbifold points, we can embed the solutions into Type IIA string
theory where they give consistent non-geometric vacua realized as modular invariant asymmetric
orbifolds.

\subsection{Modified $K3\times T^2$}
\label{modk3}

Let us first consider $K3$. The base of an elliptic fibration of $K3$ is an $S^2$. At the
$T^4/\IZ_2$ orbifold point, there are four $D_4$ singularities in the base (see \fref{tetrah}).
The purely geometric $D_4$ monodromies are
\be
 \mathcal{M}_{D_4} =  (\mathbbm{-1}_{2\times 2}) \oplus \mathbbm{1}_{2\times 2} \ \in \ SL(2)_\tau \times SL(2)_\rho
\ee
\comment{
\be
  \mathcal{M}_{D_4} =
  \left( \begin{array}{rrrr}
  -1 & 0 & 0 & 0 \\
  0 & -1 & 0 & 0 \\
  0 & 0 & \ 1 & 0 \\
  0 & 0 & 0 & \ 1
  \end{array}
  \right)
\ee}
By changing the monodromies by hand, it is possible to construct non-geometric spaces. In
\cite{Hellerman:2002ax}, $K3$ was modified into the union of two half K3's which we denote by
$\widetilde{K3}$. This non-geometric space has two ordinary $D_4$'s and two non-geometric $D'_4$
singularities with monodromies
\be
 \mathcal{M}_{D'_4} = \mathbbm{1} \oplus (\mathbbm{-1}) 
\ee
\comment{
\be
  \mathcal{M}_{D'_4} =
  \left( \begin{array}{rrrr}
  1 & 0 & 0 & 0 \\
  0 & \ 1 & 0 & 0 \\
  0 & 0 & -1 & 0 \\
  0 & 0 & 0 & -1
  \end{array}
  \right)
\ee}
If we had changed one or three $D_4$s into $D'_4$, then the monodromy at infinity would not be
trivial. In fact, it would be $\mathcal{M}_{D_4} \cdot \mathcal{M}_{D'_4}
=\mathcal{M}_{(-1)^{F_L}}$. This means that the $T^2 \times T^2$ fiber is orbifolded everywhere
in the base by the $\IZ_2$ action which inverts the fiber coordinates. In principle, this could
be interpreted as an overall orbifolding by $(-1)^{F_L}$ which moves us from Type IIA to IIB.
However, it is not clear what should happen to the odd number of $D_4$ and $D'_4$ singularities
as they don't have a trivial monodromy at infinity in IIB either. Therefore, we do not consider
such examples any further.

Let us now compactify further and consider $K3\times T^2$ or $\widetilde{K3}\times T^2$. The
base is $S^2\times S^1$ where the second factor is the base of the two-torus as described in
Section \ref{onedimsec}. The relevant monodromies are embedded in the $SL(4)$ duality group as
follows
\be
 \mathcal{M}_{D_4} =  \textrm{diag}(-1,-1,1,1) \qquad \mathcal{M}_{D'_4} =  \textrm{diag}(1,1,-1,-1)
\ee
Since in lower dimension the duality group is larger, one can consider another $D_4$-like
monodromy
\be
  \mathcal{M}_{D''_4} = \textrm{diag}(1,-1,-1,1)
\ee
which is not in the $SL(2)\times SL(2)$ subgroup of $SL(4)$, and thus it was not possible for
the case of $T^2$ compactifications. In principle, we can have spaces with monodromies
\be
  (2 \times D_4) + (2 \times D''_4) \quad \textrm{or} \quad  (2 \times D'_4) + (2 \times D''_4)
\ee
These are T-dual to each other by an $x^7 - x^{10}$ flip. Thus, it is enough to consider the
first one which is geometric since the monodromies act only in the upper-left  $SL(3)$ subsector
of $SL(4)$. However, this space is not Calabi-Yau. Supersymmetry suggests that in the base,
parallel lines\footnote{Parallelism makes sense in the context of $D_4$ singularities since the
base has a flat metric.} of singularities should have the same monodromies (possibly up to a
factor of $(-1)^{F_L}$ as in the case of $\widetilde{K3}\times T^2$). This is not the case for
this space. A way to explicitly see the absence of supersymmetry is to exhibit the total space
as the $(\IR \times T^5) / \langle \alpha, \beta \rangle$ orbifold,
\bea
  \alpha: && (x,\theta_1,\theta_2 \, | \,  \theta_3,\theta_4,\theta_5) \mapsto
(L-x,\theta_1,-\theta_2 \, |  \, {-\theta_3},-\theta_4,\theta_5) \\
  \beta: && (x,\theta_1,\theta_2 \, | \,  \theta_3,\theta_4,\theta_5) \mapsto ( -x,\theta_1,-\theta_2 \, | \,  \theta_3, -\theta_4,-\theta_5)
\eea
Here $x$, $\theta_{1,2}$ are coordinates on the base and $\theta_{3,4,5}$ are coordinates on the
fiber. $x$ is non-compact and $\theta_i$ are periodic. The orbifold group $\langle \alpha, \beta
\rangle$ also contains the element
\be
 \alpha \beta: \ (x,\theta_1,\theta_2 \, | \,  \theta_3,\theta_4,\theta_5) \mapsto
(x+L,\theta_1,\theta_2 \, | \, {-\theta_3},\theta_4,-\theta_5)
\ee
which breaks supersymmetry because it projects out the gravitini.

We see that by considering conjugate $D_4$ singularities, in the above reducible case we do not
obtain any other supersymmetric examples than those already considered in
\cite{Hellerman:2002ax} even if the duality group is extended. Hence, we move on to threefolds
in the next section.

\subsection{Non-geometric $T^6 / \IZ_2 \times \IZ_2$}

\label{nongeot6}

\begin{figure}[ht]
\begin{center}
  \includegraphics[totalheight=7.0cm,angle=0,origin=c]{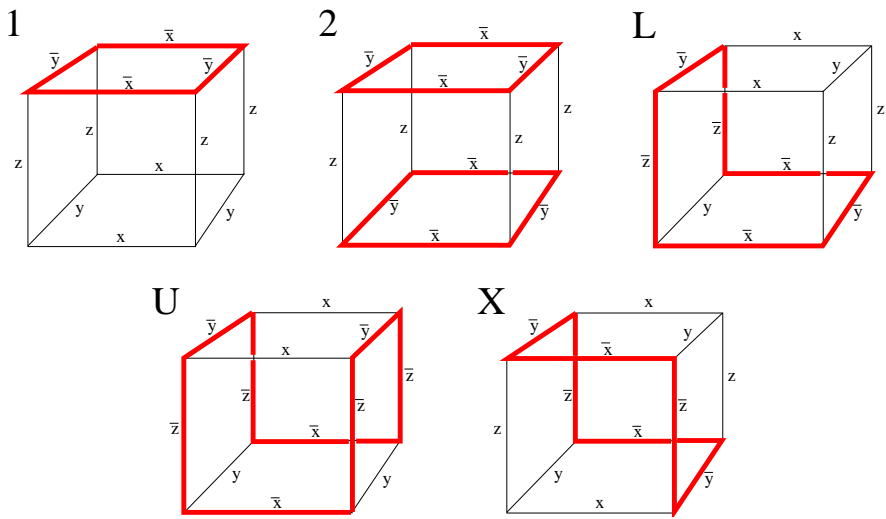}
  \caption{Almost non-geometric $T^6/\IZ_2\times\IZ_2$ spaces. Monodromies are modified along the red loops. We refer to the models as one-plaquette,
two-plaquette, ``L'', ``U'' and ``X'', respectively.}
  \label{allcubes}
\end{center}
\end{figure}

Let us consider the orbifold $T^6 / \IZ_2 \times \IZ_2$ that we described in detail in Section
\ref{threedim}. \fref{cube_mono} shows the monodromies of the singular edges. These monodromies
have the following $SL(4)$ representations,
\be
  x = \textrm{diag}(1,-1,-1,1) \quad  y = \textrm{diag}(-1,1,-1,1) \quad  z = \textrm{diag}(-1,-1,1,1)
\ee
\comment{
\be
  x =
  \left( \begin{array}{rrrr}
  1 & 0 & 0 & 0 \\
  0 & -1 & 0 & 0 \\
  0 & 0 & -1 & 0 \\
  0 & 0 & 0 & \ 1
  \end{array}
  \right)
  \quad
  y =
  \left( \begin{array}{rrrr}
  -1 & 0 & 0 & 0 \\
  0 & \ 1 & 0 & 0 \\
  0 & 0 & -1 & 0 \\
  0 & 0 & 0 & \ 1
  \end{array}
  \right)\quad
  z =
  \left( \begin{array}{rrrr}
  -1 & 0 & 0 & 0 \\
  0 & -1 & 0 & 0 \\
  0 & 0 & \ 1 & 0 \\
  0 & 0 & 0 & \ 1
  \end{array}
  \right)
\ee }
These are of course geometric since they only act on the first three coordinates. How can we
deform the orbifold into something non-geometric? There are three more $D_4$ type singularities
that we can use. They have the following monodromies,
\be
  \bar x \equiv -x
  \qquad
  \bar y  \equiv -y \qquad
  \bar z  \equiv -z
\ee
These all invert the $x^{10}$ coordinate. A simple modification of $T^6/\IZ_2\times\IZ_2$ is
possible by replacing the original monodromies by $\bar x,\bar y$ or $\bar z$. The junction
condition says that an even number of negative signs should meet at each vertex. Therefore,
consistent monodromy assignments are given by switching signs along loops. There are five
theories obtained this way as shown in \fref{allcubes}. Since these simple spaces have a
geometric total space at this orbifold point of their moduli space, we call them ``almost
non-geometric''.

\subsection{Asymmetric orbifolds}
\label{asod}

\begin{figure}[ht]
\begin{center}
  \includegraphics[totalheight=5cm,angle=0,origin=c]{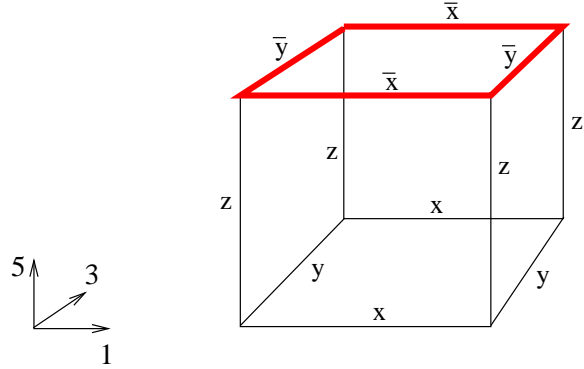}
  \caption{Simple non-geometric $T^6/\IZ_2\times\IZ_2$.}
  \label{sngex}
\end{center}
\end{figure}

In the previous section, we changed the monodromies by hand and obtained ``almost
non-geometric'' spaces. In particular, monodromies in the loops contained the extra action of
$(-1)^{F_L}$, which reverses the signs of all RR-charges,
\be
 x \cdot \mathcal{M}_{(-1)^{F_L}} = \bar x \qquad y  \cdot \mathcal{M}_{(-1)^{F_L}} = \bar y \qquad
   z  \cdot \mathcal{M}_{(-1)^{F_L}}= \bar z
\ee
where
\be
  \mathcal{M}_{(-1)^{F_L}} = \textrm{diag}(-1,-1,-1,-1)
\ee
Hence, we can realize the non-geometric spaces of the previous section as asymmetric orbifolds
\cite{Narain:1986qm, Mueller:1986yr} (see also \cite{Dine:1997ji, Dabholkar:1998kv,
Blumenhagen:2000fp, Gaberdiel:2002jr, Aoki:2004sm, Kakushadze:1996hi}). We consider the simple
example of \fref{sngex}: the
one-plaquette model. 

If we parametrize the $T^6$ torus by angles $\theta_i$, then the original $\IZ_2\times\IZ_2$
orbifold group action is generated by
\bea
  \alpha : (\theta_1,\theta_2,\theta_3,\theta_4,\theta_5,\theta_6) \mapsto
  (-\theta_1,-\theta_2,-\theta_3,-\theta_4,\theta_5,\theta_6) \\
  \beta : (\theta_1,\theta_2,\theta_3,\theta_4,\theta_5,\theta_6) \mapsto
  (-\theta_1,-\theta_2,\theta_3,\theta_4,-\theta_5,-\theta_6)
\eea
The base coordinates can be chosen to be $(\theta_1,\theta_3,\theta_5)$. The singular edges
along these directions have monodromies $(x,y,z)$, respectively.

Now the example of \fref{sngex} has modified monodromies. In particular, edges on the top of the
cube have monodromies which include $(-1)^{F_L}$. We use the same trick as in Section
\ref{modk3}: let us choose the vertical $x_5$ coordinate to be non-compact and then compactify
it with an asymmetric action,
\bea
&  \alpha :&  \ (\theta_1,\theta_2,\theta_3,\theta_4,x_5,\theta_6) \mapsto
  (-\theta_1,-\theta_2,-\theta_3,-\theta_4,x_5,\theta_6) \\
&   \beta_1 :& \ (\theta_1,\theta_2,\theta_3,\theta_4,x_5,\theta_6) \mapsto
  (-\theta_1,-\theta_2,\theta_3,\theta_4,-x_5,-\theta_6) \\
& \beta_2 :& \ 
 (\theta_1,\theta_2,\theta_3,\theta_4,x_5,\theta_6) \mapsto
  (-\theta_1,-\theta_2,\theta_3,\theta_4,L-x_5,-\theta_6)  \ \times \ (-1)^{F_L}
\eea
This realizes the example as an asymmetric orbifold. The Type IIA spectrum is computed in
Appendix~\ref{aspectrum2}. It has $\mathcal{N}=1$ supersymmetry with a gravity multiplet, 16
vector multiplets and 71 chiral multiplets.

The theory is consistent since decorating $D_4$ singularities with $(-1)^{F_L}$ does not destroy
modular invariance. In the Green-Schwarz formalism, adding $(-1)^{F_L}$ changes the boundary
conditions for the four complex left-moving fermionic coordinates as
\be
   D_4: (++ --)  \quad \longrightarrow \quad D_4 \times (-1)^{F_L}: (-- ++)
\ee
Hence, the energy of the twisted sector ground state does not change and thus level-matching is
satisfied \cite{Vafa:1986wx}. In the RNS formalism, $(-1)^{F_L}$ does not act on the world-sheet
fields and therefore the moding does not change. However, the left-moving GSO projection changes
and various generalized discrete torsion signs show up in the twisted sectors as discussed in
the Appendices. (See also related literature \cite{Aoki:2004sm, Hellerman:2006tx}.) For Abelian
orbifolds, one-loop modular invariance implies higher loop modular invariance
\cite{Vafa:1986wx}. Here we are actually considering a non-Abelian orbifold\footnote{\ldots
since $x \mapsto -x$ and $x \mapsto L-x$ do not commute.} for which level-matching is not
sufficient for consistency. Further constraints may arise if a modular transformation takes a
pair of commuting group elements $(g,h)$ into their own conjugacy class \cite{Freed:1987qk},
\be
  (g,h) \longrightarrow (g^a h^b, g^c h^d) = (p g p^{-1}, p h p^{-1})
\ee
where $a,b,c$ and $d$ are the elements of an $SL(2,\IZ)$ matrix. In this case, the path integral
with boundary conditions $(g,h)$ and $(p g p^{-1}, p h p^{-1})$ for the torus world-sheet should
give the same result. Since we only consider $D_4$ singularities, the twists of world-sheet
fermions by orbifold group elements do commute and thus non-commutativity can only come from the
action on the bosons. However, left-moving and right-moving bosons are treated symmetrically and
thus we do not get any further constraints. Therefore, one expects this model to be modular
invariant. Moreover, this theory has an alternative presentation as a $(\IZ_2)^3$ Abelian
orbifold of $T^6$ as we will see in Section \ref{dualitysec}.

The rest of the modified $T^6/\IZ_2\times\IZ_2$ spaces (\fref{allcubes}) have asymmetric
orbifold descriptions as well. These are listed in Appendix \ref{alist}. The modular invariance
argument of the previous paragraph applies to these as well. Some of the models are dual to each
other. This will be discussed in Section \ref{dualitysec}.

\subsection{Joyce manifolds}
\label{joycesec}

In Section \ref{UdualityandGtwo}, we saw how a class of non-geometric spaces can be transformed
into geometric M-theory compactifications by U-duality. Naturally, one can try to interpret
existing $G_2$ spaces from the literature as ``non-geometric'' Type IIA string theory vacua.

Let us denote the coordinates on $\IR^7$ (and $T^7$) by $x_1, x_2, x_3$ (base), $ y_1, y_2, y_3,
y_4$ (fiber). The exceptional group $G_2$ is the subgroup of $GL(7,\IR)$ which preserves the
form
\bean
  \varphi = dx_1 \wedge dy_1 \wedge dy_2 + dx_2 \wedge dy_1 \wedge dy_3 + dx_3 \wedge dy_2 \wedge
  dy_3 + dx_2 \wedge dy_2 \wedge dy_4 \\
  - dx_3 \wedge dy_1 \wedge dy_4 - dx_1 \wedge dy_3 \wedge
  dy_4 - dx_1 \wedge dx_2 \wedge dx_3 \qquad
\eean
It also preserves the orientation and the Euclidean metric on $\IR^7$ and so it is a subgroup of
$SO(7)$. In this section, we consider particular compact examples. Joyce manifolds \cite{Joyce1,
Joyce2} are (resolved) $T^7/(\IZ_2)^3$ orbifolds which preserve the calibration. We consider the
following action,
\bean
  \alpha &:& (x_1, x_2, x_3 \, | \, y_1, y_2, y_3, y_4) \mapsto (x_1, -x_2, -x_3 \, |  \,  y_1, y_2, -y_3, -y_4) \\
  \beta &:& (x_1, x_2, x_3  \, | \,  y_1, y_2, y_3, y_4) \mapsto (-x_1, x_2, A_1-x_3 \, |  \,  y_1, -y_2, y_3, -y_4) \\
  \gamma &:& (x_1, x_2, x_3 \, |  \,  y_1, y_2, y_3, y_4) \mapsto (A_2-x_1, A_3-x_2, x_3 \, |  \,  {-y_1}, y_2, y_3, -y_4)
\eean
where $A_i \in \{0, \frac{1}{2}\}$. Note that $\alpha^2=\beta^2=\gamma^2=1$ and $\alpha$,
$\beta$ and $\gamma$ commute. Some of the choices of $\vec{A}\equiv (A_1,A_2,A_3)$ are
equivalent to others by a change of coordinates. Only shifts for the base coordinates are
included since fiber shifts can't be realized by a linear transformation. (We comment on this
later in Section \ref{fibershift}.) The blow-ups of these spaces are described in \cite{Joyce2,
Joyce:book}.

These orbifolds can be interpreted as non-geometric Type II backgrounds as follows. The $T^4$
fiber coordinates are already chosen to be $\{ y_i \}$. One needs to pick a direction for the
extra $x^{10}$ circle. Theories that differ in this choice are T-dual to each other. Then,
whenever a generator contains a minus sign for the $x^{10}$ circle, a $(-1)^{F_L}$ must be
separated from its action. The geometric action is then given by inverting the fiber signs (and
omitting the extra circle). For instance, if $y_4$ is the $x^{10}$ circle, then $\alpha$ will
become
\be
\alpha_0 : (x_1, x_2, x_3 \, | \, y_1, y_2, y_3) \mapsto (x_1, -x_2, -x_3 \, | \,  {-y_1}, -y_2,
y_3)
\ee
and this geometric action will be accompanied by $(-1)^{F_L}$.

In the following, we list the spaces of different shifts and discuss their singularity
structure.

\begin{itemize}

\item
$\vec{A}=(0,0,0)$

(i) Let us first consider the $\IZ_2 \times \IZ_2$ orbifold generated by only $\alpha$ and
$\beta$. Then, by identifying $y^4$ with the extra $x^{10}$ coordinate, we obtain the model in
\fref{topbottom}. This is U-dual to the pure geometry $T^6 / \IZ_2 \times \IZ_2$ by a $y_1-y_4$
flip.

(ii) Let us now include $\gamma$. This gives the most singular example of Joyce
manifolds. The $x_i$ and $y_i$ coordinates parametrize the $S^3$ base and the $T^4$ fiber,
respectively. The $(\IZ_2)^3$ orbifold group is equally well generated by $\langle \alpha,
\beta, \alpha\beta\gamma \rangle$. It is important to note that the product $\alpha\beta\gamma$ does not act on the
base coordinates. In principle, this could be interpreted as globally orbifolding\footnote{This
interpretation would give $T^6 / \IZ_2 \times \IZ_2$ in Type IIB. This is mirror to Type IIA on
$T^6 / \IZ_2 \times \IZ_2$ with discrete torsion turned on \cite{Vafa:1994rv}.} by $(-1)^{F_L}$.
However, this leads to problems similar to those in our earlier discussion in Section
\ref{modk3}.

It is also easy to see that U-duality does not work in this case\footnote{The general fiber in a
Lagrangian fibration on any symplectic manifold is a torus. However, the general fiber for a
coassociative $G_2$ fibration is expected to be $T^4$ or $K3$ \cite{Lee:2002fa}. The adiabatic
argument for U-duality only works for the $T^4$ case \cite{Vafa:1995gm, Sen:1996na} which must
be taken into account when choosing the fiber coordinates.}. Compactifying M-theory on a $G_2$
manifold gives $\mathcal{N}=1$ supersymmetry in 4d. However, the above configuration in Type II
has $\mathcal{N}=2$ supersymmetry\footnote{Although one of the gravitini is projected out by
$(-1)^{F_L}$, it comes back in the twisted sector to give extended supersymmetry.}, and
therefore cannot be equivalent to the M-theory configuration. Thus we will not discuss this
example any further.

\item
$\vec{A}=(0,0,\frac{1}{2}) \sim (0,\frac{1}{2},0) \sim (\frac{1}{2},0,0)$

The extra identification by $\gamma$ cuts the fundamental cell of $T^6 / \IZ_2 \times \IZ_2$ in
half. The resulting base is again an $S^3$ which can be constructed as shown in \fref{halfcube}.
The non-geometric space has the same monodromies as the model in \fref{fourthex} that we already
constructed by directly modifying the monodromies of $T^6/\IZ_2\times\IZ_2$.


\begin{figure}[ht]
\begin{center}
 \hskip -5cm
  \includegraphics[totalheight=5.5cm,origin=c]{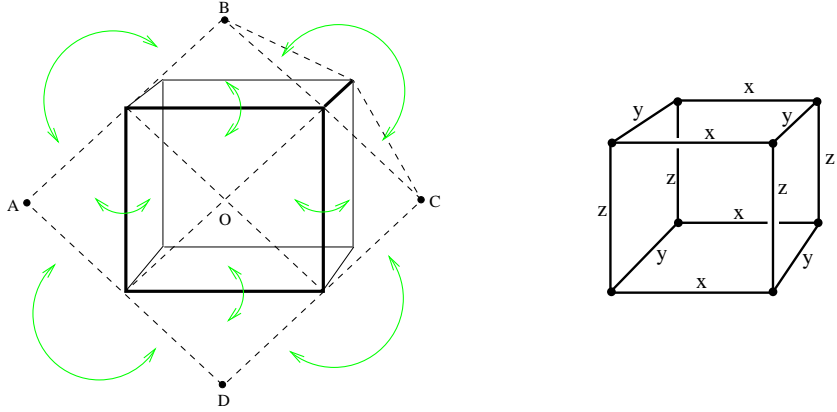}
  \vskip -4.5cm
  \hskip 8cm
  \includegraphics[totalheight=3.0cm,origin=c]{cube.eps}
  \vskip 1.5cm
  \caption{(i) Fundamental domain of the base after modding by $\gamma$: half of a rhombic dodecahedron. The arrows
show how the faces are identified.
  \ (ii) Schematic picture indicating the structure of the degenerations.}
  \label{halfcube}
\end{center}
\end{figure}

\item
$\vec{A}=(0,\frac{1}{2},\frac{1}{2}) \sim (\frac{1}{2},0,\frac{1}{2}) \sim
(\frac{1}{2},\frac{1}{2},0)$


Let us consider $\vec{A}=(0,\frac{1}{2},\frac{1}{2})$, as the others are equivalent by a
coordinate transformation. The action of $\alpha$ and $\beta$ generate $T^6 / \IZ_2 \times
\IZ_2$ as usual. The third $\IZ_2$ is generated by $\gamma$. It has a fixed edge which goes
through two parallel faces of the cube (see \fref{otherhalfcube}). The base is again an $S^3$.
(The proof of this statement goes roughly as that of $T^6 / \IZ_2 \times \IZ_2$.)

\clearpage

\begin{figure}[ht]
\begin{center}
  \hskip -5cm
  \includegraphics[totalheight=5.5cm,origin=c]{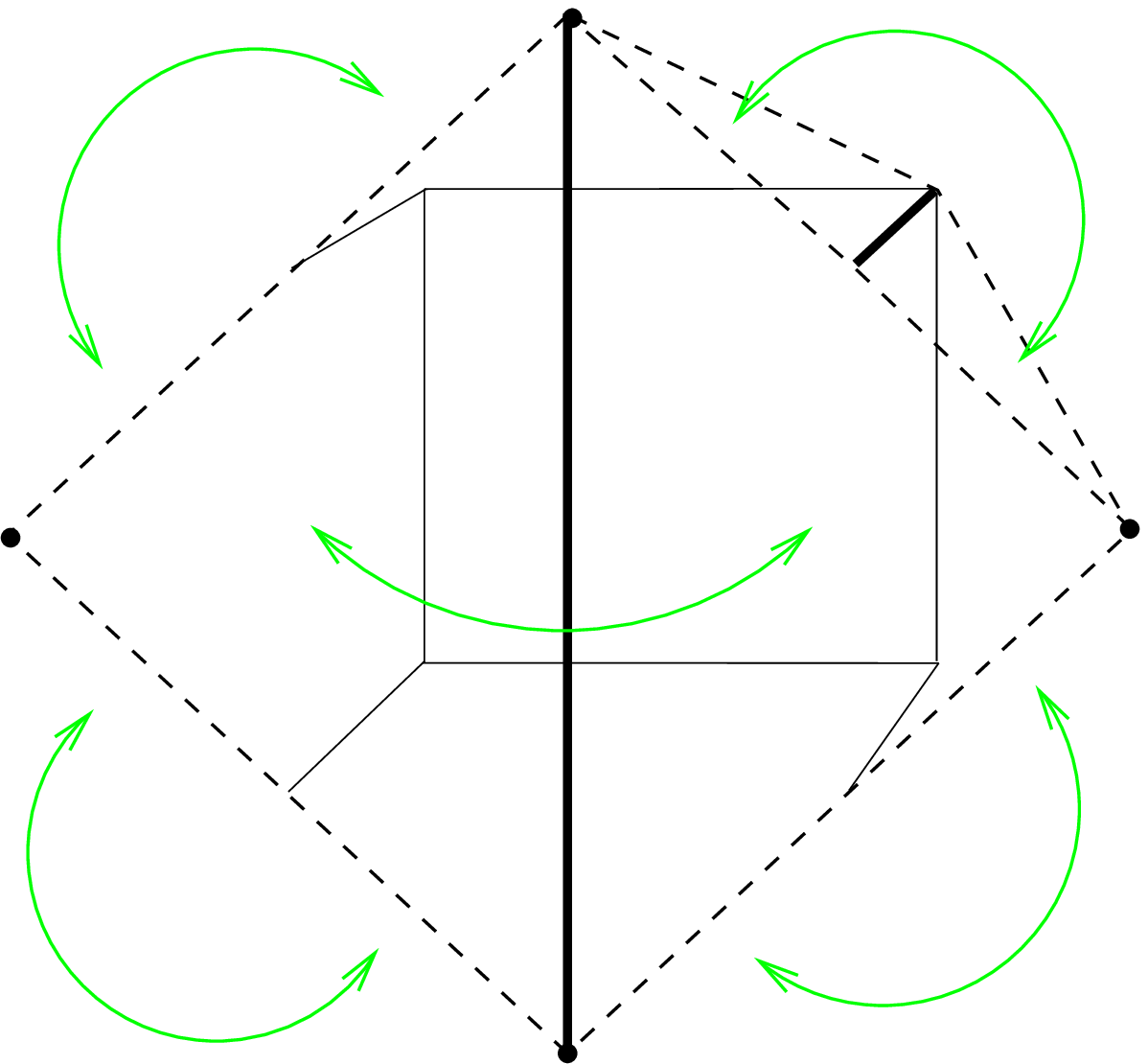}
  \vskip -4.5cm
  \hskip 8cm
  \includegraphics[totalheight=3.5cm,origin=c]{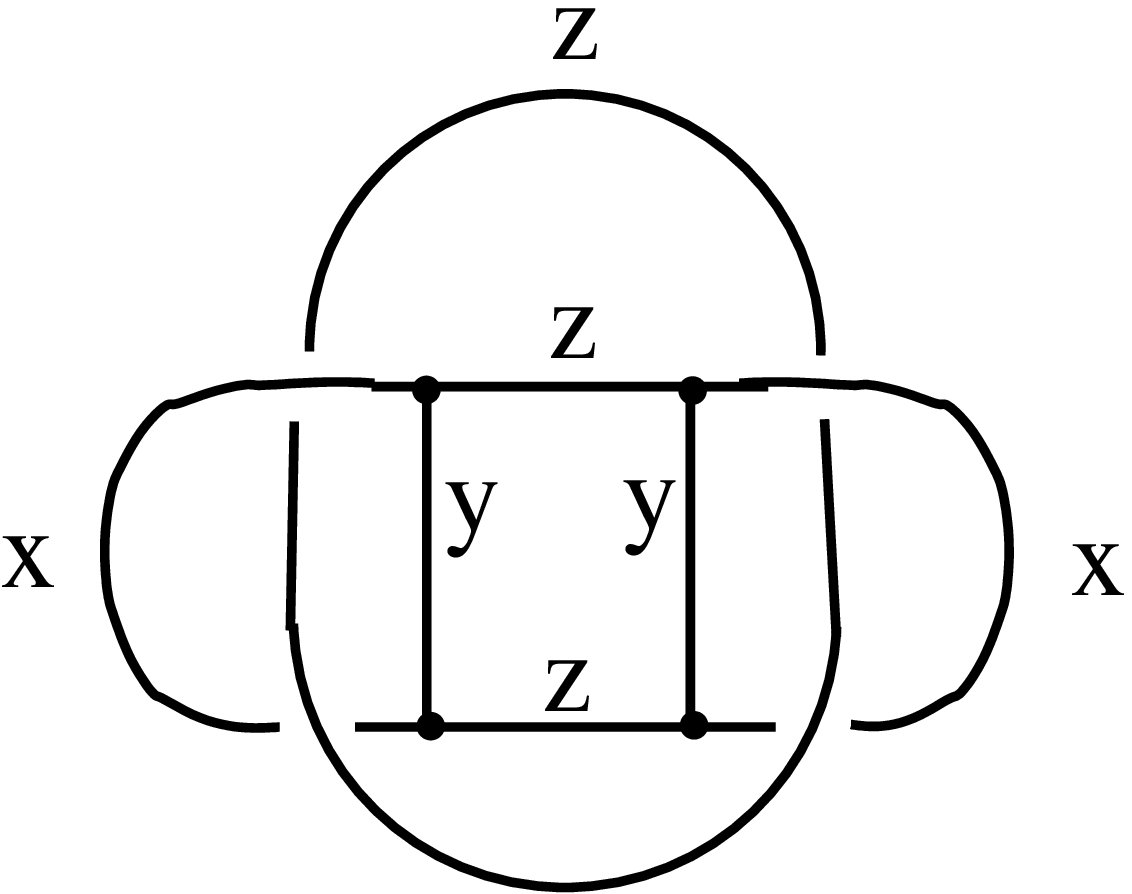}
  \vskip 1.5cm
  \caption{(i) Half of the fundamental domain after modding by $\gamma$.  \ (ii) Schematic picture.}
  \label{otherhalfcube}
\end{center}
\end{figure}



\item
$\vec{A}=(\frac{1}{2},\frac{1}{2},\frac{1}{2})$

(i) Let us first omit the action of $\gamma$. This gives a somewhat simpler space with base
depicted in \fref{trunc1}. It is the union of a truncated tetrahedron, plus a small tetrahedron.
This base can be obtained as the intersection of fundamental domains of the two commuting
$\IZ_2$ actions. Both of these domains are $S^1$ times the square (with solid edges) depicted in
\fref{t4fund}. The identification of the faces and the schematic structure of the degenerations
are shown in \fref{trunc11}.

\begin{figure}[ht]
\begin{center}
  \includegraphics[totalheight=7.5cm,angle=270,origin=c]{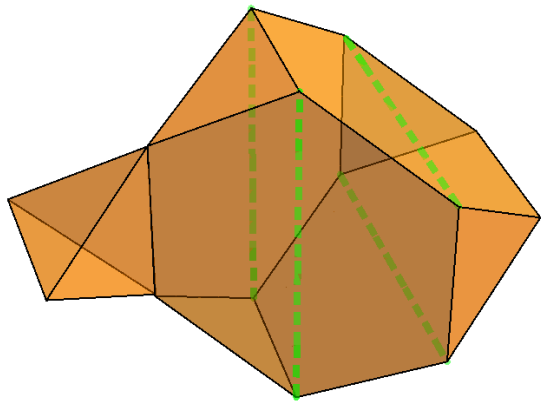}
  \caption{ The base of $T^6/(\IZ_2)^2$ where the generators of $\IZ_2$'s include coordinate shifts.
  Four non-intersecting $D_4$ strings (dashed green lines in the middle of hexagons) curve the space into an $S^3$.
  See Figure 36 in Appendix H for a pattern that can be cut out.}  
  \label{trunc1}
\end{center}
\end{figure}

\begin{figure}[ht]
\begin{center}
  \hskip -7cm
  \includegraphics[totalheight=5.5cm,origin=c]{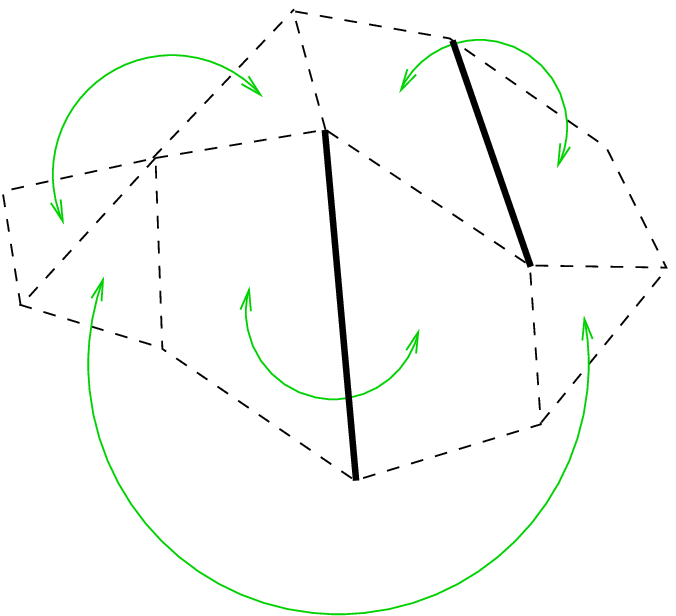}
  \vskip -5cm
  \hskip 8cm
  \includegraphics[totalheight=3.5cm,origin=c]{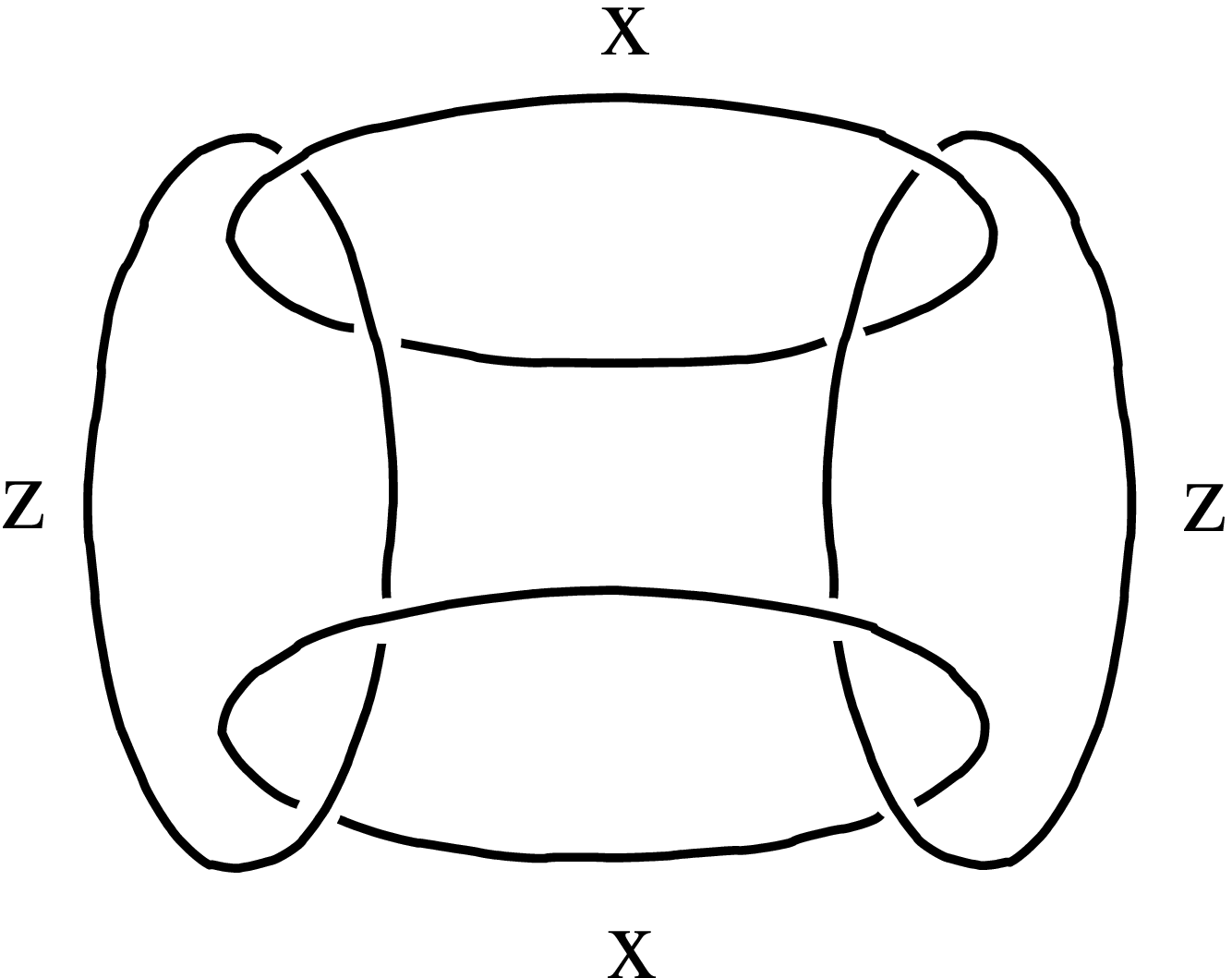}
  \vskip 2.0cm
  \caption{(i) The base can be constructed by gluing the truncated tetrahedron (dashed lines) to itself along with a small tetrahedron.
  It is easy to check that the $D_4$ strings (solid lines) have $180^\circ$ deficit angle whereas the dashed lines are non-singular.
\ (ii) Schematic picture.  The truncated tetrahedron example can roughly be understood as four
linked rings of $D_4$ singularities.
  All of the rings are penetrated by two other rings which curve the space into a cylinder as they have tension 12.
  This forces the string to come back to itself.}
  \label{trunc11}
\end{center}
\end{figure}

(ii) Let us now include $\gamma$ as well. The coordinate shifts in the $\IZ_2$ actions make sure
that the fixed edges do not intersect. The structure of the base is shown in \fref{kocka}.


\begin{figure}[ht]
\begin{center}
 \hskip -5cm
  \includegraphics[totalheight=4.5cm,origin=c]{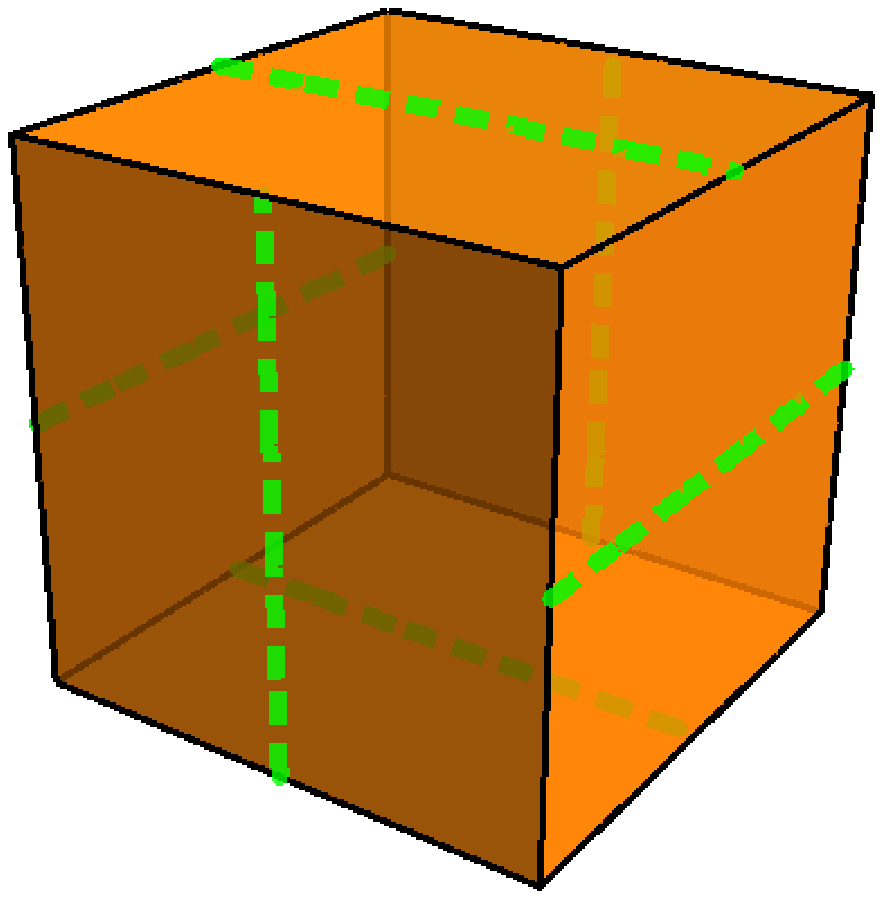}
  \vskip -4cm
  \hskip 7cm
  \includegraphics[totalheight=3.5cm,origin=c]{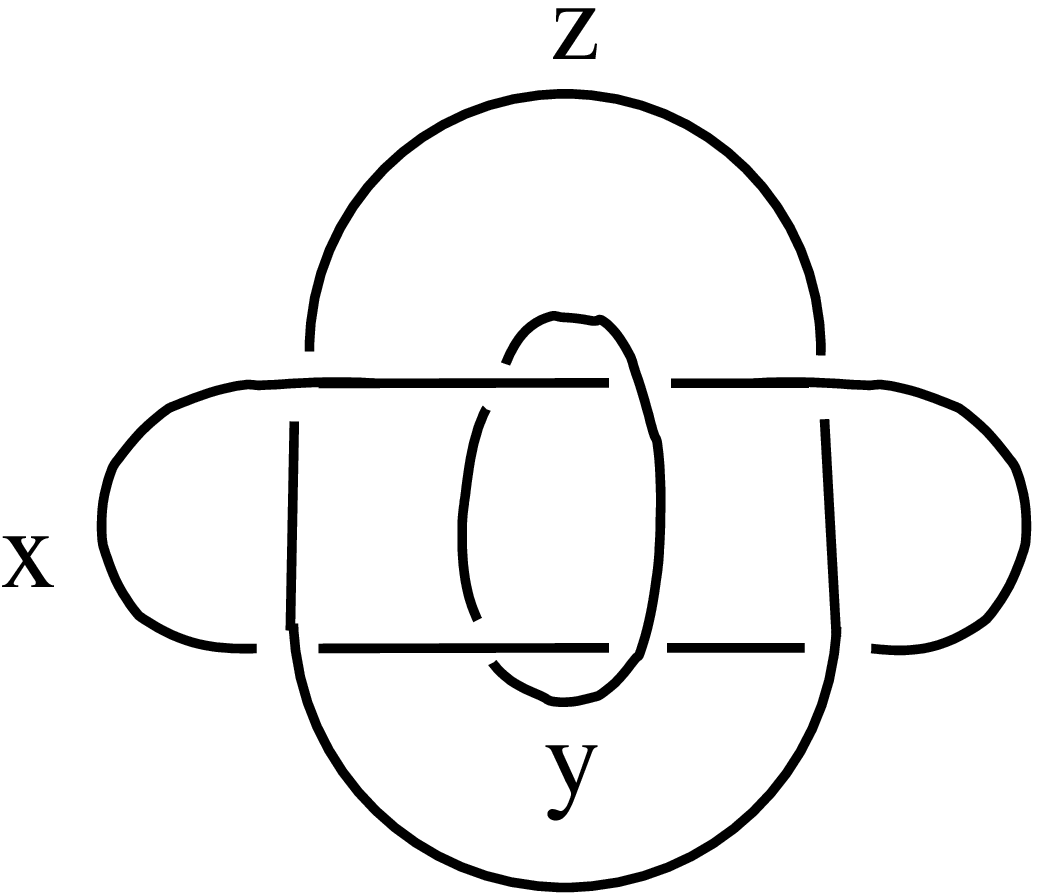}
  \vskip 1cm
  \caption{(i) The base of the $(\frac{1}{2},\frac{1}{2},\frac{1}{2})$ Joyce orbifold. There are
six strings located on the faces of a cube. These faces are folded up which generates the
$180^\circ$ deficit angles.
  \ (ii) Schematic picture. The degenerations form three rings of $D_4$ singularities.}
  \label{kocka}
\end{center}
\end{figure}

\end{itemize}

\clearpage
\subsection{Dualities between models}
\label{dualitysec}

The two-plaquette model can be realized as $T^6/\IZ_2\times\IZ_2$ by the following orbifold
action\footnote{The action of $\alpha$ creates four parallel edges of the singular cube in the
base. Then, $\beta$ and $\alpha\beta$ generate $4+4$ edges with $(-1)^{F_L}$. These give the two
``red plaquettes'' (see \fref{allcubes}).},
\bean
  \alpha : (\theta_1,\theta_2,\theta_3,\theta_4,\theta_5,\theta_6) & \mapsto &
  (-\theta_1,-\theta_2,-\theta_3,-\theta_4,\theta_5,\theta_6) \\
 \beta :
 (\theta_1,\theta_2,\theta_3,\theta_4,x_5,\theta_6)  & \mapsto &
  (-\theta_1,-\theta_2,\theta_3,\theta_4,-\theta_5,-\theta_6)  \ \times \ (-1)^{F_L}
\eean
Performing a single T-duality on $\theta_6$ turns $\beta$ into
\be
 \tilde\beta :
 (\theta_1,\theta_2,\theta_3,\theta_4,x_5,\theta_6)   \mapsto
  (-\theta_1,-\theta_2,\theta_3,\theta_4,-\theta_5,-\theta_6)
\ee
and keeps $\alpha$ intact\footnote{In the $T^2$ fiber language, the duality exchanges $\tau$ and
$\rho$ and therefore takes a $D'_4$ singularity into $D_4$.}. We thus learn that Type IIA on the
two-plaquette model is dual to Type IIB on $T^6/\IZ_2\times\IZ_2$. The details of the spectrum
computation is presented in Appendix \ref{aspectrum}.

\begin{figure}[ht]
\begin{center}
  \includegraphics[totalheight=4.5cm,origin=c]{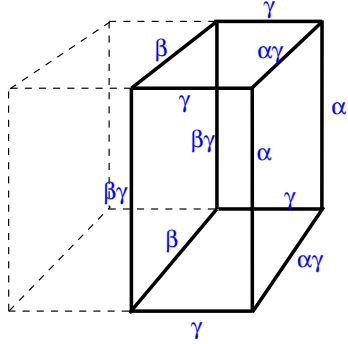}
  \caption{Monodromies of the one-shift Joyce orbifold.}
  \label{cube_dual}
\end{center}
\end{figure}

Another duality is provided by considering the $\vec A=(\frac{1}{2},0,0)$ Joyce orbifold,
\begin{displaymath}
\begin{array}{llllrrrrrrrr}
  \alpha &:& (x_1, x_2, x_3 \, | \, y_1, y_2, y_3, y_4) & \mapsto ( & x_1, &  -x_2, &  -x_3 &  \, |  \, &   y_1,  & y_2,  & -y_3, &  -y_4) \\
  \beta &:& (x_1, x_2, x_3  \, | \,  y_1, y_2, y_3, y_4) &  \mapsto ( & -x_1,  & x_2,  & \frac{1}{2}-x_3 &  \, |  \,  &  y_1, &  -y_2, &  y_3,  &-y_4) \\
  \gamma &:& (x_1, x_2, x_3 \, |  \,  y_1, y_2, y_3, y_4) &  \mapsto ( & -x_1, &  -x_2,  & x_3  & \, |  \,   & {-y_1},  & y_2,  & y_3,  & -y_4)
\end{array}
\end{displaymath}
The monodromies of the singularities in the base are shown in \fref{cube_dual} (see also
\fref{halfcube}). The action of $\alpha$ and $\gamma$ creates the usual cubic structure and
$\beta$ cuts the cube in half.

This $G_2$ orbifold can be interpreted as a Type IIA background in more than one way depending
on which coordinate we choose for the $x^{10}$ circle. As discussed in the previous section, a
minus sign in the $x^{10}$ direction is interpreted as $(-1)^{F_L}$ (this interpretation is
accompanied by an inversion of fiber signs). From \fref{cube_dual} it is clear that $x^{10}=y_2$
or $y_3$ gives the one-plaquette model since in these cases $\beta$ or $\alpha$, respectively,
will contain $(-1)^{F_L}$. On the other hand, choosing $x^{10}=y_1$ or $y_4$ gives model ``U''.
Since relabeling $x^{10}$ is an element of the $SL(4)$ T-duality group, these backgrounds are
T-dual to each other. The spectrum is computed in Appendix \ref{aspectrum2}.

\clearpage
\subsection{U-duality and affine monodromies}
\label{fibershift}

For usual orbifolds, it is known that the untwisted sector contains information about the
singular space, whereas the twisted sectors describe resolutions (or deformations
\cite{Vafa:1994rv, Gaberdiel:2004vx}) thereof. It is typically said that string theory ``knows''
about the non-singular resolution and the number of the various particles are determined by the
Hodge numbers. Here we can see this happening in a more general setup. In M-theory, the number
of $\mathcal{N}=1$  vector and chiral multiplets are respectively determined by the $b_2$ and
$b_3$ Betti numbers of the $G_2$-manifold. When U-duality works, one should obtain the same
massless spectrum from the asymmetric (non-geometric) orbifold of Type IIA.

Joyce \cite{Joyce1, Joyce2} computed Betti numbers for blown-up $T^7/(\IZ_2)^3$ examples. These
examples, however, contained $1/2$ shifts also in directions that were interpreted as fiber
coordinates in the previous section\footnote{The notation $b_i$ and $c_i$ is from \cite{Joyce2}.
These constants should not be confused with the Betti numbers.},
\begin{displaymath}
\begin{array}{llllrrrrrrrr}
  \alpha &:& (x_1, x_2, x_3 \, | \, y_1, y_2, y_3, y_4) & \mapsto ( & x_1, &  -x_2, &  -x_3 &  \, |  \, &   y_1,  & y_2,  & -y_3, &  -y_4) \\
  \beta &:& (x_1, x_2, x_3  \, | \,  y_1, y_2, y_3, y_4) &  \mapsto ( & -x_1,  & x_2,  & {b_2}-x_3 &  \, |  \,  &  y_1, &  -y_2, &  y_3,  & b_1-y_4) \\
  \gamma &:& (x_1, x_2, x_3 \, |  \,  y_1, y_2, y_3, y_4) &  \mapsto ( & c_5-x_1, &  c_3-x_2,  & x_3  & \, |  \,   & {-y_1},  & y_2,  & y_3,  & c_1-y_4)
\end{array}
\end{displaymath}
These shifts are recommended, otherwise one encounters ``bad singularities'' which can't easily
be resolved. If interpreted as a fibration, the monodromies acting on $T^4$ are affine
transformations which also include half-shifts for some of the fiber coordinates. Although these
orbifolds can readily be interpreted as non-geometric backgrounds for Type IIA, the naive
U-duality map does not necessarily work and the spectrum does not match with that of M-theory.

In Appendix \ref{joycespectrum}, we discuss the cases of two Joyce manifolds, with two and three
shifts $(b_1, b_2, c_1, c_3, c_5) = (0,\frac{1}{2},\frac{1}{2},0,0,0)$ and
$(0,\frac{1}{2},\frac{1}{2},\frac{1}{2},0,0)$. Naive U-duality works well for the three shift
example and one obtains the same spectrum from the non-geometric compactification. However, the
two shift example gives a different spectrum from what we expect from the Betti numbers of the
$G_2$-manifold\footnote{An ambiguity is immediately discovered by noticing that a redefinition
the fiber coordinates $\tilde y \equiv y+1/4$ changes the naive interpretation of $(-1)^{F_L}$
as $\textrm{diag}(-1,-1,-1,-1)$. The new monodromy action for $(-1)^{F_L}$ will now include
$1/2$ shifts in the fiber. In some cases, this ambiguity can be exploited to match the IIA and
M-theory spectra.}. The puzzle can simply be resolved by choosing a different (coassociative)
fiber. Taking $\{ x_1, x_2, y_2, y_3\}$ for fiber coordinates, the $\IZ_2$ transformations have
no shifts in these directions and the non-geometric Type IIA spectrum indeed matches the
M-theory spectrum.

\newpage
\section{Compactifications with ${\bf E_n}$ singularities}

In this section, we list geometric orbifolds containing singularities other than $D_4$.
Non-geometric modifications of these orbifolds may be done similarly to the previous section.
For $D_4$ singularities, the constant shape of the fiber can be arbitrary. The main difference
in the $E_n$ case is that the fiber shape is determined by the symmetry group. In practice, this
means that in two dimensions $\tau=i$ or $\tau = e^{i\pi/3}$.

\subsection{Orbifold limits of $K3$}

Simple warm-up examples are provided by considering  $T^4 / \IZ_n$ orbifolds. These have been
analyzed from the F-theory point of view in \cite{Dasgupta:1996ij}.

\begin{figure}[ht]
\begin{center}
  \includegraphics[totalheight=3.5cm,angle=0,origin=c]{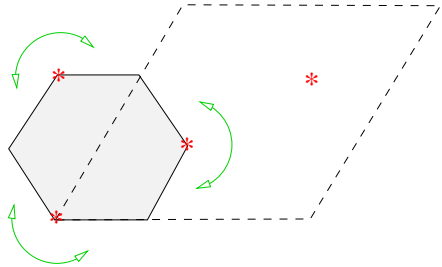}
  \caption{The base of the $T^4/\IZ_3$ orbifold contains three $E_6$ singularities. }
  \label{t4z3fund}
\end{center}
\end{figure}

\vskip 0.5cm \noindent {\bf The ${\bf T^4 / \IZ_3}$ orbifold.} The action of the generator of
the orbifold group is given by
\be
  \alpha: \, (z_1, z_2) \mapsto (e^{i\pi/3}z_1, \, e^{-i\pi/3}z_2)
\ee
which respects the torus identifications
\be
  z_i \sim z_i+1 \sim z_i+e^{i\pi/3}
\ee
The base is $T^2 / \IZ_3$ and can be parametrized by $z_1$. It contains three $E_6$
singularities of deficit angle $4\pi / 3$. The monodromy around these are given by
\be
  \mathcal{M}_{E_6} = (ST)^2 = \left( \begin{array}{cc}
    -1  & -1 \\
    1 & 0
  \end{array}
  \right)
\ee
A fundamental cell is shown in \fref{t4z3fund}.

\begin{figure}[ht]
\begin{center}
  \includegraphics[totalheight=4cm,angle=0,origin=c]{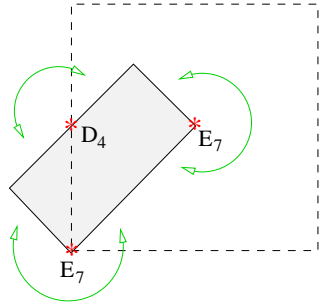}
  \caption{The base of the $T^4/\IZ_4$ orbifold contains two $E_7$ and one $D_4$ singularities. }
  \label{t4z4fund}
\end{center}
\end{figure}

\vskip 0.5cm \noindent {\bf The ${\bf T^4 / \IZ_4}$ orbifold.} The generator of $\IZ_4$ is given
by
\be
  \alpha: \, (z_1, z_2) \mapsto (i z_1, \, -i z_2)
\ee
with the torus identifications
\be
  z_i \sim z_i+1 \sim z_i+i
\ee
The base is $T^2 / \IZ_4$. This orbifold contains two $E_7$ and one $D_4$ singularity. They have
deficit angles $3\pi/2$ and $\pi$, respectively. The $E_7$ and $D_4$ monodromies are given by
\be
  \mathcal{M}_{E_7} = S = \left( \begin{array}{cc}
    0  & -1 \\
    1 & 0
  \end{array}
  \right)\qquad  \mathcal{M}_{D_4} = \left( \begin{array}{cc}
    -1  & 0 \\
    0 & -1
  \end{array}
  \right)
\ee
A fundamental cell is shown in \fref{t4z4fund}.

\vskip 0.5cm \noindent {\bf The ${\bf T^4 / \IZ_6}$ orbifold.} The base is $T^2 / \IZ_6$. This
orbifold contains $E_8$, $E_6$ and $D_4$ singularities. The $E_8$ monodromy is given by
\be
  \mathcal{M}_{E_8} = ST = \left( \begin{array}{cc}
    0  & -1 \\
    1 & 1
  \end{array}
  \right)
\ee

\begin{figure}[ht]
\begin{center}
  \includegraphics[totalheight=3.5cm,angle=0,origin=c]{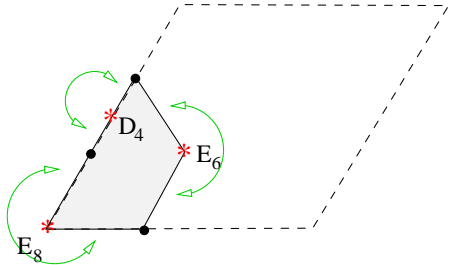}
  \caption{The base of the $T^4/\IZ_6$ orbifold contains $E_8$, $E_6$ and $D_4$ singularities.
  The three black dots denote one non-singular point.}
  \label{t4z6fund}
\end{center}
\end{figure}

A fundamental cell is shown in \fref{t4z6fund}.

\subsection{Example: $T^6 / \IZ_3$}

\begin{figure}[ht]
\begin{center}
  \includegraphics[totalheight=5.5cm,origin=c]{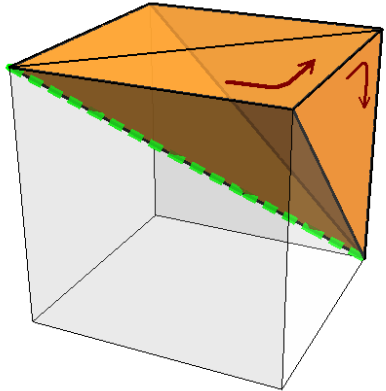}
  \caption{The base of $T^6 / \IZ_3$. The green line shows the $E_6$ singularity. Six triangles bound the domain.
  Two triangles touching the singular green line are identified by folding. Two triangles should be identified according to the orientation
  given by the arrows. The remaining two triangles are identified in a similar fashion.}
  \label{kocka_e6}
\end{center}
\end{figure}

We continue by discussing three dimensional examples. The simplest one is $T^6 / \IZ_3$. This is
created by orbifolding the square $T^6$ by cyclic permutations of (complex) coordinates
\be
  \alpha: \, (z_1, z_2, z_3) \mapsto (z_2, z_3, z_1)
\ee
Clearly, this action preserves the holomorphic volume form,
\be
  \Omega = dz_1 \wedge dz_2 \wedge dz_3
\ee
and the \kahler form
\be
 \omega = \sum_i dz_i \wedge d\bar z_i
\ee
Let us now choose the real parts of $z_i$ for the base coordinates. Before orbifolding, the base
is a cube as shown in \fref{kocka_e6}. The fixed loci of $\alpha$ are at $z_1=z_2=z_3$ that is
along a diagonal. The cube has a $\IZ_3$ symmetry about this diagonal, and thus the orbifolding
procedure respects the torus identifications.

Since $\IZ_3 \subset SU(2)$, this example preserves $\mathcal{N}=4$ supersymmetry in four
dimensions. By making the identifications of the bounding triangles, one can check that the only
singularity is $E_6$. It is along the diagonal which gives a closed loop in the base. Since
there are no other gravitating strings to curve the space, this is a good sign that the space
factorizes. In particular, we do not expect it to be an $S^3$.

\clearpage
\subsection{Example: $T^6 / \Delta_{12} $}

\label{d12}

A more complicated example is gained by orbifolding $T^6 /(\IZ_2)^2$ by the above described
cyclic permutations. These permutations do not commute with the sign flips and together they
give $\Delta_{12} \subset SU(3)$. This group has the faithful representation described by the
following matrices (see \cite{Greene:1998vz}, and also \cite{Berenstein:2000mb, Hanany:1998sd,
Feng:2000af}) which act on the $(z_1, z_2, z_3)$ complex coordinates

\be
   \left( \begin{array}{ccc}
  (-1)^p & 0 & 0 \\
  0 & (-1)^q & 0 \\
  0 & 0 & (-1)^{p+q}
  \end{array}
  \right) \qquad
   \left( \begin{array}{ccc}
  0 & 0 & (-1)^p \\
  (-1)^q & 0 & 0 \\
  0 & (-1)^{p+q} & 0
  \end{array}
  \right) \qquad
   \left( \begin{array}{ccc}
  0 & (-1)^p & 0 \\
  0 & 0 & (-1)^q \\
  (-1)^{p+q} & 0 & 0
  \end{array}
  \right)
\ee

It can be generated by two elements,
\bean
  \alpha: \, (z_1, z_2, z_3) &\mapsto & (z_2, z_3, z_1) \\
  \beta: \, (z_1, z_2, z_3) &\mapsto & (-z_1, -z_2, z_3)
\eean

\begin{figure}[ht]
\begin{center}
  \hskip -6cm
  \includegraphics[totalheight=7cm,origin=c]{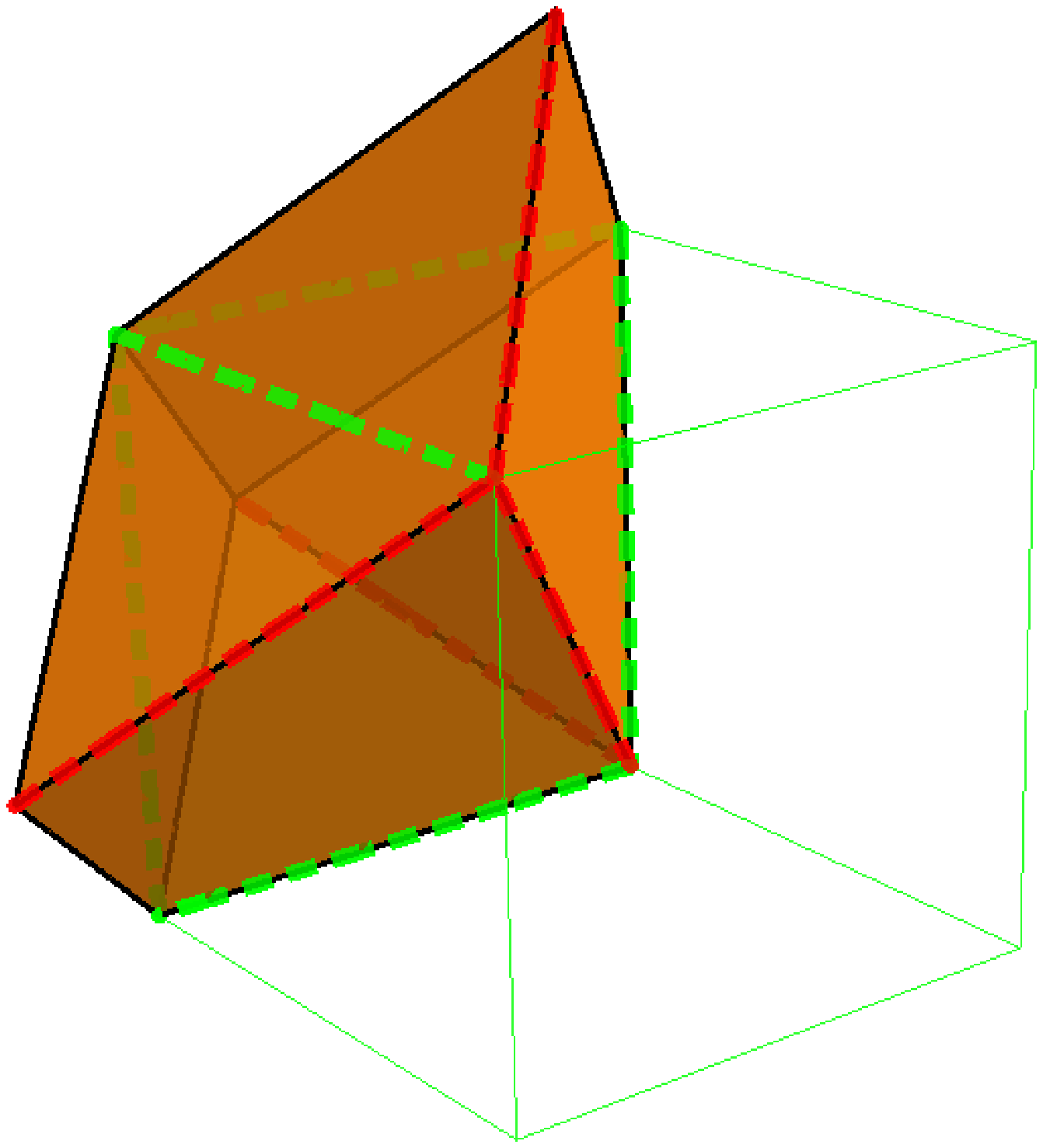}
  \vskip -5.5cm
  \hskip 8cm
  \includegraphics[totalheight=3.5cm,origin=c]{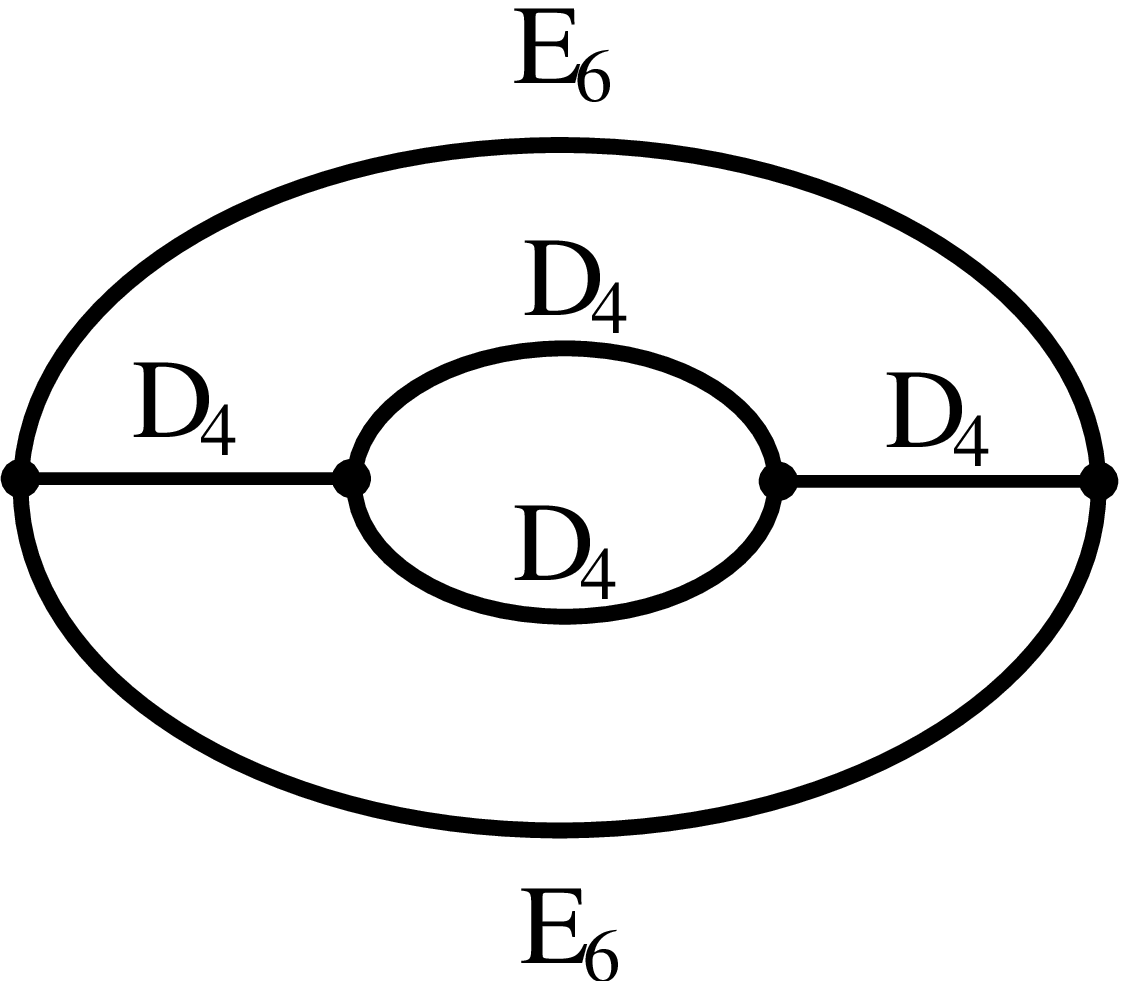}
  \vskip 2cm
  \caption{(i) The base of $T^6 /\Delta_{12}$. The red and green lines indicate $E_6$, $D_4$ singularities, respectively. The other edges are non-singular.
  The solid green cube indicates the $D_4$ singularities of the original $T^6 / (\IZ_2)^2$ orbifold. (ii) Schematic picture describing
  the topology of the singular lines. See Appendix H for building this polyhedron at home.}
  \label{rhombic_e6}
\end{center}
\end{figure}

The fundamental domain is shown in \fref{rhombic_e6}. There are two $E_6$ and four $D_4$
singularities in the base. They meet in $E_6$-$E_6$-$D_4$ and $D_4$-$D_4$-$D_4$ vertices. The
solid angle around these vertices are $\pi/3$ and $\pi$, respectively. The base is topologically
an $S^3$.

\clearpage
\subsection{Example: $T^6 / (\IZ_2)^2 \times \IZ_4 $}

Another example is obtained from  $T^6 / (\IZ_2)^2$ by further orbifolding it by $\IZ_4$. This
is possible because the rhombic dodecahedron has fourfold symmetry axes. These are the axes of
the green cube in \fref{cube_fold}.

\begin{figure}[ht]
\begin{center}
  \hskip -7cm
  \includegraphics[totalheight=6cm,origin=c]{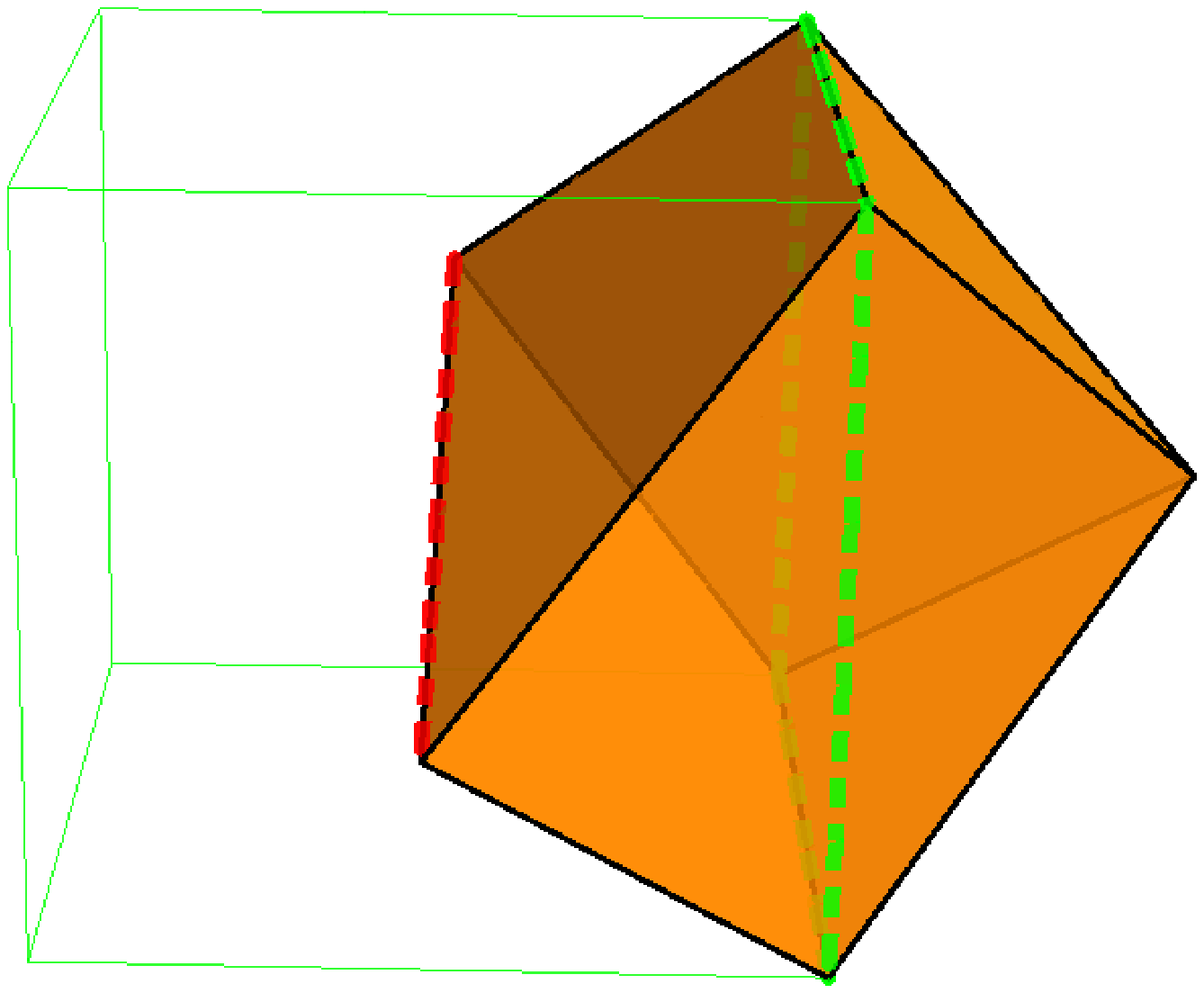}
  \vskip -4.8cm
  \hskip 8cm
  \includegraphics[totalheight=3.0cm,origin=c]{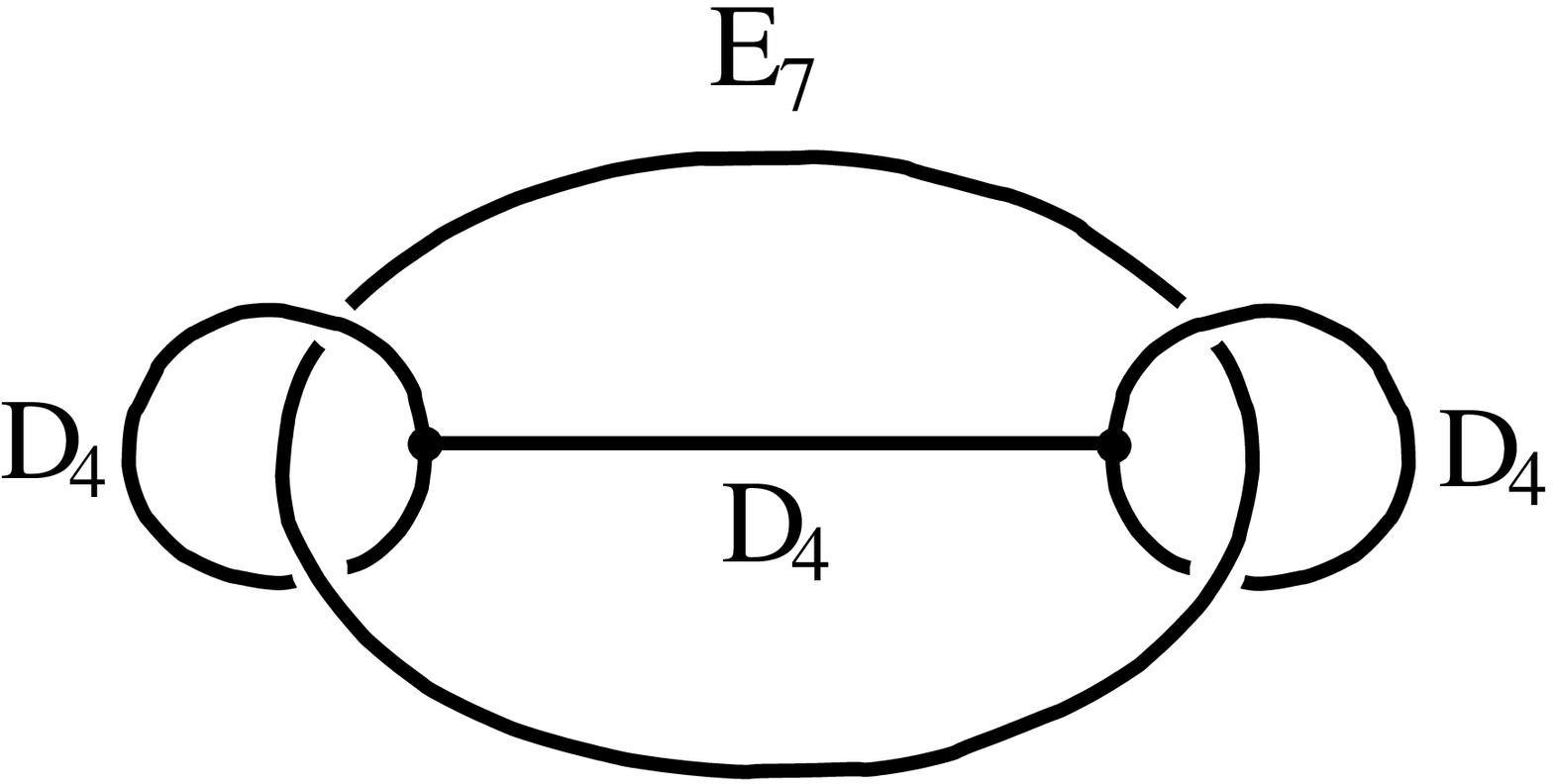}
  \vskip 2cm
  \caption{(i) The base of $T^6 / (\IZ_2)^2 \times \IZ_4 $. The red and green lines indicate $E_7$, $D_4$ singularities, respectively. The other edges are non-singular.
  The solid green cube indicates the $D_4$ singularities of the original $T^6 / (\IZ_2)^2$ orbifold. (ii) Schematic picture describing
  the topology of the singular lines.}
  \label{rhombic_e7}
\end{center}
\end{figure}

The resulting base is shown in \fref{rhombic_e7}. There is one $E_7$ line which is topologically
a circle. In contrast to the $T^6 / \IZ_3$ example, this happens because the other $D_4$
singularities curve the base and make this contractible loop a geodesic. The base only contains
familiar $D_4$-$D_4$-$D_4$ vertices.

\subsection{Example: $T^6 / \Delta_{24} $ }

Our final example can be constructed by first taking $T^6$. Its base is a cube with opposite
faces identified. We now place $E_7$ singularities on the twelve edges of the cube. We also add
diagonal $E_6$ singularities as in Section \ref{d12}. These are realized by the following
matrices which act on $(z_1, z_2, z_3)$ complex coordinates
\be
  \alpha_{E_6} = \left( \begin{array}{ccc}
  0 & 1 & 0 \\
  0 & 0 & 1 \\
  1 & 0 & 0
  \end{array}
  \right) \qquad \qquad
  \beta_{E_7} = \left( \begin{array}{ccc}
  0 & -1 & 0 \\
  1 & 0 & 0 \\
  0 & 0 & 1
  \end{array}
  \right)
\ee
These generate the $\Delta_{24}$ group. Compared to $\Delta_{12}$, it also contains odd
permutations of the coordinates. Since odd permutations come with an odd number of minus signs,
the volume form is again invariant.

\begin{figure}[ht]
\begin{center}
  \hskip -7cm
  \includegraphics[totalheight=6.5cm,origin=c]{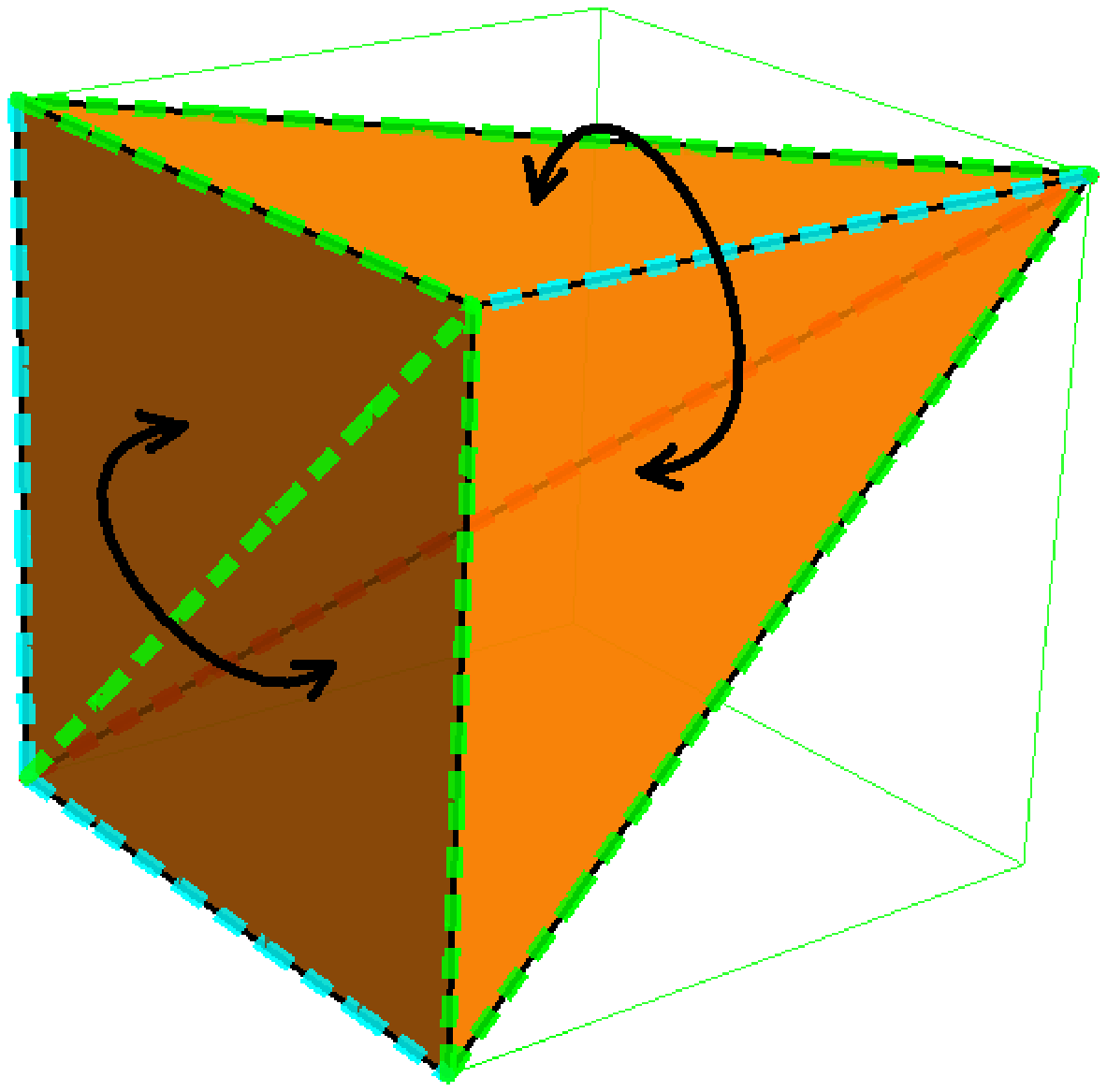}
  \vskip -4.8cm
  \hskip 8cm
  \includegraphics[totalheight=3.0cm,origin=c]{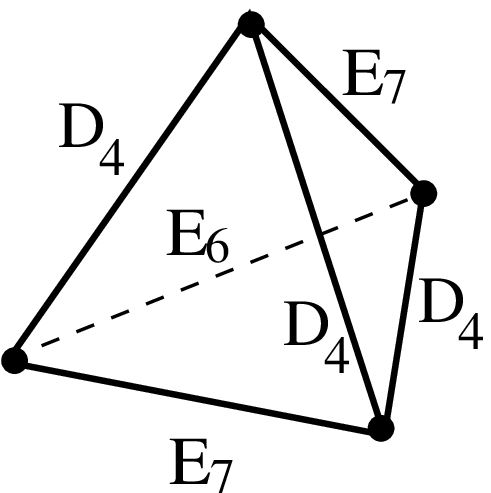}
  \vskip 2cm
  \caption{(i) The base of $T^6 / \Delta_{24} $. The cyan, red and green lines indicate $E_7$, $E_6$ and $D_4$ singularities, respectively.
  (ii) Schematic picture describing the topology of the singular lines. }
  \label{rhombic_d24}
\end{center}
\end{figure}

In \fref{rhombic_d24}, the resulting base is shown. The green cube around the base is $1/8$ or
the original base of $T^6$. The faces should be folded as indicated by the arrows. The rear
faces touching $E_6$ should be also folded. This gives an $S^3$ with curvature concentrated in
the singular lines (see the right-hand side of the figure). The base contains two types of
composite vertices. One is an intersection of $E_7$, $E_6$ and $D_4$ edges. The other one comes
from the collision of an $E_7$ and two $D_4$ singularities.

\subsection{Non-geometric modifications}


Having discussed the geometric structure of the fibrations with exceptional singularities, we
can try to modify them into non-geometric spaces. Similarly to the examples in Section
\ref{nongeot6}, closed loops of $D_4$, $E_7$ and $E_8$ singularities\footnote{Since the
monodromy of $E_6$ is an order three modular transformation, adding a sign would make it order
six.} may be decorated with the action of $(-1)^{F_L}$. For example, $E_8={\tiny
\mtwo{0}{-1}{1}{1} \oplus \mtwo{1}{0}{0}{1}}$ with $(-1)^{F_L}$ has the same monodromy as a
composite of $A_2={\tiny \mtwo{0}{1}{-1}{-1}}$ and a $D'_4$ (which acts on the other $T^2
\subset T^4$). The tension four $A_2$ and the tension six $D'_4$ give the original deficit angle
of the tension ten $E_8$ (see \ref{twodimsec} for the Kodaira classification of singularities).

The simplest example is to add $(-1)^{F_L}$ to the $D_4$ and one of the $E_7$ singularities of
the $T^4/\IZ_4$ orbifold (\fref{t4z4fund}), or instead decorate both $E_7$ singularities. The
$T^4/\IZ_6$ orbifold (\fref{t4z6fund}) can similarly be modified by adding $(-1)^{F_L}$ to the
$D_4$ and the $E_8$ singularities. By performing a single T-duality in the fiber, the
$T^4/\IZ_4$ monodromies can be changed to act on $SL(2)_\rho$ instead of $SL(2)_\tau$. The
resulting Type IIB theory has a $D'_4=D_4 \times (-1)^{F_L}$ and two $E'_7$ singularities. The
$E'_7$ corresponds to a double T-duality and thus the background is globally non-geometric, even
though it has a geometric dual.

\begin{figure}[ht]
\begin{center}
  \includegraphics[totalheight=4cm,origin=c]{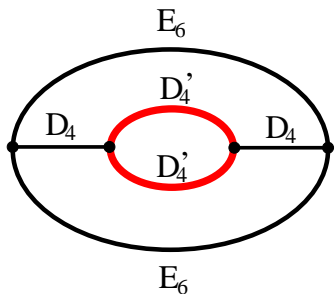}
  \caption{Non-geometric $T^6 /\Delta_{12}$. The red lines indicate extra $(-1)^{F_L}$ factors.}
  \label{rhombic_e6_red}
\end{center}
\end{figure}

\begin{figure}[ht]
\begin{center}
  \includegraphics[totalheight=3.5cm,origin=c]{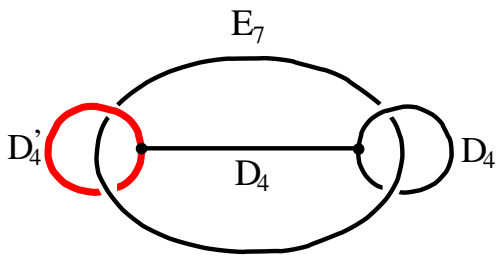}
  \caption{Non-geometric $T^6 / (\IZ_2)^2 \times \IZ_4 $.}
  \label{rhombic_e7_red}
\end{center}
\end{figure}

\begin{figure}[ht]
\begin{center}
  \includegraphics[totalheight=3.5cm,origin=c]{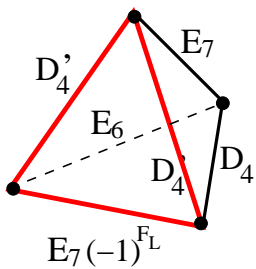}
  \caption{Non-geometric $T^6 / \Delta_{24}$. }
  \label{rhombic_d24_red}
\end{center}
\end{figure}

Turning to the three dimensional examples, $(-1)^{F_L}$ can be added to the $D_4$ loop of $T^6
/\Delta_{12}$ as shown in \fref{rhombic_e6_red}. This is obtained by orbifolding the last
example in \fref{allcubes}. The $D_4$ loops or the $E_7$ loop of $T^6 / (\IZ_2)^2 \times \IZ_4 $
can similarly be modified. An example is shown in \fref{rhombic_e7_red} where a single $D_4$ has
been changed into $D'_4$ corresponding to the first example of \fref{allcubes}. A single
T-duality on the geometric $T^6 / (\IZ_2)^2 \times \IZ_4$ gives Type IIB with a circle of $E'_7$
and thus the dual background is non-geometric. $T^6 / \Delta_{24} $ can similarly be modified
(\fref{rhombic_d24_red}).

These spaces can serve as perturbative string backgrounds. The consistency of these vacua,
however, needs further investigation.

\newpage
\section{Chiral Scherk-Schwarz reduction}
\label{chirals}

In previous sections, we studied non-geometric spaces mainly by using a $(-1)^{F_L}$ monodromy
around singular loci in the base. Another possibility is to have this transformation in the
fiber as a Wilson line. Fields still do not depend on the fiber coordinates, and in this sense
this is a (chiral) Scherk-Schwarz reduction.

\subsection{One dimension}

Let us consider Type IIA compactified on a circle with $(-1)^{F_L}$ Wilson line. This will be a
one-dimensional fiber. The configuration breaks half of the supersymmetry keeping sixteen
right-moving supercharges. In M-theory, $(-1)^{F_L}$ is described as reflection of $x^{11}$.
Hence, the background lifts to M-theory as compactification on a Klein bottle
\cite{Dabholkar:1996pc}.

An important feature of the background that one can try to exploit in the construction of
non-geometric spaces is that T-duality on the circle takes Type IIA to IIA (not IIB as usual)
\cite{Gutperle:2000bf, Hellerman:2005ja, Aharony:2007du}. Although the duality switches between
the $SO(8)$ spinor and conjugate spinor representations in the right-moving sector, it also
exchanges the untwisted and twisted R/NS sectors \cite{Hellerman:2005ja}. Therefore, when the
circle decompactifies, the two massless 10d gravitini have different chiralities and thus the
theory is still Type IIA.

At self-dual radius, the bosonic string has additional massless states and one obtains the gauge
group $SU(2)\times SU(2)$. In Type II strings, these extra states are destroyed by the GSO
projection and one is left with $U(1)\times U(1)$ only. With the above Wilson line, however, an
extended $SU(2)\times U(1)$ gauge symmetry is obtained. In the effective theory, T-duality is
part of the $SU(2)$ gauge group and thus a T-duality monodromy can be regarded as a Wilson line.

A simple two-dimensional non-geometric space is obtained by compactifying on another base circle
with a monodromy that is a T-duality on the fiber circle. The consistency of this model has to
be further investigated.

\subsection{Two dimensions}

These ideas can be generalized by considering $T^n$ compactifications and turning on a
$(-1)^{F_L}$ Wilson line. This still preserves half of the supersymmetry.  In order to glue
spaces, only those monodromies can be considered which preserve Wilson lines, that is the ``spin
structure'' of the $T^n$ fiber. Therefore, the perturbative duality group will be a proper
subgroup of $O(n,n,\IZ)$.

In the following, we consider the simplest examples where the base is taken to be two
dimensional and is parametrized by the complex coordinate $z$. The shape of the two-torus fiber
is described by the $\tau$ complex parameter. We take the Wilson line\footnote{ The case of
Wilson lines turned on for both fiber circles is the same since a modular $T$ transformation
converts the $(-,-)$ spin structure into $(-,+)$. } to be along the real direction denoted by
$x^9$. Then, along this coordinate axis, a single T-duality is possible. Applying the Buscher
rules, this duality is mirror symmetry for the two-torus fiber.

\begin{figure}[ht]
\begin{center}
  \includegraphics[totalheight=6cm,angle=0,origin=c]{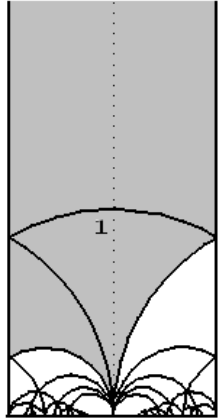}
  \caption{Fundamental domain (gray area) for the action of the $\Gamma_0(2)$ on the upper half-plane.}
\label{fundomgamma}
\end{center}
\end{figure}

Let us denote the components of an arbitrary $SL(2,\IZ)$ element $M$ by
\be
  M=\left( \begin{array}{cc}
 a & b \\
 c & d
  \end{array}
  \right), \qquad ad-bc=1
\ee
Geometric transformations must preserve the $(-,+)$ spin structure. If $(x,y)\in \IZ^2$ denotes
the homotopy class of a one-cycle, then this constraint is equivalent to
\be
  (-1)^{x} =   (-1)^{ax+by}
\ee
that is
\be
 (a-1)x+by = 0 \quad (\textrm{mod} \ 2)
\ee
Since $y$ is arbitrary, $b$ must be even. Then, $\textrm{det}\, M=1$ forces $a$ (and $d$) to be
odd and the above equation is satisfied. Therefore, the geometric part of the duality group is
the $\Gamma_0(2) \subset SL(2)$ congruence subgroup of index three. A maximal subgroup of it is
$\Gamma(2)$ that contains matrices with even off-diagonal elements.  $\Gamma_0(2)$ can be
generated by $\Gamma(2)$ and the $TST^{-1}$ transformation which exchanges two cycles in the
fiber. Its fundamental domain is shown in \fref{fundomgamma}. The full duality group contains
another copy of $\Gamma_0(2)$ for $\rho$, and a single T-duality along $x^9$.

A geometric $K3$ fibration\footnote{ This $K3$ fibration has been used in the literature
(\cite{Berglund:1998va}, see also \cite{Bershadsky:1998vn}) to describe F-theory duals of 8d CHL
strings \cite{Chaudhuri:1995fk, Chaudhuri:1995bf}. Nine dimensional CHL strings are defined by
taking $E_8 \times E_8$ heterotic strings and orbifolding by a $\IZ_2$ action which shifts the
ninth coordinate and interchanges the two $E_8$ factors. For a recent study of the moduli space
of nine dimensional theories with sixteen supercharges, see \cite{Aharony:2007du}. } with such
restricted transformations can be described by \cite{Berglund:1998va}
\be
  y^2 + x^4 +x^2 w^2 f_4(z) + w^4 g_8(z) =0
\ee
where $(x,y,w) \in \IC P^2_{1,2,1}$ and $f_4, g_8$ are holomorphic sections of degree 4 and 8,
respectively. The $j$-function is given by
\be
  j(\tau) = \frac{(f_4^2+12g_8)^3}{108 \, g_8(-f_4^2+4g_8)^2}
\ee
The discriminant of the elliptic fibration vanishes generically at 16 points out of which 8 are
double zeros. The moduli space is ten dimensional, in contrast to the 18 dimensional space of
the cubic Weierstrass equation.

As explained in Appendix B of \cite{Berglund:1998va}, the types of possible degenerations are
$A_n$, $D_n$ and $E_7$. The $K3$ geometry can reach the $T^4/\IZ_2$ orbifold limit where four
$D_4$ singularities close the base into an $S^2$. The orbifold is then generated by
\begin{displaymath}
\begin{array}{llllrrrr}
  \alpha &:& (x_1, x_2 \, | \,  y_1, y_2) & \mapsto ( & {-x_1}, & -x_2 \, | \,  & -y_1, & -y_2) \\
  \beta &:& (x_1, x_2 \, | \,  y_1, y_2) & \mapsto ( & {x_1}, & x_2 \, | \,  & y_1, & \frac{1}{2}+y_2)
\end{array}
\end{displaymath}
with $\beta$ containing $(-1)^{F_L}$. This is the same theory as the asymmetric orbifold limit
of the $12+12'$ model of \cite{Hellerman:2002ax}. The anomaly free $\mathcal{N}=1$ 6d spectrum
contains a supergravity multiplet, nine tensor multiplets, eight vector multiplets and twenty
hypermultiplets. The strong coupling limit is M-theory on a $\IZ_2$ orbifold\footnote{ This is
to be compared with the CHL string in six dimensions which is dual \cite{Bershadsky:1998vn} to
M-theory on
\be
  (K3 \times S^1)/ \left\{\sigma \cdot (y\rightarrow y+1/2) \right\}
\ee
by utilizing the heterotic-Type II duality \cite{Witten:1995ex}.}
\be
  (K3 \times S^1)/ \left\{ \sigma \cdot (y\rightarrow -y) \right\}
\ee
where $y\in[0,1)$ is the $S^1$ coordinate and $\sigma$ is an involution on $K3$ that acts with
eight fixed points. It preserves twelve of the harmonic $(1,1)$ forms and changes the sign of
the other eight harmonic $(1,1)$ forms. The spectrum computation \cite{Sen:1996tz} matches that
of the asymmetric orbifold.

The resolved $12+12'$ model used a doubly elliptic Weierstrass fibration over an $S^2$ base,
\be
  y^2=x^3+p_4(z)x+q_6(z) \qquad   \tilde y^2=\tilde x^3+\tilde p_4(z)\tilde x+\tilde q_6(z)
\ee
The constants in the polynomials give a 19 dimensional moduli space. In the above orbifold
limit, the complex base coordinate includes $y_2$ (which has the Wilson line). The $\Gamma_0(2)$
construction resolves the orbifold in a different `frame': it chooses a different set of base
coordinates, namely $x_1$ and $x_2$. It presumably slices out a different subspace in the full
moduli space of the model.


Finally, T-duality along the $x^9$ circle can also be considered. The $\tau(z)$ and $\rho(z)$
sections can be described by considering a doubly elliptic fibration over the base. In
\cite{Hellerman:2002ax}, the fiber tori were independent and thus $\tau(z)$ and $\rho(z)$ were
unrelated. For the present configuration with a Wilson line, however, a single T-duality can
exchange them and result in more complicated non-geometric spaces. The construction of such
backgrounds is left for future work.

\clearpage

\section{Conclusions}
\label{section_conclusions}

A perturbative vacuum of string theory is specified by a conformal field theory on the
worldsheet. Only in special cases will the CFT have a geometric description. Such cases include
flat space, Calabi-Yau and flux compactifications, which have been studied in great detail. The
development of a more systematic understanding of the set of consistent string vacua will
inevitably require the study of non-geometric compactifications.


String dualities allow for the construction of string vacua that are locally geometric but not
necessarily manifolds globally. Using this idea, we have constructed non-geometric
compactifications preserving $\mathcal{N} = 1$ supersymmetry in four dimensions. In the two
dimensional case, the Weierstrass equation with holomorphic coefficients solves the equation of
motion and allows for sharing the $\IZ_4$ and $\IZ_6$ orbifold points which is necessary for
$SU(2)$ holonomy. Since an appropriate generalization of the Weierstrass equation was not at our
disposal, we were only able to describe such spaces at the asymmetric orbifold point in their
moduli space. A strong motivation for departing the flat-base limit is that it presumably
generates a non-trivial potential for the overall volume modulus. Note that for $D_4$
singularities, the size of the fiber is an arbitrary free parameter which (typically) runs to
large volume.

Although our explicit examples were all orbifolds, in principle, it is possible to build
non-orbifold examples by means of $D_4$ and $E_n$ singularities. Since the base in this case is
flat, it could be obtained by gluing various polyhedra along their faces. By carefully choosing
the dihedral angles of the building blocks, one can create the appropriate deficit angles for
the edges. However, it is not easy to satisfy the constraints on monodromies coming from
supersymmetry and the constructions quickly get complicated.
A good step in this direction would be to find a good basis of building blocks
which suffice even to reconstruct the orbifold examples.
By the relation discussed in
Section \ref{UdualityandGtwo}, such spaces would presumably give new examples of $G_2$
manifolds.

In Appendix \ref{aspectrum2}, Type IIA string theory has been compactified in a non-geometric
way on the ``one-shift'' $T^7/(\IZ_2)^3$ orbifold down to four dimensions. The massless spectrum
is equivalent to that of the M-theory compactification on a particular resolution of this
orbifold with $(b_2, b_3)=(16,71)$. The orbifold has, however, numerous other resolutions with
very different Betti numbers \cite{Joyce:book}. It would be interesting to see whether these
other resolutions arise in Type IIA by the introduction of discrete torsion (and possibly
NS5-branes).

In the other direction, the result of section 3.4 shows that a general $T^3$-fibration with
$SO(3,3,\IZ)$ T-duality monodromies has a globally geometric M-theory dual. This is striking
given the difficulty of describing such creatures from the string theory point of view. More
generic constructions with $SL(5)$ monodromies presumably have no duality frame where they are
globally geometric.

In this paper, we have focussed on compactifications where the monodromy group was a subgroup of
the perturbative duality group. There is no obstacle in principle to the extension of the
monodromy group to include the full $SL(5)$ U-duality group\footnote{An early attempt to
geometrize such examples was made in \cite{Kumar:1996zx}.}. In this manner one can extend these
techniques to include in the compactification Ramond-Ramond fields, D-branes and orientifolds,
and presumably to find vacua with no massless scalars. In Appendices C and D we build confidence
that such objects can be treated consistently in the semiflat approximation by rederiving from
this viewpoint the Hanany-Witten brane-creation effect and the duality between M-theory on
$T^5/\IZ_2$ and type IIB on K3. Although we studied vacua of Type II string theory, the
discussion can be applied to heterotic strings as well where the duality group $O(16+d, d)$ is
much larger \cite{Flournoy:2004vn}.

Another interesting direction is the study of leaving the large complex structure limit. Our
special flat-base examples had a worldsheet description as modular invariant asymmetric
orbifolds. However, in the generic case, this powerful tool is missing. Any available tools,
such as the gauged linear sigma model \cite{Witten:1993yc}, should be brought to bear on this
problem.

In \cite{Greene:1989ya} it is proved for the $T^2$-fibered case that a solution in the semiflat
approximation determines an exact solution.  While the power of holomorphy is lacking in the
$T^3$-fibered case, the physical motivation for this statement \cite{Hellerman:2002ax} remains.
The idea is that the violations of the semiflat approximation are localized in the base, and we
have a microscopic description of the degenerations, as D-branes or NS-branes or as parts of
well-understood CY manifolds or orbifolds or U-duality images of these things.

It is expected that the singular edges in the base transform into ribbon graphs as we move away
from the semi-flat limit \cite{Joyce:2000ru, dave}. It seems possible that one can construct
local (in the base) invariants of the fibration which give `NUT charges'  \cite{Hull:1997kt}.
These invariants, which are analogous to the number of seven-branes in the stringy cosmic
strings construction, appear in the \cite{Shelton:2005cf} mirror-symmetry-covariant
superpotential.

\vskip 0.5cm

{\bf Acknowledgements}

We thank Allan Adams, Henriette Elvang, Mark Hertzberg, Bal\'azs K\H{o}m\H{u}ves, Vijay Kumar,
Albion Lawrence, Ruben Minasian, Dave Morrison, Washington Taylor and Alessandro Tomasiello for
discussions and comments on the draft. JM acknowledges early conversations on related matters
with S. Hellerman in 2003. This work is supported in part by funds provided by the U.S.
Department of Energy (D.O.E.) under cooperative research agreement DE-FG0205ER41360.

\appendix
\section{Appendix: Flat-torus reduction of type IIA to seven dimensions}

\label{redapp}

The following discussion is based on \cite{Maharana:1992my}. Let us consider the action for the
massless NS-NS fields of type II strings (in any number of dimensions)
\be
  S=\int dx \int dy \sqrt{-\hat g} \cdot e^{-\hat \phi}  \, [ R(\hat g) + \partial_\mu\hat \phi\partial^\mu\hat \phi - \frac{1}{12}\hat H_{\mu\nu\rho}\hat H^{\mu\nu\rho} ].
\ee
The $x$ coordinates label so-far-noncompact directions, and $y$ are coordinates on a $T^d$. We
want to reduce the theory and eliminate the $y$ coordinates. Let $\mu, \nu, \ldots$ and $\alpha,
\beta, \ldots$ label the corresponding indices. Taking the following ans\"atze,
\be
  \hat g=:\left( \begin{array}{cc}
  g_{\mu\nu} + A^\gamma_\mu A_{\nu\gamma} & \ A_{\mu\beta}   \\
  A_{\nu\alpha}  & \ G_{\alpha\beta}
  \end{array}
  \right)
\ee
\be
  \phi := \hat \phi-\frac 1 2 \textrm{log det} \,  G_{\alpha\beta}
\qquad   F^{(1)\alpha}_{\mu\nu} := \partial_\mu A^\alpha_\nu - \partial_\nu A^\alpha_\mu
\ee
\be
  H_{\mu\nu\alpha} := \hat H_{\mu\nu\alpha}-A^\beta_\mu\hat H_{\beta\nu\alpha}-A^\beta_\nu \hat
  H_{\mu\beta\alpha}
\ee
one obtains the following terms after reduction
\be
  S=\int dx \sqrt{-g}  \, e^{-\phi} \mathcal{L}
\ee
with $\mathcal{L}=\mathcal{L}_1+\mathcal{L}_2+\mathcal{L}_3+\mathcal{L}_4$ and
\bea
  \mathcal{L}_1 &=& R + \partial_\mu \phi \partial^\mu \phi \\
  \mathcal{L}_2 &=& \frac{1}{4}(\partial_\mu G_{\alpha\beta} \partial^\mu G^{\alpha\beta} -
  G^{\alpha\beta}G^{\gamma\delta} \partial_\mu B_{\alpha\gamma} \partial^\mu B_{\beta\delta}) \\
   \mathcal{L}_3 &=& -\frac 1 4 g^{\mu\rho}g^{\nu\lambda} (G_{\alpha\beta} F^{(1)\alpha}_{\mu\nu}
  F^{(1)\beta}_{\rho\lambda} + G^{\alpha\beta} H_{\mu\nu\alpha} H_{\rho\lambda\beta}) \\
  \mathcal{L}_4 &=& -\frac{1}{12}H_{\mu\nu\rho} H^{\mu\nu\rho}
\eea
In order to see the $SO(d,d,\IZ)$ symmetry, one introduces the $2d\times 2d$ matrix
\be
  M=\left( \begin{array}{cc}
 G^{-1}  & G^{-1}B \\
 BG^{-1} & \ \ G-BG^{-1}B
  \end{array}
  \right)
\ee
This symmetric matrix is in $SO(d,d)$, that is
\be
  M^T  \eta  M = \eta  \qquad
\qquad
  \eta\equiv\left( \begin{array}{cc}
 0  & 1_{2\times 2} \\
 1_{2\times 2} & 0
  \end{array}
  \right)
\ee
$M$ is positive definite which can be seen as follows. First notice that the above properties of
$M$ imply that the eigenvalues are present with their reciprocals,
\be
  M\vec{v} = \lambda \vec{v} \qquad \Longrightarrow \qquad   M(\eta \vec{v}) = \lambda^{-1} (\eta \vec{v})
\ee
Let us now turn off the B-field. The eigenvalues of $M(B=0)$ are simply the eigenvalues of $G$
and the reciprocals: $\lambda_i$ and $1/\lambda_i$, all positive. As we turn on the B-field, we
do not expect any singularities in the eigenvalues since $M$ is quadratic in $B$. Therefore the
eigenvalues remain positive.

Let us introduce
\be
  H_{\mu\nu\alpha}=\partial_\mu B_{\nu\alpha} - \partial_\nu B_{\mu\alpha} =: F^{(2)}_{\mu\nu \alpha}
\ee
we can collect the field strength in the following $SO(d,d)$ vector
\be
  \mathcal{F}^i_{\mu\nu} := \binom{ F^{(1)\alpha}_{\mu\nu}}{ F^{(2)}_{\mu\nu \beta}}
\ee
With these ingredients, one can explicitly see the $SO(3,3)$ invariance of the Lagrangian.
$\mathcal{L}_1$ is trivially invariant. The kinetic terms can be written as
\be
   \mathcal{L}_2 = \frac 1 8 \textrm{Tr}(\partial_\mu M^{-1} \partial^\mu M)
\ee
Also,
\be
  \mathcal{L}_3= -\frac 1 4 \mathcal{F}^i_{\mu\nu} (M^{-1})_{ij} \mathcal{F}^{\mu\nu j}
\ee
which is invariant. Since $H_{\mu\nu\rho}$ does not change under the duality group,
$\mathcal{L}_4$ is also invariant.

\section{Appendix: Semi-flat vs. exact solutions}
\label{sfvs}

In this section we compare the exact supergravity solutions to the semi-flat description. We
study the approximation through the example of an NS5-brane. NS5-branes are parametrically
heavier than D-branes\footnote{D-branes have a tension $T_\textrm{Dp-brane}=1/{g_s (l_s)^{p+1}}$
where $g_s$ is the string coupling and $l_s$ is the string length. On the other hand, NS5-branes
have tension $T_\textrm{NS5-brane}=1/{(g_s)^2 (l_s)^{6}}$ which is much larger at weak coupling.
} and they curve spacetime even to zeroth approximation.

\vskip 0.5cm \noindent {\bf Semi-flat approximation.}  In order for the semi-flat machinery to
work, we need to compactify the transverse space. The transverse space is now a two-torus fiber
over a complex $z$-plane. The NS5-brane can be described by a $\rho \mapsto \rho+1$ monodromy
around a singular point in the $z$-plane. This can be achieved by the following solution
\cite{Greene:1989ya}
\be
  \rho(z) \sim \frac{1}{2\pi i} \textrm{log}(z)
  \label{sfsol}
\ee
As we approach the brane ($|z|\rightarrow 0$), the torus fiber decompactifies,
\be
  V_\textrm{fiber} \sim \rho_2 \sim -\textrm{log}|z|
\ee
Since the eight dimensional dilaton can be set to constant \cite{Hellerman:2002ax}, the ten
dimensional dilaton is (Appendix \ref{redapp})\footnote{There is a slight change of variables
compared to Appendix \ref{redapp} which includes $\phi \rightarrow 2\phi$.}
\be
   2\phi = \frac 1 2 \textrm{log det} \,  G_{\alpha\beta}
\ee
that is
\be
  e^{2\phi} = V_\textrm{fiber}
\ee
and the dilaton grows near the origin\footnote{The \kahler potential for such 2d semi-flat
solutions can be computed and is given in \cite{Greene:1989ya, Loftin:2004qu}.}.

 \vskip 0.5cm \noindent {\bf Exact NS5 solution.} If $x^m$
are transverse to the NS5-brane and $x^\mu$ are tangent to it, then the exact non-compact
classical solution in the string frame is given by (see \cite{Polchinski:1998rr}, page 183)
\bea
  G_{mn} = e^{2\phi} \, \delta_{mn} &\qquad & G_{\mu\nu} = \eta_{\mu\nu} \\
  H_{mnp} = - \epsilon_{mnp}^q \partial_q \phi &\qquad &
  e^{2\phi} = e^{2\phi(\infty)} + \frac{1}{2\pi^2 r^2}
\eea
where $r^2 = \sum (x^m)^2$. The geometry has an infinite throat: the origin $x^m = 0$ is at
infinite distance and the angular $S^3$ approaches an asymptotic constant size. In the Einstein
frame ($G^\textrm{Einstein}_{\mu\nu} = e^{-\phi/2} \, G^\textrm{string}_{\mu\nu}$), the
singularity is at finite distance. There is a growing dilaton in the throat and string
perturbation theory eventually breaks down. At infinity, the metric asymptotes to flat space.

\vskip 0.5cm

In order to derive the semi-flat solution from the exact one, we need to compactify the latter
on a two-torus. In the covering space, this amounts to placing an infinite number of NS5-branes
in a 2d lattice $\Lambda$ in the 4d transverse space. For sake of simplicity, we take this to be
a square lattice. Since the branes are BPS, the solution comes from simple
superposition\footnote{Note that the sum was made convergent by subtracting an infinite
constant. For two dimensional lattices, this constant does not depend on $x$. This is the same
trick that one uses in the definition of Weierstrass's elliptic function. Elliptic functions
have been generalized to higher dimensions (see \cite{Saclioglu:1995de} and references therein,
\cite{Hellerman:unp}).},
\be
  e^{2\phi(x)} = e^{2\phi(\infty)} + \frac{1}{2\pi^2} \sum_{(n,m)\in \Lambda} \left( \frac{1}{ |x-x_{n,m}|^2} - \frac{1}{ |x_{n,m}|^2} \right)
\ee
where $x$ and $x_{n,m}$ are four-vectors, the latter one denoting the positions of the lattice
points parametrized by two integers $n$ and $m$.

If we neglect distances smaller than the lattice spacing, then we obtain
\be
  e^{2\phi(z,\bar z)} = e^{2\phi(\infty)} + \frac{1}{2\pi^2} \sum_{n,m} \left( \frac{1}{z\bar z + n^2+m^2} - \frac{1}{n^2+m^2} \right)
\ee
where we introduced a complex $z$ coordinate perpendicular to the 2d lattice. This is the base
coordinate. This expression can now be compared to the semi-flat solution. If we denote $r
\equiv |z|$, then taking the derivative w.r.t $r$ gives for the semi-flat solution
\be
  \partial_r \, e^{2\phi_\textrm{semi-flat}} =  \partial_r \, V_\textrm{fiber} \sim - \frac{1}{r}
\ee
For our expression,
\be
  \partial_r \, e^{2\phi(z,\bar z)} = -\frac{1}{2\pi^2} \sum_{n,m} \frac{2r}{r^2 + n^2+m^2} \approx -\int du \, dv \frac{2r}{r^2 + u^2+v^2} \sim -\frac{1}{r}
\ee
where we approximated the sum by an integral. Thus, we reproduced the semi-flat dilaton from the
exact one.

Although the semi-flat approximation reproduces the qualitative features of the exact solution,
due to the partially compactified transverse space, we have to deal with another problem. The
function (\ref{sfsol}) is valid only close to the brane. A semi-flat solution that extends to
the entire complex plane may be given by means of the $j$-function. This solution suffers from
the problem of orbifold points as we described in Section \ref{twodimsec} and it is not possible
to describe a single NS5-brane consistently. The exact solution avoids this problem by roughly
speaking going to the limit very close to the brane (in the $z$-plane) compared to the size of
the fiber. Hence, any possible orbifold points are pushed to an infinite distance (in the base)
and thus are invisible.


\begin{table}[htdp]
\begin{center}
\begin{tabular}{|c | c |}
\hline  {\bf example} &  {\bf potential} \\ \hline
  \textrm{flat space} &  $V(\vec{x}) = \frac{1}{|\vec{x}|}$ \\
  \textrm{Taub-NUT} &  $V(\vec{x}) = 1 + \frac{1}{|\vec{x}|}$ \\
  \textrm{Eguchi-Hanson} &  $V(\vec{x}) = \frac{1}{|\vec{x}-\vec{x_1}|} + \frac{1}{|\vec{x}-\vec{x_2}|}$ \\
\hline
\end{tabular}
\end{center}
\caption{Some well-known examples for the Gibbons-Hawking ansatz.}
\end{table}

\comment{ A Taub-NUT metric is described by
\be
  V(\vec{x}) = 1 + \frac{1}{|\vec{x}|}
\ee
The Eguchi-Hanson metric is given by
\be
  V(\vec{x}) = \frac{1}{|\vec{x}-\vec{x_1}|} + \frac{1}{|\vec{x}-\vec{x_2}|}
\ee
Flat space is
\be
  V(\vec{x}) = \frac{1}{|\vec{x}|}
\ee
}

\noindent {\bf The Gibbons-Hawking ansatz.} A perhaps more visual comparison of the
semi-flat and exact metrics is possible through the ansatz\footnote{Every 4d hyper-\kahler
metric with a (triholomorphic) Killing vector can be written in this form
\cite{Gibbons:1987sp}.},
\be
  ds^2 = V(\vec{x}) d\vec{x}^2 + V(\vec{x})^{-1} \, (dt + \vec{A}\cdot d\vec{x})^2
\ee
where $\vec{A}$ is given through $\vec{\nabla} \times \vec{A} =  \vec{\nabla} V$. This defines a
circle fibration over a three dimensional base parametrized by $\vec{x}$. A semi-flat solution
for a degenerating fiber is given through
\be
  \tau(z) \sim \frac{1}{2\pi i} \textrm{log} (z)
  \label{sfsolg}
\ee
This corresponds to \cite{Ooguri:1996me}
\be
  V = \tau_2 = -\frac{1}{2\pi} \textrm{log} \,|z|
\ee
\be
  A_x = \tau_1 = \frac{1}{4\pi i} \textrm{log}(z/\bar z) \qquad A_z=0 \qquad A_{\bar z} = 0
\ee
where the 3d $\vec{x}$ space has been decomposed into $(x, z, \bar z)$. This gives a singular
metric which is translationally invariant in the $x$ direction.
\begin{figure}[ht]
\begin{center}
  \includegraphics[totalheight=4.0cm,angle=0,origin=c]{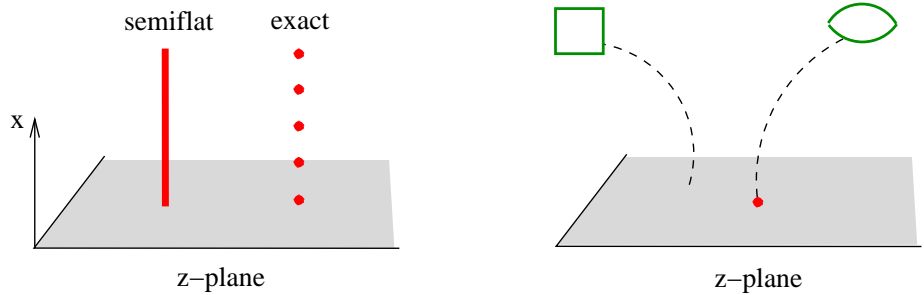}
  \caption{(i) Comparing semi-flat and exact metrics for around degenerating fibers.
  The base is 3d, parametrized by the periodic $x$ coordinate and the complex $z$-plane. The red line / red dots indicate where the $S^1$ fiber vanishes.
Translational invariance of the semi-flat solution is replaced by periodicity of the exact
metric in the $x$ direction. \ (ii) The same (exact) metric from a different viewpoint. The
horizontal direction in the torus fiber is the $x$ coordinate. The torus pinches at the
degeneration point (red dot) in the 2d base.  Topologically, the singular fiber is an $S^2$ with
two points glued together. This replaces the degenerating $\tau_2 \rightarrow \infty$ torus of
the semi-flat solution.}
  \label{insta}
\end{center}
\end{figure}

The exact non-singular hyper-\kahler metric is not translationally invariant, but periodic.
\be
V = \frac{1}{4\pi} \sum_{n=-\infty}^\infty \left( \frac{1}{\sqrt{(x-n)^2 + z\bar z}}
-\frac{1}{|n|} \right) +\textrm{const.}
\ee
In  \fref{insta} (i), red dots indicate where this potential is singular. An $S^3$ (the
``throat'') close to such a degeneration point can be seen as a Hopf-fibration with fiber $t$
above the $S^2$ surrounding the singularity. On the right-hand side of \fref{insta}, we see
again the SYZ-like fibration with a two-torus fiber above the complex plane. The torus fiber
degenerates into a ``pillow''\footnote{{\it Torus} was the Latin word for a torus-shaped
cushion.} above a codimension two locus.

It is possible to glue an approximate $K3$ metric from 24 such patches. For details, see
\cite{GrossWilson}. A more conventional way to obtain a smooth approximate metric is to start
with the singular $T^4 / \IZ_2$ orbifold and blow up the 16 fixed points. This is possible by
cutting out a small neighborhood (whose boundary is homeomorphic to $\IR \IP^3$) around the
fixed point and gluing there an Eguchi-Hanson space. This is the unique smooth hyper-\kahler
metric which asymptotes to $\IC^2 / \IZ_2$.

\newpage

\section{Appendix: The Hanany-Witten effect from the semiflat approximation}
\label{hwapp}

The Hanany-Witten effect \cite{Hanany:1996ie} is an interesting phenomenon of brane
creation. It can be used to construct brane configurations that realize four dimensional
$\mathcal{N}=1$ supersymmetric gauge theories \cite{ Elitzur:1997fh} which exhibit Seiberg
duality. This duality relates two different gauge theories which give the same infrared physics
\cite{Seiberg:1994pq}. In string theory, it is realized in a very geometric way: as the branes
move around, new branes appear which can change the rank of the gauge group in the 4d
theory\footnote{See related works \cite{Katz:1996fh, Ooguri:1997ih, Witten:1997sc,
Elitzur:1997hc, Feng:2000mi, Klebanov:2000hb, Cachazo:2001sg, Beasley:2001zp, Feng:2001bn,
Berenstein:2002fi, Franco:2005rj, Hanany:2005ss} and references therein.}.

The Hanany-Witten setup co tains D4-branes stretched between NS5-branes in the presence of
D6-branes. The D6, D4 and NS5 are magnetically charged under $C^{(1)}$, $C^{(3)}$ and $B$,
respectively. The branes that we are going to use have the following orientations
\begin{table}[htdp]
\begin{center}
\begin{tabular}{c| cccc |ccc| ccc}
  & 0 & 1 &  2 &  3 & 4 & 5 & 6 & 7 & 8 & 9 \\
\hline\hline {\bf NS5} & x & x & x & x & x & &  &  & & x \\
\hline {\bf D6} & x & x & x & x &  & x & & x & x &  \\
\hline {\bf D4} & x & x & x & x & & & x & & &
\end{tabular}
\end{center}
\caption{Branes in the Hanany-Witten setup. $456$ are the base, $789$ are the fiber
coordinates.}
\end{table}

\begin{figure}[ht]
\begin{center}
  \includegraphics[totalheight=4cm,angle=0,origin=c]{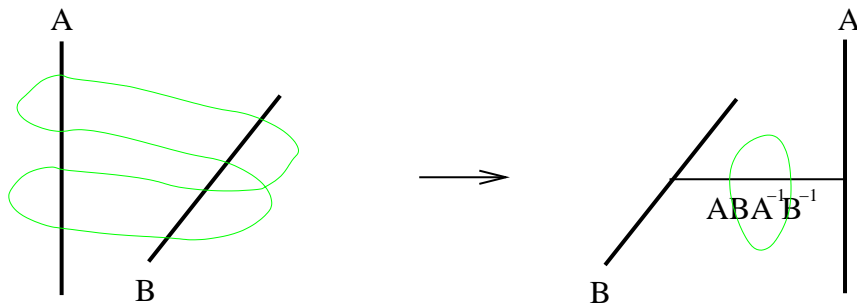}
  \caption{Hanany-Witten brane creation mechanism. $A$ and $B$ are the monodromies of the NS5- and D6-branes, respectively.
  As the two branes pass through each other, a new brane appears with a monodromy around the green circle (see right-hand side).
  This monodromy can be easily computed in the original configuration (left-hand side) where the green path was a deformed loop around the two branes.
  The result is $A B A^{-1} B^{-1}$ which is simply the monodromy of a D4-brane.}
  \label{HWs}
\end{center}
\end{figure}

Let us now consider an NS5-brane and a D6-brane as in the left-hand side of \fref{HWs}. As the
two branes pass though each other, a new brane is created. In order to verify that it is indeed
a D4-brane, we need to determine its monodromy. Ramond-Ramond charges $( C_7, C_8, C_9,
C_{789})$ transform in the dual fundamental representation of $SL(4)$, which means that we have
to use the transposed monodromy matrices.

The NS5 monodromy,
\be
  A = \mathcal{M}_\textrm{NS5} =
  \left( \begin{array}{rrrr}
  1 & 0 & 0 & 0 \\
  0 & 1 & 0 & 0 \\
  0 & 0 & 1 & 1 \\
  0 & 0 & 0 & 1
  \end{array}
  \right)
\ee
and the D6 monodromy\footnote{The D4- and D6-brane monodromies can be realized linearly if we
include the $x^{11}$ M-theory circle in the discussion.}
\be
  B: \ C_9 \mapsto C_9 + 1
\ee

\begin{figure}[ht]
\begin{center}
  \includegraphics[totalheight=5cm,angle=0,origin=c]{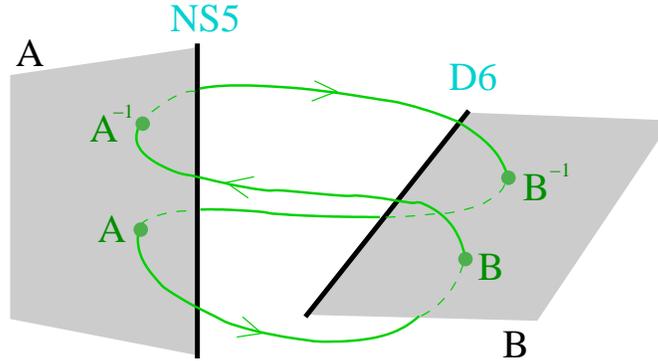}
  \caption{Determining the monodromy around the green loop by means of 2d branch cuts. }
  \label{HWs2}
\end{center}
\end{figure}

From these ingredients, we need to determine the monodromy along the green loop in the left-hand
side of \fref{HWs}. A moment's thought will convince the reader that it is the (group-theoretic)
commutator, $A B A^{-1} B^{-1}$. This is best seen by choosing ``branch cut planes'' that start
from the branes and studying how the green loop intersects these (see \fref{HWs2}).

Finally, the commutator can be computed
\be
  A B A^{-1} B^{-1}: \ C_{789} \mapsto C_{789} + 1
\ee
which is the monodromy of a D4-brane\footnote{A similar observation about monodromies has
recently been made by 't Hooft in \cite{tHooft:2008kk}}.

\newpage
\section{Appendix: Type IIA on ${\bf T^5/\IZ_2}$ and Type IIB on ${\bf S^1\times K3}$}
\label{t5app}

Type IIA string theory on $T^5/\IZ_2$ has been conjectured to be equivalent to Type IIB on
$S^1\times K3$ \cite{Witten:1995em, Dasgupta:1995zm}. We study this equivalence by means of the
semi-flat machinery.

The IIA orientifold $T^5/\IZ_2$ is generated by the action $\alpha \cdot \Omega$ that reverses
the sign of all the circles\footnote{An Op-plane has charge $2^{p-5}$ and is generated by the
action of $
  \left\{ \begin{array}{ll}
  R_{9-p} \Omega & \textrm{if $p=0,1$ (mod 4)}\\
  R_{9-p} \Omega (-1)^{F_L} & \textrm{if $p=2,3$ (mod 4).}
  \end{array}
  \right.
$ },
\be
  \alpha : (x^5, x^6, x^7, x^8, x^9) \mapsto (-x^5, -x^6, -x^7, -x^8, -x^9)
\ee
and changes the parity of the world-sheet. There are 32 D4-branes located at the fixed points
which cancel the RR tadpoles.

We cast the geometry in a fibration structure as follows. The 2d base coordinates will be $x^5$
and $x^6$. Since $\alpha$ inverts these coordinates, the base becomes $T^2/\IZ_2$, \ie a
semi-flat $S^2$. Over this sphere there is a $T^3$ fiber parametrized by $x^7$, $x^8$ and $x^9$.
The fiber degenerates at four points in the base. These singularities have a deficit angle of
$180^\circ$ (tension six, like $D_4$). The monodromy around the singular points in the base is
then
\be
  \mathcal{M} = R_{789} \cdot \Omega  
\ee
where $R_i$ is reflection of the i-th coordinate, $\Omega$ is the world-sheet parity
transformation. This monodromy already includes the monodromies of eight D4-branes which cancel
the RR-charges of the O4-plane. This is analogous to the orientifold limit of F-theory where the
O7-plane monodromy is
\be
  \mathcal{M}_{\textrm{O7}^{-}} = -T^{-4} \qquad \textrm{with} \quad T = \mathcal{M}_{\textrm{D7-brane}}=
  \left( \begin{array}{rr}
  1 & 1   \\
  0 & 1
  \end{array}
  \right)
\ee
which combines with the monodromy of the four D7-branes ($T^4$) to cancel the RR-charge. The
final monodromy matrix is then diagonal.

\begin{figure}[ht]
\begin{center}
  \includegraphics[totalheight=4cm,angle=0,origin=c]{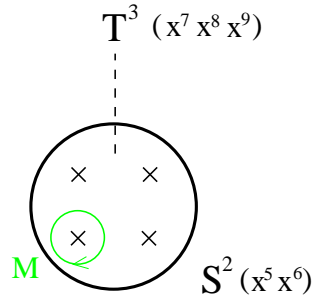}
  \caption{$T^5 / \IZ_2$ as a fibration over $S^2$. The geometric $T^3$ fiber gets promoted to $T^5$
  by adding $x^{10}$ and the M-theory circle $x^{11}$. The monodromy $ \mathcal{M}$ then acts on this $T^5$. }
  \label{t5pic}
\end{center}
\end{figure}

What is $ \mathcal{M}$ explicitly? Type IIB orientifolds at strong coupling can be described by
F-theory. As already mentioned in Section \ref{UdualityandGtwo}, the torus fiber of F-theory is
analogous to the $x^{11}-x^{10}$ coordinates in our case. Therefore, perturbative dualities are
not sufficient for determining the monodromy around the orientifold fixed points and the
$x^{11}$ coordinate should be included in the discussion. In the following, we determine the
$5\times 5$ U-duality monodromy matrices in the basis of $x^7-x^8-x^9-x^{11}-x^{10}$.

Reflection has an immediate interpretation in the vector representation of $SO(3,3)$. Since
inversion of a coordinate exchanges the two spinors that have different chirality, only an even
number of reflections give a symmetry of Type IIA. For instance, reflection of both $x^{7}$ and
$x^{8}$ can be represented by
\be
  R_{78} = \textrm{diag}(-1,-1,+1,-1,-1,+1) \in SO(3,3)
\ee
which inverts the momentum and the winding as well. It has the spinor representation
\be
  R_{78} = \textrm{diag}(-1,-1,+1,+1) \in SL(4)
\ee
Odd number of reflections must be accompanied with other internal symmetries. However, we can
still determine the corresponding monodromies, keeping in mind that we have to combine them with
other symmetries. The relevant reflection is that of $x^7-x^8-x^9$ which gives
\be
  R_{789} =  \textrm{diag}(-1, -1, -1, +1, -1)
\ee
The inversion of $x^{10}$ comes about because the three coordinate reflections change the sign
of the Levi-Civita pseudotensor that we used to convert the three-form field into a vector.

Let us consider the world-sheet parity transformation $\Omega$. In the Type IIA language, under
this transformation the metric, the dilaton and the Ramond-Ramond one-form are even, whereas the
B-field and the three-form are odd. In M-theory language this means that the three-form switches
sign. From this, one can determine the $\Omega$ monodromy to be
\be
  \Omega =  \, \pm \, \textrm{diag}(+1, +1, +1, +1, -1)
\ee
In order to fix the overall sign, let us consider an O6 orientifold plane. It is obtained by the
action of $R_{7 \, 8 \, 9 \, 11}$ which reduces to $(-1)^{F_L} \Omega R_{789}$ in the IIA limit
(\cite{Sen:1997kz}, see also \cite{Vafa:1995gm,Kachru:2001je}).

Since $(-1)^{F_L}$ changes the signs of the RR-fields, the corresponding monodromy is easy to
determine,\footnote{This transformation effectively reflects the $x^{11}$ coordinate. In
principle there could be an overall sign, but this can be fixed by remembering the $SL(4)$
representation of $(-1)^{F_L}$ which is simply $-\mathbbm{1}_{4\times 4}$.}
\be
  (-1)^{F_L} =  \textrm{diag}(-1, -1, -1, +1, -1)
\ee

Since
\be
  R_{7 \, 8 \, 9 \, 11} = \textrm{diag}(-1, -1, -1, -1, +1)
\ee
we obtain\footnote{This reflects the well-known fact that conjugation by $S$-duality takes
$(-1)^{F_L}$ to $\Omega$ in Type IIB.}
\be
  \Omega =  \textrm{diag}(-1, -1, -1, -1, +1)
\ee

From these ingredients we can now write down the explicit form of the monodromy in the
$T^5/\IZ_2$ orientifold of Type IIA,
\be
   \mathcal{M} = R_{789} \cdot \Omega = \textrm{diag}(+1, +1, +1, -1, -1)
\ee
where we immediately recognize the monodromy of a (conjugate) $D_4$ singularity
\be
   \mathcal{M} = U \underbrace{\textrm{diag}(-1, -1, +1, +1, +1)}_{R_{\tilde 7\tilde 8}} U^{-1}
\ee
Therefore by flipping $x^7-x^{11}$ and $x^8-x^{10}$, we obtain Type IIA on K3.\footnote{For the
weak coupling limit where Type IIA is defined, we need $R_{\widetilde{11}} = R_7 \rightarrow
0$.} The volume of the $T^2$ fiber is the inverse of the volume of the original $T^3$ fiber,
\be
  vol(T^2) = R_{\tilde 7} R_{\tilde 8} =  R_{11} R_{10} = R_{11} \frac{1}{R_{7} R_{8}R_{9}R_{11}}
  = \frac{1}{R_{7} R_{8}R_{9}} = \frac{1}{vol(T^3)}
\ee
Here we used the fact that the $T^5$ torus has unit volume in appropriate units.

Finally, by T-dualizing the spectator $x^9$ ($R_{\tilde 9} = 1/R_9$), we arrive at the final
equivalence,
\be
  \textrm{Type IIA on} \ T^5/\IZ_2 \ \textrm{orientifold}  \quad  \cong \quad  \textrm{Type IIB on} \
  K3\times S^1
\ee

\newpage
\section{Appendix: List of asymmetric orbifolds}

\label{alist}

In this Appendix, we list the asymmetric orbifold actions that realize the almost non-geometric
spaces of Section \ref{nongeot6}. The one-plaquette example was described in Section \ref{asod}.
We use the following trick \cite{Hellerman:2002ax}: some of the compact dimensions are
``unfolded'' and compactified back with an asymmetric action. The complexity of the model
depends on how many dimensions have to be unfolded.

\subsection*{Two-plaquette model}

\begin{figure}[ht]
\begin{center}
  \includegraphics[totalheight=4cm,angle=0,origin=c]{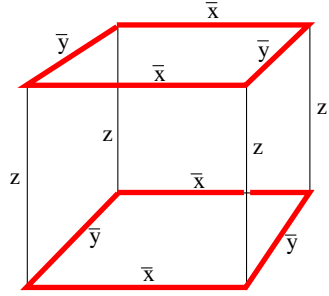}
  \caption{Almost non-geometric $T^6/\IZ_2\times\IZ_2$ which is also a Joyce orbifold. It is T-dual to $T^6/\IZ_2\times\IZ_2$.}
  \label{topbottom}
\end{center}
\end{figure}


\vskip -1cm
\bean
  \alpha : (\theta_1,\theta_2,\theta_3,\theta_4,\theta_5,\theta_6) & \mapsto &
  (-\theta_1,-\theta_2,-\theta_3,-\theta_4,\theta_5,\theta_6) \\
 \beta :
 (\theta_1,\theta_2,\theta_3,\theta_4,\theta_5,\theta_6)  & \mapsto &
  (-\theta_1,-\theta_2,\theta_3,\theta_4,-\theta_5,-\theta_6)  \ \times \ (-1)^{F_L}
\eean

\subsection*{Model ``U''}

\begin{figure}[ht]
\begin{center}
  \includegraphics[totalheight=4cm,angle=0,origin=c]{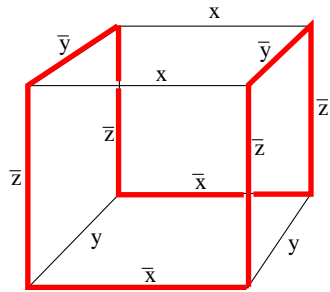}
  \caption{This modified $T^6/\IZ_2\times\IZ_2$ is also a Joyce manifold.}
  \label{fourthex}
\end{center}
\end{figure}

\vskip -1cm
\bean
 \alpha :
 (\theta_1,\theta_2,\theta_3,\theta_4,x_5,\theta_6) & \mapsto &
  (-\theta_1,-\theta_2,-\theta_3,-\theta_4,x_5,\theta_6)  \ \times \ (-1)^{F_L} \\
 \gamma_1 :
 (\theta_1,\theta_2,\theta_3,\theta_4,x_5,\theta_6) & \mapsto &
  (\theta_1,\theta_2,-\theta_3,-\theta_4,-x_5,-\theta_6)   \ \times \ (-1)^{F_L} \\
 \gamma_2 :
 (\theta_1,\theta_2,\theta_3,\theta_4,x_5,\theta_6) & \mapsto &
  (\theta_1,\theta_2,-\theta_3,-\theta_4,L-x_5,-\theta_6)
\eean

\clearpage
\subsection*{Model ``L''}

\begin{figure}[ht]
\begin{center}
  \includegraphics[totalheight=4cm,angle=0,origin=c]{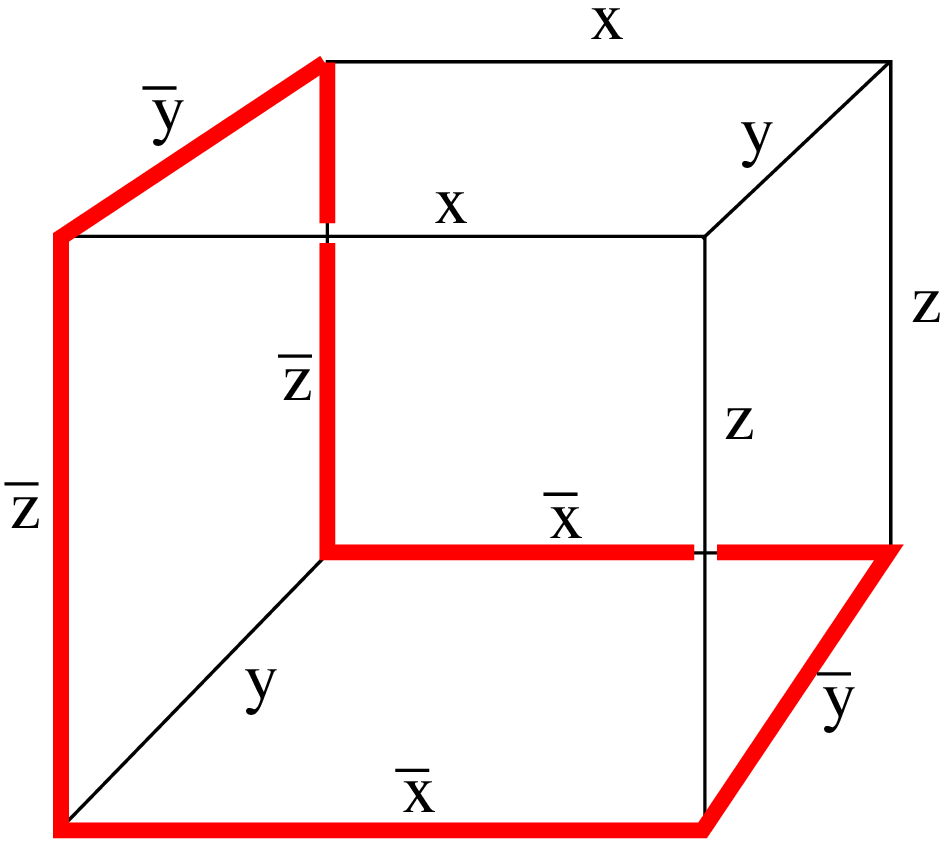}
\end{center}
\end{figure}

\vskip -1cm
\bean
 \alpha_1 :
 (x_1,\theta_2,\theta_3,\theta_4,x_5,\theta_6) & \mapsto &
  (-x_1,-\theta_2,-\theta_3,-\theta_4,x_5,\theta_6) \ \times \ (-1)^{F_L} \\
  \alpha_2 : (x_1,\theta_2,\theta_3,\theta_4,x_5,\theta_6) & \mapsto &
  (L-x_1,-\theta_2,-\theta_3,-\theta_4,x_5,\theta_6) \\
 \gamma_1 :
 (x_1,\theta_2,\theta_3,\theta_4,x_5,\theta_6) & \mapsto &
  (x_1,\theta_2,-\theta_3,-\theta_4,-x_5,-\theta_6) \ \times \ (-1)^{F_L} \\
 \gamma_2 :
 (x_1,\theta_2,\theta_3,\theta_4,x_5,\theta_6) & \mapsto &
  (x_1,\theta_2,-\theta_3,-\theta_4,L-x_5,-\theta_6)   
\eean

\subsection*{Model ``X''}

\begin{figure}[ht]
\begin{center}
  \includegraphics[totalheight=4cm,angle=0,origin=c]{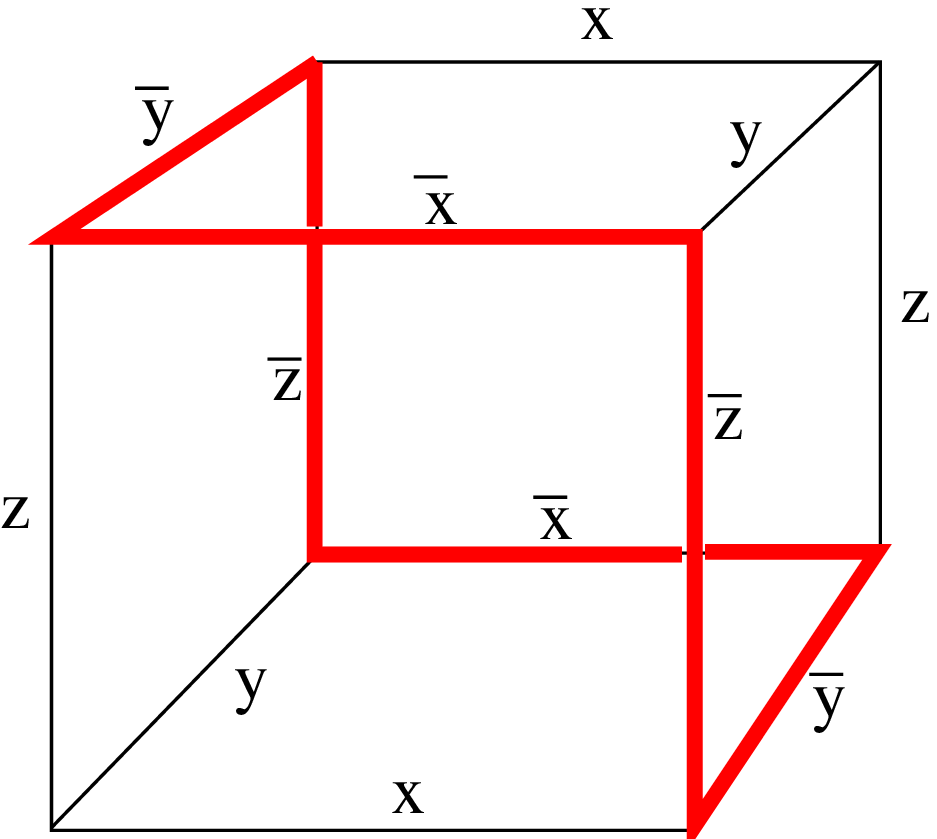}
\end{center}
\end{figure}

\vskip -1cm
\bean
 \alpha_1 :
 (x_1,\theta_2,x_3,\theta_4,x_5,\theta_6) & \mapsto &
  (-x_1,-\theta_2,-x_3,-\theta_4,x_5,\theta_6) \\
 \alpha_2 :
 (x_1,\theta_2,x_3,\theta_4,x_5,\theta_6) & \mapsto &
  (L-x_1,-\theta_2,-x_3,-\theta_4,x_5,\theta_6)   \ \times \ (-1)^{F_L}  \\
 \alpha_3 :
 (x_1,\theta_2,x_3,\theta_4,x_5,\theta_6) & \mapsto &
  (-x_1,-\theta_2,L-x_3,-\theta_4,x_5,\theta_6)   \ \times \ (-1)^{F_L}  \\
 \alpha_4 :
 (x_1,\theta_2,x_3,\theta_4,x_5,\theta_6) &  \mapsto &
  (L-x_1,-\theta_2,L-x_3,-\theta_4,x_5,\theta_6) \\
 \gamma_1 :
 (x_1,\theta_2,x_3,\theta_4,x_5,\theta_6)  & \mapsto &
  (x_1,\theta_2,-x_3,-\theta_4,-x_5,-\theta_6) \\
 \gamma_2 :
 (x_1,\theta_2,x_3,\theta_4,x_5,\theta_6) & \mapsto &
  (x_1,\theta_2,L-x_3,-\theta_4,-x_5,-\theta_6)   \ \times \ (-1)^{F_L}  \\
 \gamma_3 :
 (x_1,\theta_2,x_3,\theta_4,x_5,\theta_6) & \mapsto &
  (x_1,\theta_2,-x_3,-\theta_4,L-x_5,-\theta_6)   \ \times \ (-1)^{F_L}  \\
 \gamma_4 :
 (x_1,\theta_2,x_3,\theta_4,x_5,\theta_6) & \mapsto &
  (x_1,\theta_2,L-x_3,-\theta_4,L-x_5,-\theta_6)
\eean

\comment{
\newpage

{\bf Modular invariance. } For non-Abelian orbifolds, level-matching is not sufficient to
guarantee modular invariance at higher loops. Further constraints can arise if a modular
transformation takes a pair of commuting group elements $(g,h)$ into their own conjugacy class
\cite{Freed:1987qk},
\be
  (g,h) \longrightarrow (g^a h^b, g^c h^d) = (p g p^{-1}, p h p^{-1})
\ee
where $a,b,c$ and $d$ are the elements of an $SL(2,\IZ)$ matrix. In this case, the path integral
with boundary conditions $(g,h)$ and $(p g p^{-1}, p h p^{-1})$ for the torus world-sheet should
give the same result. We prove that when this happens we also have $(p g p^{-1}, p h p^{-1}) =
(g,h)$ for the examples in this Appendix. This means that we obtain no constraints other than
level-matching.

In the Green-Schwarz formalism, the boundary condition on the world-sheet fermions are

It is immediately clear that only the $x_i$ non-compact dimensions are relevant to our
discussion since non-commutativity of the elements comes from action on these coordinates. Such
an action can be a reflection or a translation. There are two non-trivial cases:

(i) $p g p^{-1} \ne g$ and $p h p^{-1} = h$.

(ii) $p g p^{-1} \ne g$ and $p h p^{-1} \ne h$.

}

\newpage
\section{Appendix: Spectrum of ${\bf T^6/\IZ_2 \times \IZ_2}$ and the two-plaquette model}

\label{aspectrum}

For comparison and as a warm-up exercise, in this Appendix we review the details of the
computation of massless spectra for ${T^6/\IZ_2 \times \IZ_2}$ \cite{Vafa:1994rv} and for its
T-dual, the two-plaquette model. We will work in the RNS formalism. The compactification
preserves $\mathcal{N}=2$ supersymmetry in four dimensions. The $\mathcal{N}=2$ multiplets are
listed in the Table \ref{tablen2}.


\begin{table}[htdp]
\begin{center}
\begin{tabular}{|c | c | c | c |}
\hline  
  {\bf hypermultiplet} & 2 fermions, \ 4 scalars \\
  {\bf vector multiplet} & vector, \ 2 fermions, \ 2 scalars \\
  {\bf supergravity multiplet} & graviton, \ 2 gravitini, \ vector\\
\hline
\end{tabular}
\end{center}
\caption{Massless $\mathcal{N}=2$ multiplets in four dimensions (Weyl fermions and real
scalars).} \label{tablen2}
\end{table}

\noindent
 {\bf The massless spectrum of $T^6/\IZ_2 \times \IZ_2$}

\begin{itemize}

\item {\bf untwisted sector}

The spectrum contains the states that are invariant under the orbifold projection. Each of the
orbifold group generators invert four spacetime coordinates. The action should also be specified
on the fermions. We use the following convention\footnote{Other conventions give the same
spectrum but preserve different supercharges.},
\be
  \alpha: (\pi, -\pi, 0) \quad
  \beta: (-\pi, 0, \pi) \quad
  \alpha\beta: (0, -\pi, \pi)
\ee

The action of these elements on the states are summarized in the following table,

\begin{table}[htdp]
\begin{center}
\begin{tabular}{|c | c | c | c |}
\hline  {\bf sector} &  {\bf state} & {\bf $\alpha$, $\beta$, $\alpha\beta$ charge}   \\
\hline
  $NS$ &  $\psi^{\mu}_{-1/2} |0; k\rangle $  & $+ + +$   \\
     &  $\psi^{4,5}_{-1/2} |0; k\rangle $  & $- - +$   \\
     &  $\psi^{6,7}_{-1/2} |0; k\rangle $  & $- + -$  \\
     &  $\psi^{8,9}_{-1/2} |0; k\rangle $  & $+ - -$   \\
\hline
  $R$ &  $|\textrm{\tiny gso}+++ \rangle,\ |\textrm{\tiny gso}--- \rangle $  & $+ + +$    \\
    &  $|\textrm{\tiny gso}++- \rangle,\ |\textrm{\tiny gso}--+ \rangle  $  & $+ - -$    \\
    &  $|\textrm{\tiny gso}+-+ \rangle,\ |\textrm{\tiny gso}-+- \rangle $  & $- + -$    \\
    &  $|\textrm{\tiny gso}+-- \rangle,\ |\textrm{\tiny gso}-++ \rangle  $  & $- - +$    \\
\hline
\end{tabular}
\end{center}
\caption{Untwisted NS and R sectors. In the R sector, only the spins of compact complex
dimensions are indicated. The remaining one is determined by the GSO projection as indicated by
the ``{\tiny gso}'' label. This depends on whether it's the left or right R sector.}
\end{table}

\begin{table}[htdp]
\begin{center}
\begin{tabular}{|c | c | c | c |}
\hline  {\bf sector} &  {\bf fields}  \\ \hline
  $NS_- / NS_-$   & $G_{\mu\nu}$, $B_{\mu\nu}$, dilaton, 12 real scalars   \\
  $NS_- / R_+ $   & gravitino, 7 Weyl fermions     \\
  $R_- / NS_- $   & gravitino, 7 Weyl fermions  \\
  $R_- / R_+ $    & 4 vectors, 4 cx. scalars  \\
\hline
\end{tabular}
\end{center}
\caption{Untwisted sectors. The signs show the matter GSO projection (which, due to the
superghost contributions, differ from the full GSO in the NS sectors).}
\end{table}

These states combine into 1 supergravity multiplet, 3 vector multiplets and 4 hypermultiplets
according to Table~\ref{tablen2}.

\vskip 4cm

\item {\bf twisted sectors}

There are $16+16+16=48$ fixed tori under $\alpha$, $\beta$ and $\alpha\beta$. The zero-point
energies vanish and both NS and R sectors have zero modes in the twisted and untwisted
directions, respectively.

\begin{table}[htdp]
\begin{center}
\begin{tabular}{|c | c | c | c |}
\hline  {\bf sector} &  {\bf state} & {\bf $\alpha$, $\beta$, $\alpha\beta$ charge}   \\
\hline
\hline
  $NS$ &  $| \, . \ . + +  \rangle$  & $-i, \ i, \ 1$    \\
     &  $| \, . \ . - -  \rangle$  & $i, \ -i, \ 1$    \\
\hline
  $R$ &  $|\textrm{\tiny gso} + . \ . \  \rangle$  & $i, \ -i, \ 1$    \\
    &  $|\textrm{\tiny gso} - . \ . \  \rangle$  & $-i, \ i, \ 1$    \\
\hline
\end{tabular}
\end{center}
\caption{$(\alpha\beta)$-twisted NS and R sectors. The other twisted sectors are analogous. The
dots indicate half-integer moded oscillators which generate massive states.}
\end{table}

\begin{table}[htdp]
\begin{center}
\begin{tabular}{|c | c | c | c |}
\hline  {\bf sector} &  {\bf fields}  \\ \hline
  $NS_+ / NS_+$   &  cx. scalar   \\
  $NS_+ / R_- $   & Weyl fermion   \\
  $R_+ / NS_+ $   & Weyl fermion  \\
  $R_+ / R_- $   & vector \\
\hline
\end{tabular}
\end{center}
\caption{Each twisted sector gives an $\mathcal{N}=2$ vector multiplet.}
\end{table}

These states give 48 vector multiplets\footnote{If we take Type IIB instead, then we get complex
scalars in the R--R sectors which then combine into 48 hypermultiplets. Turning on discrete
torsion has the same effect: in each twisted sector it changes the sign of projection for the
other two non-trivial group elements of $\IZ_2 \times \IZ_2$. For example, in the
$(\alpha\beta)$-twisted sector, the surviving R--R states must be even under $\alpha\beta$ (as
in the case without torsion), but {\it odd} under the $\alpha$ and $\beta$ transformations.}.

\end{itemize}

\comment{ In the various sectors, the (matter) GSO projection is given by
\bea
  \textrm{untwisted}: \quad NS_- / NS_- \quad   NS_- / R_+ \quad  R_- / NS_- \quad  R_- / R_+ \\
  \textrm{twisted}: \quad NS_+ / NS_+ \quad   NS_+ / R_- \quad  R_+ / NS_+ \quad  R_+ / R_-
\eea}
Vertex operators are local with respect to the eight supercharges
\bea
  e^{-\varphi/2} e^{\frac{i}{2}(H_0\pm H_1) \pm \frac{i}{2}(H_2+H_3+H_4)} \\
  e^{-\tilde \varphi/2} e^{\frac{i}{2}(\tilde H_0\pm \tilde H_1) \pm \frac{i}{2}(\tilde H_2+\tilde H_3+\tilde H_4)}
\eea

In $\mathcal{N}=2$ language, this gives altogether one supergravity, 51 vector and 4
hypermultiplets. This is consistent with the expectations that Type IIA compactified on a
Calabi-Yau should result in $h^{1,1}=51$ vector multiplets and $h^{2,1}+1=4$ hypermultiplets
\cite{Vafa:1994rv}.

In $\mathcal{N}=1$ language, a hyper is two chirals, a vector is a vector+chiral, and the
supergravity multiplet is a gravity + gravitino multiplet as seen from Table \ref{tablen1}.
Therefore, we obtain a gravity, a gravitino, 51 vector and 59 chiral multiplets.


\begin{table}[htdp]
\begin{center}
\begin{tabular}{|c | c | c | c |}
\hline  
  {\bf chiral multiplet} & fermion, \ 2 scalars \\
  {\bf vector multiplet} & vector, \ fermion \\
  {\bf gravitino multiplet} & gravitino, \ vector\\
  {\bf gravity multiplet} & graviton, \ gravitino \\
\hline
\end{tabular}
\end{center}
\caption{Massless $\mathcal{N}=1$ multiplets in four dimensions (Weyl fermions and real
scalars).} \label{tablen1}
\end{table}


\noindent
 {\bf The massless spectrum of the two-plaquette model.}
The theory is described in Section \ref{nongeot6} (see \fref{allcubes} for the singularity
structure). As discussed in Section \ref{dualitysec}, Type IIA on this background is T-dual to
Type IIB on ordinary $T^6/\IZ_2 \times \IZ_2$. As a further exercise, we compute the spectrum.
The theory is defined as a $\IZ_2 \times \IZ_2$ orbifold generated by $\alpha$ and $\beta$,
similarly to the previous section. In this case, however, $\alpha$ includes the action of
$(-1)^{F_L}$. In the RNS formalism, $(-1)^{F_L}$ does not act directly on the worldsheet fields.
It changes the sign of the left-moving spin-fields and hence acts as charge conjugation on
RR-fields. The GSO projection is switched in left-moving sectors twisted by $\mathcal{R}\cdot
(-1)^{F_L}$ compared to sectors twisted by $\mathcal{R}$ only. (Here $\mathcal{R}$ inverts four
spacetime coordinates.) This can be deduced using   the equivalence of the RNS and Green-Schwarz
formalisms. In the latter description, $(-1)^{F_L}$ changes the sign of the $\theta^a$
left-moving world-sheet spinor fields in the light-cone gauge (for a review, see \eg
\cite{Dabholkar:1997zd}).

We again need to impose GSO and orbifold invariance. In the twisted sectors, an ambiguity
arises: one can keep even or odd states under the action of a certain $\IZ_2$ generator. The
choices are constrained by modular invariance. The signs are shown in Table \ref{signtable}.

\begin{table}[htdp]
\begin{center}
\begin{tabular}{| c | c   c   c |}
\hline  & $\alpha$-twisted &  ${\bf \beta}$-twisted  &  ${\bf \alpha\beta}$-twisted   \\ \hline
  $P_\alpha$ & $-$ & $+$ & $-$ \\
  $P_\beta$ & $+$ & $+$ & $+$ \\
  $P_{\alpha\beta}$ & $-$ & $+$ & $-$ \\
\hline
\end{tabular}
\end{center}
\caption{Signs of projections in various twisted sectors.}
\label{signtable}
\end{table}

The signs on the diagonal are directly related to the coloring of the edges (\fref{allcubes}).
Perhaps the off-diagonal signs can be encoded in faces of the SYZ graph.

By the logic of Section \ref{dualitysec}, successive T-dualities attach $2\times 2$ blocks of
minus signs to Table \ref{signtable}. T-duality on a $T^3$ then produces Table \ref{signtable2}
which is precisely the choice of discrete torsion identified by \cite{Vafa:1994rv} in the mirror
of ${T^6/\IZ_2 \times \IZ_2}$ (see also \cite{Gaberdiel:2004vx, Douglas:1998xa}).

\begin{table}[htdp]
\begin{center}
\begin{tabular}{| c | c   c   c |}
\hline  & $\alpha$-twisted &  ${\bf \beta}$-twisted  &  ${\bf \alpha\beta}$-twisted   \\ \hline
  $P_\alpha$ & $+$ & $-$ & $-$ \\
  $P_\beta$ & $-$ & $+$ & $-$ \\
  $P_{\alpha\beta}$ & $-$ & $-$ & $+$ \\
\hline
\end{tabular}
\end{center}
\caption{Assignment of signs for discrete torsion.} \label{signtable2}
\end{table}

\begin{itemize}

\item {\bf untwisted sector:} Same result as untwisted $T^6/\IZ_2 \times \IZ_2$.

\item {\bf $\beta$ twisted sectors}

There are $16$ fixed tori under $\beta$. The zero-point energies vanish and the GSO projection
is the same as that of the $T^6/\IZ_2 \times \IZ_2$ twisted sectors. In particular, we have
$+/+$ in the NS/NS sector and $+/-$ in the R/R sector. The orbifold projection preserves
$\alpha$ even and $\beta$ even states. Due to $(-1)^{F_L}$, $\alpha$ and $\alpha\beta$ have an
extra minus sign in the left R sector. The left and right states combine to give 16
hypermultiplets.

\begin{table}[htdp]
\begin{center}
\begin{tabular}{|c | c | c | c |}
\hline  {\bf sector} &  {\bf state} & {\bf $\alpha$, $\beta$, $\alpha\beta$ charge}   \\
\hline \hline
  $NS$ &  $| \, .  +  . \ +  \rangle$  & $i, \ 1, \ i$    \\
     &  $| \, . -  . \ -  \rangle$  & $-i, \ 1, \ -i$    \\
\hline
  $R_{left}$ &  $|+ .  + . \  \rangle$  & $i, \ 1, \ i$    \\
    &  $|- .  - . \  \rangle$  & $-i, \ 1, \ -i$    \\
\hline
  $R_{right}$ &  $|- .  + . \  \rangle$  & $-i, \ 1, \ -i$    \\
    &  $|+ .  - . \  \rangle$  & $i, \ 1, \ i$    \\
\hline
\end{tabular}
\end{center}
\caption{$\beta$-twisted NS and R sectors. }
\end{table}

\item {\bf $\alpha$ and $\alpha\beta$ twisted sectors}

There are $16+16$ fixed tori under the two group elements. The GSO projection is $+/-$ in the
NS/NS sector and $-/-$ in the R/R sector. The orbifold projection preserves $\alpha$ odd and
$\beta$ even states, \ie the twisted-sector vacuum has $\alpha$-charge $(-1)$. These twisted
sectors give $16+16=32$ hypermultiplets.

\clearpage

\begin{table}[htdp]
\begin{center}
\begin{tabular}{|c | c | c | c |}
\hline  {\bf sector} &  {\bf state} & {\bf $\alpha$, $\beta$, $\alpha\beta$ charge}   \\
\hline \hline
  $NS_{left}$ &  $| \, . \ . + -  \rangle$   & $-i, \ -i, \ -1$    \\
     &  $| \, . \ . - +  \rangle$   & $i, \ i, \ -1$    \\
\hline
  $NS_{right}$ &  $| \, . \ . + +  \rangle$   & $-i, \ i, \ 1$    \\
     &  $| \, . \ . - -  \rangle$   & $i, \ -i, \ 1$    \\
\hline
  $R_{left}$ &  $|- + \ . \ . \  \rangle$  & $-i, \ -i, \ -1$    \\
    &  $|+ - \ . \ . \  \rangle$  & $i, \ i, \ -1$    \\
\hline
  $R_{right}$ &  $|- + \ . \ . \  \rangle$  & $i, \ -i, \ 1$    \\
    &  $|+ - \ . \ . \  \rangle$  & $-i, \ i, \ 1$    \\
\hline
\end{tabular}
\end{center}
\caption{$(\alpha\beta)$-twisted NS and R sectors. }
\end{table}

\end{itemize}

Altogether we obtain a gravity multiplet, 52 hypermultiplets and 3 vector multiplets. Thus, the
counting reproduces the massless spectrum of Type IIB on $T^6/\IZ_2 \times \IZ_2$.

\section{Appendix: Spectrum of the one-plaquette model}

\label{aspectrum2}

The background is flat space divided by
\begin{displaymath}
\begin{array}{lllrl}
  \alpha :  (\theta_1,\theta_2,\theta_3,\theta_4,\theta_5,\theta_6) & \mapsto &
  (-\theta_1,-\theta_2,-\theta_3,-\theta_4, & +\theta_5, & +\theta_6) \\
   \beta_1 : (\theta_1,\theta_2,\theta_3,\theta_4,\theta_5,\theta_6) & \mapsto &
  (-\theta_1,-\theta_2,+\theta_3,+\theta_4, & -\theta_5, & -\theta_6) \\
 \beta_2 : (\theta_1,\theta_2,\theta_3,\theta_4,\theta_5,\theta_6) & \mapsto &
  (-\theta_1,-\theta_2,+\theta_3,+\theta_4, & \frac{1}{2}-\theta_5, & -\theta_6) \ \times \ (-1)^{F_L}
\end{array}
\end{displaymath}
This gives $\mathcal{N}=1$ supersymmetry as we will see.  Note that $\beta_2 \cdot\beta_1$
defines a $(-1)^{F_L}$ Wilson-line for the $\theta_5$ base coordinate\footnote{For generic
circle radius, the resulting states are massive however. Massless states arise from the sector
of zero momentum and winding. For further details, the reader is referred to
\cite{Hellerman:2005ja}.}. The signs of the projection in the twisted sectors we employ are
given in Table \ref{signtable3}. They are motivated by the logic of the previous example.

\begin{table}[htdp]
\begin{center}
\begin{tabular}{| c | c c c c c |}
\hline  & \ \ \ $\alpha$ &  \ \ ${\beta_1}$  & \ \ ${ \beta_2}$  &  \ ${ \alpha\beta_1}$  & \  ${ \alpha\beta_2}$    \\
\hline
  $P_\alpha$                  &  \ \ \ $+$      &  \ \ $+$       &  \ \ $+$      & $+$       & $+$      \\
  $P_{\beta_1}$               &  \ \ \ $+$      &  \ \ $+$       &  \ \ $\cdot$  & $+$       & $\cdot$  \\
  $P_{\beta_2}$               &  \ \ \ $+$      &  \ \ $\cdot$   &  \ \ $-$      & $\cdot$   & $-$      \\
  $P_ {\alpha\beta_1}$        &  \ \ \ $+$      &  \ \ $+$       &  \ \ $\cdot$  & $+$       & $\cdot$  \\
  $P_ {\alpha\beta_2}$        &  \ \ \ $+$      &  \ \ $\cdot$   &  \ \ $-$      & $\cdot$   & $-$      \\
\hline
\end{tabular}
\end{center}
\caption{Assignment of phases for the twisted sectors (columns). Dots indicate signs that do not
affect the spectrum calculation. The group elements that are not listed here have no non-trivial
fixed loci.} \label{signtable3}
\end{table}


\begin{itemize}

\item {\bf untwisted sector}

We need to carry out a projection on the invariant subspace. On the left NS states, $(-1)^{F_L}$
acts trivially. Therefore, the NS/NS and NS/R sectors are the same as those of $T^6/\IZ_2 \times
\IZ_2$.

The R/NS and R/R sectors on the other hand do not contribute anything because there is no
massless state invariant under both $\beta_1$ and $\beta_2$. In particular, this means that half
of the gravitini are projected out compared to $T^6/\IZ_2 \times \IZ_2$. We will see that no
extra gravitini arise in the twisted sectors and hence only $\mathcal{N}=1$ supersymmetry is
preserved in four dimensions. The fields combine into a gravity multiplet and seven chiral
multiplets.

\begin{table}[htdp]
\begin{center}
\begin{tabular}{|c | c | c | c |}
\hline  {\bf sector} &  {\bf fields}  \\ \hline
  $NS_- / NS_-$  & $G_{\mu\nu}, B_{\mu\nu}$, dilaton, 12 real scalars   \\
  $NS_- / R_+ $   & gravitino, 7 Weyl fermions     \\
  $R_- / NS_- $   & \textrm{---}  \\
  $R_- / R_+ $    & \textrm{---}  \\
\hline
\end{tabular}
\end{center}
\caption{Untwisted closed sectors.}
\end{table}


\vskip 1cm


\item {\bf $\alpha$ twisted sectors}

There are $16$ fixed tori. Zero point energies vanish both in the NS and R sectors. These states
give 16 chiral multiplets. The R/R and R/NS sectors do not contribute because no states are
invariant under both $\beta_1$ and $\beta_2$.

\begin{table}[htdp]
\begin{center}
\begin{tabular}{|c | c | c | c |}
\hline  {\bf sector} &  {\bf fields}  \\ \hline
  $NS_+ / NS_+$   &  cx. scalar   \\
  $NS_+ / R_- $   & Weyl fermion   \\
  $R_+ / NS_+ $   & \textrm{---}  \\
  $R_+ / R_- $   & \textrm{---} \\
\hline
\end{tabular}
\end{center}
\caption{$\alpha$-twisted sector: a chiral multiplet.}
\end{table}

\clearpage


\item {\bf $\beta_1$, $\alpha\beta_1$ twisted sectors}

There are $8+8$ invariant fixed tori, respectively, because $\beta_2$ permutes them in pairs.
Zero point energies vanish both in the NS and R sectors. Each fixed locus gives an
$\mathcal{N}=2$ vector multiplet, so in $\mathcal{N}=1$ language we obtain 16 chiral multiplets
and 16 vector multiplets.

\begin{table}[htdp]
\begin{center}
\begin{tabular}{|c | c | c | c |}
\hline  {\bf sector} &  {\bf fields}  \\ \hline
  $NS_+ / NS_+$   &  cx. scalar   \\
  $NS_+ / R_- $   & Weyl fermion   \\
  $R_+ / NS_+ $   & Weyl fermion  \\
  $R_+ / R_- $   & vector \\
\hline
\end{tabular}
\end{center}
\caption{Twisted sector: a vector and a chiral multiplet.}
\end{table}


\item {\bf $\beta_2$, $\alpha\beta_2$ twisted sectors}

These $8+8$ sectors contain a twist by $(-1)^{F_L}$. The GSO projection is switched for all the
left-moving states. Zero point energies still vanish as moding is not affected by $(-1)^{F_L}$.
These states give 32 chiral multiplets.

\begin{table}[htdp]
\begin{center}
\begin{tabular}{|c | c | c | c |}
\hline  {\bf sector} &  {\bf fields}  \\ \hline
  $NS_- / NS_+$   &  cx. scalar   \\
  $NS_- / R_- $   & Weyl fermion   \\
  $R_- / NS_+ $   & Weyl fermion  \\
  $R_- / R_- $   & cx. scalar \\
\hline
\end{tabular}
\end{center}
\caption{Twisted sectors that include $(-1)^{F_L}$: two chiral multiplets. The left-moving GSO
projections are modified compared to the usual twisted sectors.}
\end{table}

\end{itemize}

\vskip 1cm

The other orbifold group elements have no fixed points. Vertex operators are local with respect
to four right-moving supercharges,
\be
  e^{-\tilde \varphi/2} e^{\frac{i}{2}(\tilde H_0\pm \tilde H_1) \pm \frac{i}{2}(\tilde H_2+\tilde H_3+\tilde H_4)}
\ee
Finally, we obtain an $\mathcal{N}=1$ gravity multiplet, 16 vector multiplets and 71 chiral
multiplets. It would be good to explicitly check modular invariance of the partition function
for this example.

\newpage
\section{Appendix: Spectra of Joyce orbifolds}

\label{joycespectrum}

In this Appendix, we describe the spectra of two seven dimensional Joyce manifolds interpreted
as non-geometric Type IIA compactifications down to four dimensions. The $T^7/(\IZ_2)^3$
orbifolds are generated by the following involutions,
\begin{displaymath}
\begin{array}{llllrrrrrrrr}
  \alpha &:& (x_1, x_2, x_3 \, | \, y_1, y_2, y_3, y_4) & \mapsto ( & x_1, &  -x_2, &  -x_3 &  \, |  \, &   y_1,  & y_2,  & -y_3, &  -y_4) \\
  \beta &:& (x_1, x_2, x_3  \, | \,  y_1, y_2, y_3, y_4) &  \mapsto ( & -x_1,  & x_2,  & b_2-x_3 &  \, |  \,  &  y_1, &  -y_2, &  y_3,  & b_1-y_4) \\
  \gamma &:& (x_1, x_2, x_3 \, |  \,  y_1, y_2, y_3, y_4) &  \mapsto ( & c_5-x_1, &  c_3-x_2,  & x_3  & \, |  \,   & {-y_1},  & y_2,  & y_3,  & c_1-y_4)
\end{array}
\end{displaymath}
where $b_1, b_2, c_1, c_3, c_5 \in \{0, \frac{1}{2}\}$ are constants. Note that
$\alpha^2=\beta^2=\gamma^2=1$ and $\alpha$, $\beta$ and $\gamma$ commute. The action preserves
the $G_2$-structure
\bean
  \varphi = dx_1 \wedge dy_1 \wedge dy_2 + dx_2 \wedge dy_1 \wedge dy_3 + dx_3 \wedge dy_2 \wedge
  dy_3 + dx_2 \wedge dy_2 \wedge dy_4 \\
  - dx_3 \wedge dy_1 \wedge dy_4 - dx_1 \wedge dy_3 \wedge
  dy_4 - dx_1 \wedge dx_2 \wedge dx_3 \qquad
\eean
The notation is the same as that of \cite{Joyce2}, but we reshuffled the coordinates to
distinguish between base and fiber directions.

\vskip 0.5cm
 {\bf Example with three shifts:} $(b_1, b_2, c_1, c_3, c_5) =
(0,\frac{1}{2},\frac{1}{2},\frac{1}{2},0,0)$

This is Example 3 in \cite{Joyce2}. The Betti numbers are computed to be $b^2=12$, $b^3=43$.
Therefore when M-theory is compactified on this manifold, 12 vector multiplets and 43 chiral
multiplets are obtained. In order to compute the U-dual Type IIA spectrum, we first need to
choose the $x^{10}$ direction. This can be chosen to be $y_1$ since none of the $\IZ_2$ actions
contain a shift in this direction. However, $\gamma$ inverts $y_1$ and thus $(-1)^{F_L}$ must be
separated from this transformation. This means that the geometric action on $T^6$ has inverted
fiber coordinates for $\gamma$,
\begin{displaymath}
\begin{array}{llllrrrrrrrr}
  \alpha_0 &:& (x_1, x_2, x_3 \, | \, y_2, y_3, y_4) & \mapsto ( & x_1,  & -x_2,  & -x_3 &  \, |  \,   & y_2,  & -y_3,  & -y_4) & \\
  \beta_0 &:& (x_1, x_2, x_3  \, | \,  y_2, y_3, y_4) &  \mapsto ( & -x_1,  & x_2,  & \frac{1}{2}-x_3  & \, |  \,  &  {-y_2},  & y_3,  & -y_4) & \\
  \gamma_0 &:& (x_1, x_2, x_3 \, |  \,  y_2, y_3, y_4) &  \mapsto ( & -x_1,  & \frac{1}{2}-x_2,  & x_3  & \, |  \,   & {-y_2},  & -y_3, &
\frac{1}{2}+y_4) & \ \times \ (-1)^{F_L}
\end{array}
\end{displaymath}
The base and fiber coordinates nicely pair up. The orbifold group preserves the volume form of
$T^6$ whose real part is obtained from $\varphi$. The untwisted sector contributes a gravity
multiplet and seven chiral multiplets similarly to the non-geometric $T^6/\IZ_2 \times \IZ_2$ in
Appendix \ref{aspectrum}. Twisted sectors arise at the fixed $T^2$ tori of $\alpha_0, \beta_0$
and $\beta_0\gamma_0$. On the set of fixed loci for an element say $\alpha_0$, the other two
group elements ($\beta_0$ and $\beta_0\gamma_0$) act freely by permuting the tori. Therefore, we
obtain $4+4+4$ two-tori each giving a vector and three chiral multiplets. This gives altogether
12 vector and 43 chiral multiplets which matches the U-dual M-theory result\footnote{This may be
a coincidence since there was a half-shift in the fiber.}.

\clearpage

\begin{table}[htdp]
\begin{center}
\begin{tabular}{|c | c | c | c |}
\hline  {\bf sector} &  {\bf fields}  \\ \hline
  $NS / NS$   &  2 cx. scalars   \\
  $NS / R $   & 2 Weyl fermions   \\
  $R / NS $   & 2 Weyl fermions  \\
  $R / R  $   & vector, cx. scalar \\
\hline
\end{tabular}
\end{center}
\caption{Twisted sectors for $(b_1, b_2, c_1, c_3, c_5) =
(0,\frac{1}{2},\frac{1}{2},\frac{1}{2},0,0)$ give a vector and three chiral multiplets.}
\end{table}


 {\bf Example with two shifts:} $(b_1, b_2, c_1, c_3, c_5) =
(0,\frac{1}{2},\frac{1}{2},0,0,0)$

This is Example 4 in \cite{Joyce2}. The Betti numbers are  $b^2=8+l$, $b^3=47-l$ where $l \in
\{0,\ldots, 8\}$. The non-trivial elements with fixed loci are $\alpha, \beta$ and $\gamma$.
These fix 48 copies of $T^3$. The group $\langle \beta,\gamma\rangle$ permutes the 16 three-tori
fixed by $\alpha$, and $\langle \alpha,\gamma\rangle$ permutes the tori fixed by $\beta$. These
give $4+4$ copies of $T^3$. However, the action of the element $\alpha\beta$ is trivial on the
tori fixed by $\gamma$. Therefore, we obtain 8 copies of $T^3/\IZ_2$. There are two
topologically distinct ways to resolve each of these singularities and the choice of $l$
distinguishes between the various cases.

Similarly to the previous example, we can try to interpret this $G_2$ space as a Type IIA
background
\begin{displaymath}
\begin{array}{llllrrrrrrrr}
  \alpha_0 &:& (x_1,   x_2,  x_3  \, | \,   y_2,   y_3,  y_4) \mapsto ( & x_1,  & -x_2,  & -x_3  & \, |  \,   & y_2,  & -y_3, &  -y_4) & \\
  \beta_0 &:& (x_1,  x_2,  x_3    \, | \,  y_2,   y_3,   y_4) \mapsto ( & -x_1,  & x_2,  & 1/2-x_3  & \, |  \,  &  {-y_2}, &  y_3, &  -y_4) & \\
  \gamma_0 &:& (x_1,  x_2,  x_3   \, |  \,  y_2,  y_3,   y_4) \mapsto ( & -x_1,  & -x_2,  & x_3  & \, |  \,   & {-y_2},  & -y_3, &
1/2+y_4) & \ \times \ (-1)^{F_L}
\end{array}
\end{displaymath}
where $\gamma_0$ also includes the action of $(-1)^{F_L}$. The untwisted sector again gives a
gravity multiplet and seven chiral multiplets. The non-trivial elements $\alpha_0, \beta_0,
\alpha_0\gamma_0$ and $\beta_0\gamma_0$ give twisted sectors with massless fields. Taking into
account the permutations by other group elements, $\alpha_0$ and $\beta_0$ give $4+4$ $T^2$
tori. These sectors each contribute a vector and three chiral multiplets.

Let us now consider the sectors that contain a twist by $(-1)^{F_L}$. As opposed to the seven
dimensional interpretation where one obtained 8 copies of $T^3/\IZ_2$, here the $16+16$ two-tori
fixed by $\alpha_0\gamma_0$ and $\beta_0\gamma_0$ are permuted by the other group elements which
gives 8 copies of $T^2$. Since each of them gives a vector and three chiral multiplets, the
spectrum does not match that of the M-theory compactification.

What went wrong? Since the monodromies contained a $1/2$ shift, there is an ambiguity in the
definition of $(-1)^{F_L}$. This can be seen by redefining the coordinates $\tilde y \equiv
y+\frac{1}{4}$ which adds half-shifts for the fiber coordinates for the action of $(-1)^{F_L}$.
The resulting monodromy is an affine transformation. We do not know how to fix this ambiguity in
the general case.

By separating the $T^7$ coordinates into $x_i$ and $y_j$, we have chosen a coassociative
four-cycle for the fiber (\ie $\varphi|_\textrm{fiber}=0$). The terms in the flat
$G_2$-structure $\varphi$ basically tell us which coordinate triples can be chosen for base
coordinates. Out of $\binom{7}{3}=35$ choices, there are precisely seven for which the $T^4$
fiber is coassociative. For some of the choices, however, an element of the $(\IZ_2)^3$ group
would be interpreted as an overall orbifolding by $(-1)^{F_L}$ in which case U-duality does not
work. For example, if we choose $x_1, y_1$ and $y_2$ for the base coordinates, then $\alpha$
inverts the four fiber coordinates everywhere and the local model as $T^4$ over a base breaks
down.

In the case of the three-shift example, the puzzle with the fiber shifts can be avoided if
instead of $x_i$, we take $x_3, y_1$ and $y_4$ for base coordinates. Then, there will be no
shifts in the fiber and we expect a perfect agreement with the M-theory spectrum. Picking $y_3$
for the $x^{10}$ coordinate, the generators have the following interpretation in Type IIA,
\begin{displaymath}
\begin{array}{llllrrrrrrrr}
  \alpha_0 &:& (x_3, y_1, y_4 \, | \, x_1, x_2, y_2) & \mapsto ( & {-x_3},  & y_1, & -y_4 &  \, | \, & {-x_1}, &  x_2, &  -y_2) & \ \times \ (-1)^{F_L}\\
  \beta_0 &:& (x_3, y_1, y_4 \, | \, x_1, x_2, y_2) & \mapsto ( & \frac{1}{2}-x_3, &  y_1, & -y_4 & \, | \, & {-x_1}, & x_2, &  -y_2) & \\
  \gamma_0 &:& (x_3, y_1, y_4 \, | \, x_1, x_2, y_2) & \mapsto ( & x_3, &  -y_1, & \frac{1}{2}-y_4 & \, | \, & {-x_1}, & -x_2, &
y_2) &
\end{array}
\end{displaymath}
Twisted sectors come from $\alpha_0$, $\beta_0$ and $\gamma_0$. Similarly to the M-theory case,
$\beta_0$ and $\gamma_0$ contributes $4+4$ fixed $T^2$. The $\alpha_0$-twisted sector gives 8
$T^2/\IZ_2$ and thus this Type IIA spectrum indeed reproduces the U-dual M-theory spectrum.

Another representation of the same model is possible by noticing that $\alpha\beta$ defines a
$(-1)^{F_L}$ Wilson line for the $x_3$ fiber coordinate. Using this Wilson line, we are left
with two generators,
\begin{displaymath}
\begin{array}{llllrrrrrrrr}
  \alpha_0 &:& (x_2, y_2, y_4 \, | \, x_1, x_3, y_1) & \mapsto ( & x_2, &  -y_2 & -y_4   &  \, | \, & {-x_1}, & {-x_3}, & y_1 ) & \ \times \ (-1)^{F_L} \\
  \gamma_0 &:& (x_2, y_2, y_4 \, | \, x_1, x_3, y_1) & \mapsto ( & -x_2,  & y_2, & \frac{1}{2}-y_4   & \, | \, & {-x_1},  & x_3, &  -y_1
) &
\end{array}
\end{displaymath}
It is an example for chiral Scherk-Schwarz reduction (see Section \ref{chirals}).


\vskip 0.5cm
 {\bf Example with one shift:} $(b_1, b_2, c_1, c_3, c_5) =
(0,\frac{1}{2},0,0,0,0)$

For too few $1/2$ shifts, the orbifold has ``bad singularities'' (intersecting fixed loci) and
the proper desingularization to a smooth $G_2$ holonomy manifold is more complicated
\cite{Joyce:book}. These spaces can, however, still be embedded in string theory.
\begin{displaymath}
\begin{array}{llllrrrrrrrr}
  \alpha_0 &:& (x_1, x_2, x_3 \, | \, y_2, y_3, y_4) & \mapsto ( & {x_1},  & -x_2, & -x_3 &  \, | \, & {y_2}, &  -y_3, &  -y_4) & \\
  \beta_0 &:& (x_1, x_2, x_3 \, | \, y_2, y_3, y_4) & \mapsto ( & {-x_1}, &  x_2, & \frac{1}{2}-x_3 & \, | \, & {-y_2}, & y_3, &  -y_4) & \\
  \gamma_0 &:& (x_1, x_2, x_3 \, | \, y_2, y_3, y_4) & \mapsto ( & {-x_1}, &  -x_2, & x_3 & \, | \, & {-y_2}, & -y_3, &
y_4) & \ \times \ (-1)^{F_L}
\end{array}
\end{displaymath}

This background is dual to the one-plaquette model. Assuming that U-duality works, the spectrum
calculation of Appendix \ref{aspectrum2} is a prediction for the Betti numbers of a resolution
of this singular $G_2$ orbifold. Indeed, $b_2=16$ and $b_3=71$ is one of the possibilities as
discussed in Section 12.5 in \cite{Joyce:book}. Some of the many remaining possibilities are
presumably connected to this model by turning on discrete torsion \cite{Gaberdiel:2004vx}.

\newpage
\section{Appendix: Polyhedron patterns}
\label{patternapp}

\begin{figure}[ht]
\begin{center}
  \includegraphics[totalheight=8.0cm,angle=90,origin=c]{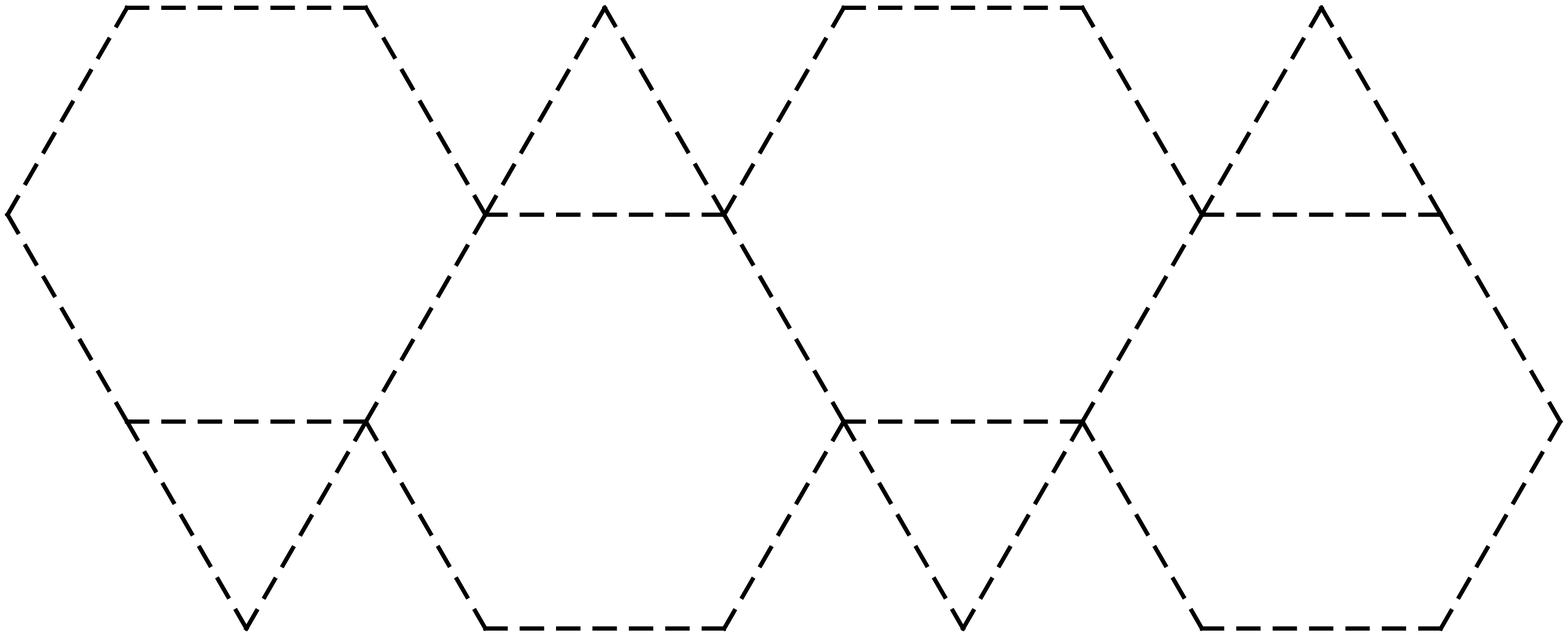} \hspace{-1cm}
  \includegraphics[totalheight=5.1cm,angle=30]{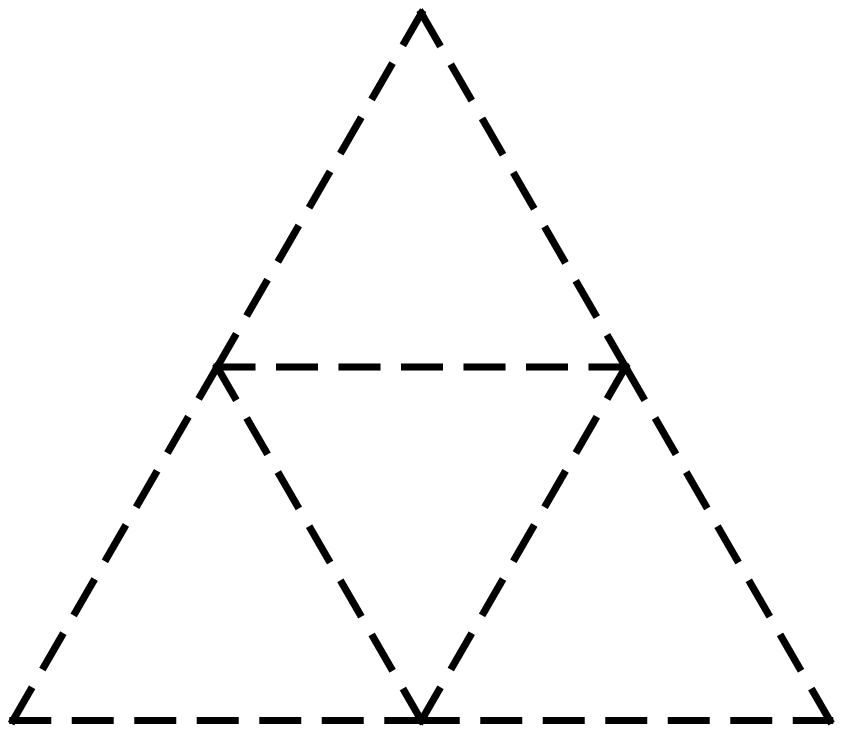}
  \caption{Truncated tetrahedron: the shifted $T^6/\IZ_2\times\IZ_2$. See Figure 15 for where to stick the nubbin.}
  \label{tthpattern}
\end{center}
\end{figure}


\begin{figure}[ht]
\begin{center}
\vskip -4cm
  \includegraphics[totalheight=10.0cm,angle=70,origin=c]{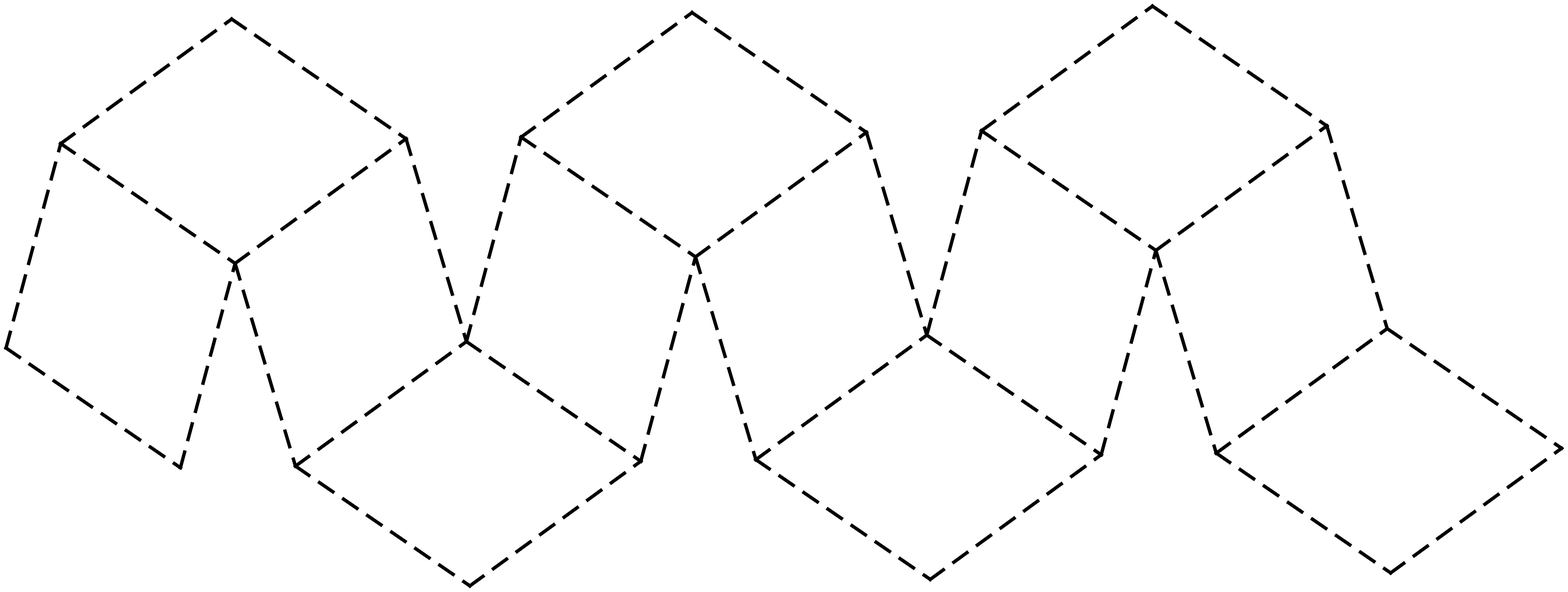}
  \vskip -1cm
  \caption{Rhombic dodecahedron: fundamental domain of $T^6/\IZ_2\times\IZ_2$, described in section 2.3.}
  \label{rdhpattern}
\end{center}
\end{figure}

\clearpage

\begin{figure}[ht]
\begin{center}
  \includegraphics[totalheight=22.0cm,angle=0,origin=c]{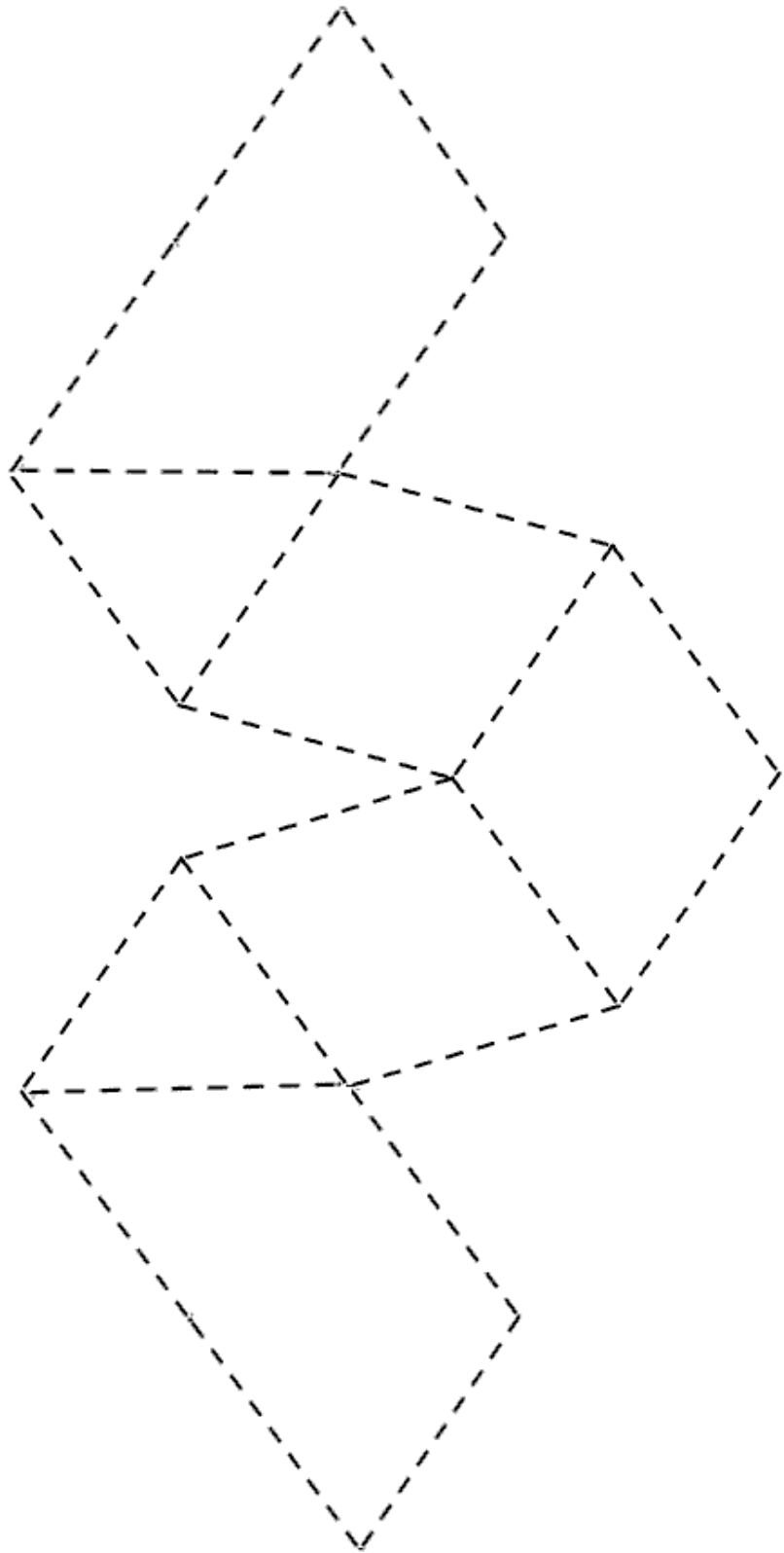}
  \caption{The $T^6 / \Delta_{12}$ fundamental domain described in section 5.3.}
  \label{re6pattern}
\end{center}
\end{figure}

\clearpage


\newpage
\bibliography{semiflat}

\providecommand{\href}[2]{#2}\begingroup\raggedright\begin{thebibliography}{10%
0}

\bibitem{Douglas:2006es}
M.~R. Douglas and S.~Kachru, {\it {Flux compactification}},  {\em Rev. Mod.
  Phys.} {\bf 79} (2007) 733--796
  [\href{http://arXiv.org/abs/hep-th/0610102}{{\tt hep-th/0610102}}].

\bibitem{Hertzberg:2007wc}
M.~P. Hertzberg, S.~Kachru, W.~Taylor and M.~Tegmark, {\it {Inflationary
  Constraints on Type IIA String Theory}},  {\em JHEP} {\bf 12} (2007) 095
  [\href{http://arXiv.org/abs/arXiv:0711.2512}{{\tt arXiv:0711.2512}}].

\bibitem{Dimopoulos:2005ac}
S.~Dimopoulos, S.~Kachru, J.~McGreevy and J.~G. Wacker, {\it {N-flation}},
  \href{http://arXiv.org/abs/hep-th/0507205}{{\tt hep-th/0507205}}.

\bibitem{Grimm:2007hs}
T.~W. Grimm, {\it {Axion Inflation in Type II String Theory}},  {\em Phys.
  Rev.} {\bf D77} (2008) 126007 [\href{http://arXiv.org/abs/0710.3883}{{\tt
  0710.3883}}].

\bibitem{Florea:2006si}
B.~Florea, S.~Kachru, J.~McGreevy and N.~Saulina, {\it {Stringy instantons and
  quiver gauge theories}},  {\em JHEP} {\bf 05} (2007) 024
  [\href{http://arXiv.org/abs/hep-th/0610003}{{\tt hep-th/0610003}}].

\bibitem{Blumenhagen:2007sm}
R.~Blumenhagen, S.~Moster and E.~Plauschinn, {\it {Moduli Stabilisation versus
  Chirality for MSSM like Type IIB Orientifolds}},  {\em JHEP} {\bf 01} (2008)
  058 [\href{http://arXiv.org/abs/0711.3389}{{\tt 0711.3389}}].

\bibitem{Shelton:2005cf}
J.~Shelton, W.~Taylor and B.~Wecht, {\it {Nongeometric flux
  compactifications}},  {\em JHEP} {\bf 10} (2005) 085
  [\href{http://arXiv.org/abs/hep-th/0508133}{{\tt hep-th/0508133}}].

\bibitem{Silverstein:2007ac}
E.~Silverstein, {\it {Simple de Sitter Solutions}},  {\em Phys. Rev.} {\bf D77}
  (2008) 106006 [\href{http://arXiv.org/abs/0712.1196}{{\tt 0712.1196}}].

\bibitem{Adams:2006kb}
A.~Adams, M.~Ernebjerg and J.~M. Lapan, {\it {Linear models for flux vacua}},
  \href{http://arXiv.org/abs/hep-th/0611084}{{\tt hep-th/0611084}}.

\bibitem{Adams:2007vp}
A.~Adams, {\it {Conformal field theory and the Reid conjecture}},
  \href{http://arXiv.org/abs/hep-th/0703048}{{\tt hep-th/0703048}}.

\bibitem{Hellerman:2002ax}
S.~Hellerman, J.~McGreevy and B.~Williams, {\it Geometric constructions of
  nongeometric string theories},  {\em JHEP} {\bf 01} (2004) 024
  [\href{http://arXiv.org/abs/hep-th/0208174}{{\tt hep-th/0208174}}].

\bibitem{Greene:1989ya}
B.~R. Greene, A.~D. Shapere, C.~Vafa and S.-T. Yau, {\it Stringy cosmic strings
  and noncompact {C}alabi-{Y}au manifolds},  {\em Nucl. Phys.} {\bf B337}
  (1990) 1.

\bibitem{Vafa:1996xn}
C.~Vafa, {\it Evidence for {F}-theory},  {\em Nucl. Phys.} {\bf B469} (1996)
  403--418 [\href{http://arXiv.org/abs/hep-th/9602022}{{\tt hep-th/9602022}}].

\bibitem{Strominger:1996it}
A.~Strominger, S.-T. Yau and E.~Zaslow, {\it Mirror symmetry is {T}-duality},
  {\em Nucl. Phys.} {\bf B479} (1996) 243--259
  [\href{http://arXiv.org/abs/hep-th/9606040}{{\tt hep-th/9606040}}].

\bibitem{Dasgupta:1996ij}
K.~Dasgupta and S.~Mukhi, {\it {F-theory at constant coupling}},  {\em Phys.
  Lett.} {\bf B385} (1996) 125--131
  [\href{http://arXiv.org/abs/hep-th/9606044}{{\tt hep-th/9606044}}].

\bibitem{GrossWilson}
M.~Gross and P.~M.~H. Wilson, {\it {Large complex structure limits of K3
  surfaces}},  {\em J. Differential Geom.} {\bf 55} (2000) 475--546
  [\href{http://arXiv.org/abs/math.DG/0008018}{{\tt math.DG/0008018}}].

\bibitem{Buscher:1987sk}
T.~H. Buscher, {\it {A Symmetry of the String Background Field Equations}},
  {\em Phys. Lett.} {\bf B194} (1987) 59.

\bibitem{Buscher:1987qj}
T.~H. Buscher, {\it {Path Integral Derivation of Quantum Duality in Nonlinear
  Sigma Models}},  {\em Phys. Lett.} {\bf B201} (1988) 466.

\bibitem{Loftin:2004qu}
J.~Loftin, S.-T. Yau and E.~Zaslow, {\it {Affine Manifolds, SYZ Geometry, and
  the Y Vertex}},  \href{http://arXiv.org/abs/math/0405061}{{\tt
  math/0405061}}.

\bibitem{Gaberdiel:1997ud}
M.~R. Gaberdiel and B.~Zwiebach, {\it {Exceptional groups from open strings}},
  {\em Nucl. Phys.} {\bf B518} (1998) 151--172
  [\href{http://arXiv.org/abs/hep-th/9709013}{{\tt hep-th/9709013}}].

\bibitem{Bershadsky:1996nh}
M.~Bershadsky {\em et.~al.}, {\it {Geometric singularities and enhanced gauge
  symmetries}},  {\em Nucl. Phys.} {\bf B481} (1996) 215--252
  [\href{http://arXiv.org/abs/hep-th/9605200}{{\tt hep-th/9605200}}].

\bibitem{Gross:1999hc}
M.~Gross, {\it Topological mirror symmetry},
  \href{http://arXiv.org/abs/math/9909015}{{\tt math/9909015}}.

\bibitem{Tomasiello:2005bp}
A.~Tomasiello, {\it {Topological mirror symmetry with fluxes}},  {\em JHEP}
  {\bf 06} (2005) 067 [\href{http://arXiv.org/abs/hep-th/0502148}{{\tt
  hep-th/0502148}}].

\bibitem{Ruan1}
W.-D. Ruan, {\it Lagrangian torus fibration of quintic {C}alabi-{Y}au
  hypersurfaces {I}: {F}ermat quintic case},
  \href{http://arXiv.org/abs/math/9904012}{{\tt math/9904012}}.

\bibitem{Joyce:2000ru}
D.~Joyce, {\it Singularities of special {L}agrangian fibrations and the {SYZ}
  conjecture},  \href{http://arXiv.org/abs/math/0011179}{{\tt math/0011179}}.

\bibitem{Sen:1997xi}
A.~Sen, {\it {String network}},  {\em JHEP} {\bf 03} (1998) 005
  [\href{http://arXiv.org/abs/hep-th/9711130}{{\tt hep-th/9711130}}].

\bibitem{Fairbairn}
W.~M. {Fairbairn}, T.~{Fulton} and W.~H. {Klink}, {\it {Finite and Disconnected
  Subgroups of $SU_{3}$ and their Application to the Elementary-Particle
  Spectrum}},  {\em Journal of Mathematical Physics} {\bf 5} (July, 1964)
  1038--1051.

\bibitem{Hull:1994ys}
C.~M. Hull and P.~K. Townsend, {\it {Unity of superstring dualities}},  {\em
  Nucl. Phys.} {\bf B438} (1995) 109--137
  [\href{http://arXiv.org/abs/hep-th/9410167}{{\tt hep-th/9410167}}].

\bibitem{Kumar:1996zx}
A.~Kumar and C.~Vafa, {\it {U-manifolds}},  {\em Phys. Lett.} {\bf B396} (1997)
  85--90 [\href{http://arXiv.org/abs/hep-th/9611007}{{\tt hep-th/9611007}}].

\bibitem{Kaloper:1999yr}
N.~Kaloper and R.~C. Myers, {\it {The O(dd) story of massive supergravity}},
  {\em JHEP} {\bf 05} (1999) 010
  [\href{http://arXiv.org/abs/hep-th/9901045}{{\tt hep-th/9901045}}].

\bibitem{Hitchin:2004ut}
N.~Hitchin, {\it {Generalized Calabi-Yau manifolds}},  {\em Quart. J. Math.
  Oxford Ser.} {\bf 54} (2003) 281--308
  [\href{http://arXiv.org/abs/math/0209099}{{\tt math/0209099}}].

\bibitem{Gualtieri:2003dx}
M.~Gualtieri, {\it Generalized complex geometry},
  \href{http://arXiv.org/abs/math/0401221}{{\tt math/0401221}}.

\bibitem{Grana:2004bg}
M.~Grana, R.~Minasian, M.~Petrini and A.~Tomasiello, {\it {Supersymmetric
  backgrounds from generalized Calabi-Yau manifolds}},  {\em JHEP} {\bf 08}
  (2004) 046 [\href{http://arXiv.org/abs/hep-th/0406137}{{\tt
  hep-th/0406137}}].

\bibitem{Kachru:2002sk}
S.~Kachru, M.~B. Schulz, P.~K. Tripathy and S.~P. Trivedi, {\it {New
  supersymmetric string compactifications}},  {\em JHEP} {\bf 03} (2003) 061
  [\href{http://arXiv.org/abs/hep-th/0211182}{{\tt hep-th/0211182}}].

\bibitem{Gurrieri:2002wz}
S.~Gurrieri, J.~Louis, A.~Micu and D.~Waldram, {\it {Mirror symmetry in
  generalized Calabi-Yau compactifications}},  {\em Nucl. Phys.} {\bf B654}
  (2003) 61--113 [\href{http://arXiv.org/abs/hep-th/0211102}{{\tt
  hep-th/0211102}}].

\bibitem{Grana:2006kf}
M.~Grana, R.~Minasian, M.~Petrini and A.~Tomasiello, {\it {A scan for new N=1
  vacua on twisted tori}},  {\em JHEP} {\bf 05} (2007) 031
  [\href{http://arXiv.org/abs/hep-th/0609124}{{\tt hep-th/0609124}}].

\bibitem{Lawrence:2006ma}
A.~Lawrence, M.~B. Schulz and B.~Wecht, {\it {D-branes in nongeometric
  backgrounds}},  {\em JHEP} {\bf 07} (2006) 038
  [\href{http://arXiv.org/abs/hep-th/0602025}{{\tt hep-th/0602025}}].

\bibitem{Dabholkar:2005ve}
A.~Dabholkar and C.~Hull, {\it {Generalised T-duality and non-geometric
  backgrounds}},  {\em JHEP} {\bf 05} (2006) 009
  [\href{http://arXiv.org/abs/hep-th/0512005}{{\tt hep-th/0512005}}].

\bibitem{Hull:2004in}
C.~M. Hull, {\it {A geometry for non-geometric string backgrounds}},  {\em
  JHEP} {\bf 10} (2005) 065 [\href{http://arXiv.org/abs/hep-th/0406102}{{\tt
  hep-th/0406102}}].

\bibitem{Flournoy:2004vn}
A.~Flournoy, B.~Wecht and B.~Williams, {\it {Constructing nongeometric vacua in
  string theory}},  {\em Nucl. Phys.} {\bf B706} (2005) 127--149
  [\href{http://arXiv.org/abs/hep-th/0404217}{{\tt hep-th/0404217}}].

\bibitem{Hull:2006va}
C.~M. Hull, {\it {Doubled geometry and T-folds}},  {\em JHEP} {\bf 07} (2007)
  080 [\href{http://arXiv.org/abs/hep-th/0605149}{{\tt hep-th/0605149}}].

\bibitem{Shelton:2006fd}
J.~Shelton, W.~Taylor and B.~Wecht, {\it {Generalized flux vacua}},  {\em JHEP}
  {\bf 02} (2007) 095 [\href{http://arXiv.org/abs/hep-th/0607015}{{\tt
  hep-th/0607015}}].

\bibitem{Becker:2006ks}
K.~Becker, M.~Becker, C.~Vafa and J.~Walcher, {\it {Moduli stabilization in
  non-geometric backgrounds}},  {\em Nucl. Phys.} {\bf B770} (2007) 1--46
  [\href{http://arXiv.org/abs/hep-th/0611001}{{\tt hep-th/0611001}}].

\bibitem{Maharana:1992my}
J.~Maharana and J.~H. Schwarz, {\it Noncompact symmetries in string theory},
  {\em Nucl. Phys.} {\bf B390} (1993) 3--32
  [\href{http://arXiv.org/abs/hep-th/9207016}{{\tt hep-th/9207016}}].

\bibitem{Brace:1998xz}
D.~Brace, B.~Morariu and B.~Zumino, {\it {T-duality and Ramond-Ramond
  backgrounds in the matrix model}},  {\em Nucl. Phys.} {\bf B549} (1999)
  181--193 [\href{http://arXiv.org/abs/hep-th/9811213}{{\tt hep-th/9811213}}].

\bibitem{Schwarz:1998qj}
A.~S. Schwarz, {\it Morita equivalence and duality},  {\em Nucl. Phys.} {\bf
  B534} (1998) 720--738 [\href{http://arXiv.org/abs/hep-th/9805034}{{\tt
  hep-th/9805034}}].

\bibitem{Brace:1998ku}
D.~Brace, B.~Morariu and B.~Zumino, {\it {Dualities of the matrix model from
  T-duality of the Type II string}},  {\em Nucl. Phys.} {\bf B545} (1999)
  192--216 [\href{http://arXiv.org/abs/hep-th/9810099}{{\tt hep-th/9810099}}].

\bibitem{Cremmer:1979up}
E.~Cremmer and B.~Julia, {\it The {$SO(8)$} supergravity},  {\em Nucl. Phys.}
  {\bf B159} (1979) 141.

\bibitem{Sezgin:1982gi}
E.~Sezgin and A.~Salam, {\it {Maximal extended supergravity theory in
  seven-dimensions}},  {\em Phys. Lett.} {\bf B118} (1982) 359.

\bibitem{Cremmer:1997ct}
E.~Cremmer, B.~Julia, H.~Lu and C.~N. Pope, {\it Dualisation of dualities.
  {I}},  {\em Nucl. Phys.} {\bf B523} (1998) 73--144
  [\href{http://arXiv.org/abs/hep-th/9710119}{{\tt hep-th/9710119}}].

\bibitem{Gukov:2002jv}
S.~Gukov, S.-T. Yau and E.~Zaslow, {\it Duality and fibrations on {$G_2$}
  manifolds},  \href{http://arXiv.org/abs/hep-th/0203217}{{\tt
  hep-th/0203217}}.

\bibitem{Narain:1986qm}
K.~S. Narain, M.~H. Sarmadi and C.~Vafa, {\it Asymmetric orbifolds},  {\em
  Nucl. Phys.} {\bf B288} (1987) 551.

\bibitem{Mueller:1986yr}
M.~T. Mueller and E.~Witten, {\it Twisting toroidally compactified heterotic
  strings with enlarged symmetry groups},  {\em Phys. Lett.} {\bf B182} (1986)
  28.

\bibitem{Dine:1997ji}
M.~Dine and E.~Silverstein, {\it {New M-theory backgrounds with frozen
  moduli}},  \href{http://arXiv.org/abs/hep-th/9712166}{{\tt hep-th/9712166}}.

\bibitem{Dabholkar:1998kv}
A.~Dabholkar and J.~A. Harvey, {\it {String islands}},  {\em JHEP} {\bf 02}
  (1999) 006 [\href{http://arXiv.org/abs/hep-th/9809122}{{\tt
  hep-th/9809122}}].

\bibitem{Blumenhagen:2000fp}
R.~Blumenhagen, L.~Gorlich, B.~Kors and D.~Lust, {\it {Asymmetric orbifolds,
  noncommutative geometry and type I string vacua}},  {\em Nucl. Phys.} {\bf
  B582} (2000) 44--64 [\href{http://arXiv.org/abs/hep-th/0003024}{{\tt
  hep-th/0003024}}].

\bibitem{Gaberdiel:2002jr}
M.~R. Gaberdiel and S.~Schafer-Nameki, {\it {D-branes in an asymmetric
  orbifold}},  {\em Nucl. Phys.} {\bf B654} (2003) 177--196
  [\href{http://arXiv.org/abs/hep-th/0210137}{{\tt hep-th/0210137}}].

\bibitem{Aoki:2004sm}
K.~Aoki, E.~D'Hoker and D.~H. Phong, {\it {On the construction of asymmetric
  orbifold models}},  {\em Nucl. Phys.} {\bf B695} (2004) 132--168
  [\href{http://arXiv.org/abs/hep-th/0402134}{{\tt hep-th/0402134}}].

\bibitem{Kakushadze:1996hi}
Z.~Kakushadze and S.~H.~H. Tye, {\it {Asymmetric orbifolds and grand
  unification}},  {\em Phys. Rev.} {\bf D54} (1996) 7520--7544
  [\href{http://arXiv.org/abs/hep-th/9607138}{{\tt hep-th/9607138}}].

\bibitem{Vafa:1986wx}
C.~Vafa, {\it {Modular Invariance and Discrete Torsion on Orbifolds}},  {\em
  Nucl. Phys.} {\bf B273} (1986) 592.

\bibitem{Hellerman:2006tx}
S.~Hellerman and J.~Walcher, {\it {Worldsheet CFTs for flat monodrofolds}},
  \href{http://arXiv.org/abs/hep-th/0604191}{{\tt hep-th/0604191}}.

\bibitem{Freed:1987qk}
D.~S. Freed and C.~Vafa, {\it {Global anomalies on orbifolds}},  {\em Commun.
  Math. Phys.} {\bf 110} (1987) 349.

\bibitem{Joyce1}
D.~D. Joyce, {\it Compact {R}iemannian 7-manifolds with holonomy {$G_2$}. {I}},
   {\em J. Diff. Geom.} {\bf 43} (1996) 291--328.

\bibitem{Joyce2}
D.~D. Joyce, {\it Compact {R}iemannian 7-manifolds with holonomy {$G_2$}.
  {II}},  {\em J. Diff. Geom.} {\bf 43} (1996) 329--375.

\bibitem{Joyce:book}
D.~D. Joyce, {\it {Compact manifolds with special holonomy}},  {\em Oxford
  University Press} (2000).

\bibitem{Vafa:1994rv}
C.~Vafa and E.~Witten, {\it On orbifolds with discrete torsion},  {\em J. Geom.
  Phys.} {\bf 15} (1995) 189--214
  [\href{http://arXiv.org/abs/hep-th/9409188}{{\tt hep-th/9409188}}].

\bibitem{Lee:2002fa}
J.-H. Lee and N.~C. Leung, {\it {Geometric structures on $G_2$ and
  Spin(7)-manifolds}},  \href{http://arXiv.org/abs/math/0202045}{{\tt
  math/0202045}}.

\bibitem{Vafa:1995gm}
C.~Vafa and E.~Witten, {\it Dual string pairs with {N=1} and {N=2}
  supersymmetry in four dimensions},  {\em Nucl. Phys. Proc. Suppl.} {\bf 46}
  (1996) 225--247 [\href{http://arXiv.org/abs/hep-th/9507050}{{\tt
  hep-th/9507050}}].

\bibitem{Sen:1996na}
A.~Sen, {\it {Duality and Orbifolds}},  {\em Nucl. Phys.} {\bf B474} (1996)
  361--378 [\href{http://arXiv.org/abs/hep-th/9604070}{{\tt hep-th/9604070}}].

\bibitem{Gaberdiel:2004vx}
M.~R. Gaberdiel and P.~Kaste, {\it {Generalised discrete torsion and mirror
  symmetry for $G_2$ manifolds}},  {\em JHEP} {\bf 08} (2004) 001
  [\href{http://arXiv.org/abs/hep-th/0401125}{{\tt hep-th/0401125}}].

\bibitem{Greene:1998vz}
B.~R. Greene, C.~I. Lazaroiu and M.~Raugas, {\it {D-branes on nonabelian
  threefold quotient singularities}},  {\em Nucl. Phys.} {\bf B553} (1999)
  711--749 [\href{http://arXiv.org/abs/hep-th/9811201}{{\tt hep-th/9811201}}].

\bibitem{Berenstein:2000mb}
D.~Berenstein, V.~Jejjala and R.~G. Leigh, {\it {D-branes on singularities: New
  quivers from old}},  {\em Phys. Rev.} {\bf D64} (2001) 046011
  [\href{http://arXiv.org/abs/hep-th/0012050}{{\tt hep-th/0012050}}].

\bibitem{Hanany:1998sd}
A.~Hanany and Y.-H. He, {\it {Non-Abelian finite gauge theories}},  {\em JHEP}
  {\bf 02} (1999) 013 [\href{http://arXiv.org/abs/hep-th/9811183}{{\tt
  hep-th/9811183}}].

\bibitem{Feng:2000af}
B.~Feng, A.~Hanany, Y.-H. He and N.~Prezas, {\it {Discrete torsion, non-Abelian
  orbifolds and the Schur multiplier}},  {\em JHEP} {\bf 01} (2001) 033
  [\href{http://arXiv.org/abs/hep-th/0010023}{{\tt hep-th/0010023}}].

\bibitem{Dabholkar:1996pc}
A.~Dabholkar and J.~Park, {\it {Strings on Orientifolds}},  {\em Nucl. Phys.}
  {\bf B477} (1996) 701--714 [\href{http://arXiv.org/abs/hep-th/9604178}{{\tt
  hep-th/9604178}}].

\bibitem{Gutperle:2000bf}
M.~Gutperle, {\it {Non-BPS D-branes and enhanced symmetry in an asymmetric
  orbifold}},  {\em JHEP} {\bf 08} (2000) 036
  [\href{http://arXiv.org/abs/hep-th/0007126}{{\tt hep-th/0007126}}].

\bibitem{Hellerman:2005ja}
S.~Hellerman, {\it {New type II string theories with sixteen supercharges}},
  \href{http://arXiv.org/abs/hep-th/0512045}{{\tt hep-th/0512045}}.

\bibitem{Aharony:2007du}
O.~Aharony, Z.~Komargodski and A.~Patir, {\it {The Moduli Space and M(atrix)
  Theory of 9d N=1 Backgrounds of M/String Theory}},  {\em JHEP} {\bf 05}
  (2007) 073 [\href{http://arXiv.org/abs/hep-th/0702195}{{\tt
  hep-th/0702195}}].

\bibitem{Berglund:1998va}
P.~Berglund, A.~Klemm, P.~Mayr and S.~Theisen, {\it {On type IIB vacua with
  varying coupling constant}},  {\em Nucl. Phys.} {\bf B558} (1999) 178--204
  [\href{http://arXiv.org/abs/hep-th/9805189}{{\tt hep-th/9805189}}].

\bibitem{Bershadsky:1998vn}
M.~Bershadsky, T.~Pantev and V.~Sadov, {\it {F-theory with quantized fluxes}},
  {\em Adv. Theor. Math. Phys.} {\bf 3} (1999) 727--773
  [\href{http://arXiv.org/abs/hep-th/9805056}{{\tt hep-th/9805056}}].

\bibitem{Chaudhuri:1995fk}
S.~Chaudhuri, G.~Hockney and J.~D. Lykken, {\it {Maximally supersymmetric
  string theories in $D<10$}},  {\em Phys. Rev. Lett.} {\bf 75} (1995)
  2264--2267 [\href{http://arXiv.org/abs/hep-th/9505054}{{\tt
  hep-th/9505054}}].

\bibitem{Chaudhuri:1995bf}
S.~Chaudhuri and J.~Polchinski, {\it {Moduli space of CHL strings}},  {\em
  Phys. Rev.} {\bf D52} (1995) 7168--7173
  [\href{http://arXiv.org/abs/hep-th/9506048}{{\tt hep-th/9506048}}].

\bibitem{Witten:1995ex}
E.~Witten, {\it {String theory dynamics in various dimensions}},  {\em Nucl.
  Phys.} {\bf B443} (1995) 85--126
  [\href{http://arXiv.org/abs/hep-th/9503124}{{\tt hep-th/9503124}}].

\bibitem{Sen:1996tz}
A.~Sen, {\it {M-Theory on $(K3 \times S^1)/\IZ_2$}},  {\em Phys. Rev.} {\bf
  D53} (1996) 6725--6729 [\href{http://arXiv.org/abs/hep-th/9602010}{{\tt
  hep-th/9602010}}].

\bibitem{Witten:1993yc}
E.~Witten, {\it {Phases of N = 2 theories in two dimensions}},  {\em Nucl.
  Phys.} {\bf B403} (1993) 159--222
  [\href{http://arXiv.org/abs/hep-th/9301042}{{\tt hep-th/9301042}}].

\bibitem{dave}
D.~{Morrison,} {unpublished}.

\bibitem{Hull:1997kt}
C.~M. Hull, {\it {Gravitational duality, branes and charges}},  {\em Nucl.
  Phys.} {\bf B509} (1998) 216--251
  [\href{http://arXiv.org/abs/hep-th/9705162}{{\tt hep-th/9705162}}].

\bibitem{Polchinski:1998rr}
J.~Polchinski, {\it {String theory. Vol. 2: Superstring theory and beyond}},
  {\em Cambridge University Press} (1998).

\bibitem{Saclioglu:1995de}
C.~Saclioglu, {\it {Generalization of Weierstrassian elliptic functions to
  $\IR^n$}},  {\em J. Phys.} {\bf A29} (1996) L17--L22
  [\href{http://arXiv.org/abs/hep-th/9506060}{{\tt hep-th/9506060}}].

\bibitem{Hellerman:unp}
S.~{Hellerman,} {unpublished}.

\bibitem{Gibbons:1987sp}
G.~W. Gibbons and P.~J. Ruback, {\it {The Hidden Symmetries of Multicenter
  Metrics}},  {\em Commun. Math. Phys.} {\bf 115} (1988) 267.

\bibitem{Ooguri:1996me}
H.~Ooguri and C.~Vafa, {\it {Summing up D-instantons}},  {\em Phys. Rev. Lett.}
  {\bf 77} (1996) 3296--3298 [\href{http://arXiv.org/abs/hep-th/9608079}{{\tt
  hep-th/9608079}}].

\bibitem{Hanany:1996ie}
A.~Hanany and E.~Witten, {\it Type {IIB} superstrings, {BPS} monopoles, and
  three-dimensional gauge dynamics},  {\em Nucl. Phys.} {\bf B492} (1997)
  152--190 [\href{http://arXiv.org/abs/hep-th/9611230}{{\tt hep-th/9611230}}].

\bibitem{Elitzur:1997fh}
S.~Elitzur, A.~Giveon and D.~Kutasov, {\it {Branes and N = 1 duality in string
  theory}},  {\em Phys. Lett.} {\bf B400} (1997) 269--274
  [\href{http://arXiv.org/abs/hep-th/9702014}{{\tt hep-th/9702014}}].

\bibitem{Seiberg:1994pq}
N.~Seiberg, {\it {Electric - magnetic duality in supersymmetric nonAbelian
  gauge theories}},  {\em Nucl. Phys.} {\bf B435} (1995) 129--146
  [\href{http://arXiv.org/abs/hep-th/9411149}{{\tt hep-th/9411149}}].

\bibitem{Katz:1996fh}
S.~H. Katz, A.~Klemm and C.~Vafa, {\it {Geometric engineering of quantum field
  theories}},  {\em Nucl. Phys.} {\bf B497} (1997) 173--195
  [\href{http://arXiv.org/abs/hep-th/9609239}{{\tt hep-th/9609239}}].

\bibitem{Ooguri:1997ih}
H.~Ooguri and C.~Vafa, {\it {Geometry of N = 1 dualities in four dimensions}},
  {\em Nucl. Phys.} {\bf B500} (1997) 62--74
  [\href{http://arXiv.org/abs/hep-th/9702180}{{\tt hep-th/9702180}}].

\bibitem{Witten:1997sc}
E.~Witten, {\it {Solutions of four-dimensional field theories via M- theory}},
  {\em Nucl. Phys.} {\bf B500} (1997) 3--42
  [\href{http://arXiv.org/abs/hep-th/9703166}{{\tt hep-th/9703166}}].

\bibitem{Elitzur:1997hc}
S.~Elitzur, A.~Giveon, D.~Kutasov, E.~Rabinovici and A.~Schwimmer, {\it {Brane
  dynamics and N = 1 supersymmetric gauge theory}},  {\em Nucl. Phys.} {\bf
  B505} (1997) 202--250 [\href{http://arXiv.org/abs/hep-th/9704104}{{\tt
  hep-th/9704104}}].

\bibitem{Feng:2000mi}
B.~Feng, A.~Hanany and Y.-H. He, {\it {D-brane gauge theories from toric
  singularities and toric duality}},  {\em Nucl. Phys.} {\bf B595} (2001)
  165--200 [\href{http://arXiv.org/abs/hep-th/0003085}{{\tt hep-th/0003085}}].

\bibitem{Klebanov:2000hb}
I.~R. Klebanov and M.~J. Strassler, {\it {Supergravity and a confining gauge
  theory: Duality cascades and chiSB-resolution of naked singularities}},  {\em
  JHEP} {\bf 08} (2000) 052 [\href{http://arXiv.org/abs/hep-th/0007191}{{\tt
  hep-th/0007191}}].

\bibitem{Cachazo:2001sg}
F.~Cachazo, B.~Fiol, K.~A. Intriligator, S.~Katz and C.~Vafa, {\it {A geometric
  unification of dualities}},  {\em Nucl. Phys.} {\bf B628} (2002) 3--78
  [\href{http://arXiv.org/abs/hep-th/0110028}{{\tt hep-th/0110028}}].

\bibitem{Beasley:2001zp}
C.~E. Beasley and M.~R. Plesser, {\it {Toric duality is Seiberg duality}},
  {\em JHEP} {\bf 12} (2001) 001
  [\href{http://arXiv.org/abs/hep-th/0109053}{{\tt hep-th/0109053}}].

\bibitem{Feng:2001bn}
B.~Feng, A.~Hanany, Y.-H. He and A.~M. Uranga, {\it {Toric duality as Seiberg
  duality and brane diamonds}},  {\em JHEP} {\bf 12} (2001) 035
  [\href{http://arXiv.org/abs/hep-th/0109063}{{\tt hep-th/0109063}}].

\bibitem{Berenstein:2002fi}
D.~Berenstein and M.~R. Douglas, {\it {Seiberg duality for quiver gauge
  theories}},  \href{http://arXiv.org/abs/hep-th/0207027}{{\tt
  hep-th/0207027}}.

\bibitem{Franco:2005rj}
S.~Franco, A.~Hanany, K.~D. Kennaway, D.~Vegh and B.~Wecht, {\it {Brane dimers
  and quiver gauge theories}},  {\em JHEP} {\bf 01} (2006) 096
  [\href{http://arXiv.org/abs/hep-th/0504110}{{\tt hep-th/0504110}}].

\bibitem{Hanany:2005ss}
A.~Hanany and D.~Vegh, {\it {Quivers, tilings, branes and rhombi}},  {\em JHEP}
  {\bf 10} (2007) 029 [\href{http://arXiv.org/abs/hep-th/0511063}{{\tt
  hep-th/0511063}}].

\bibitem{tHooft:2008kk}
G.~'t~Hooft, {\it {A locally finite model for gravity}},
  \href{http://arXiv.org/abs/0804.0328}{{\tt 0804.0328}}.

\bibitem{Witten:1995em}
E.~Witten, {\it Five-branes and {M}-theory on an orbifold},  {\em Nucl. Phys.}
  {\bf B463} (1996) 383--397 [\href{http://arXiv.org/abs/hep-th/9512219}{{\tt
  hep-th/9512219}}].

\bibitem{Dasgupta:1995zm}
K.~Dasgupta and S.~Mukhi, {\it {Orbifolds of M-theory}},  {\em Nucl. Phys.}
  {\bf B465} (1996) 399--412 [\href{http://arXiv.org/abs/hep-th/9512196}{{\tt
  hep-th/9512196}}].

\bibitem{Sen:1997kz}
A.~Sen, {\it A note on enhanced gauge symmetries in {M-} and string theory},
  {\em JHEP} {\bf 09} (1997) 001
  [\href{http://arXiv.org/abs/hep-th/9707123}{{\tt hep-th/9707123}}].

\bibitem{Kachru:2001je}
S.~Kachru and J.~McGreevy, {\it {M-theory on manifolds of $G_2$ holonomy and
  Type IIA orientifolds}},  {\em JHEP} {\bf 06} (2001) 027
  [\href{http://arXiv.org/abs/hep-th/0103223}{{\tt hep-th/0103223}}].

\bibitem{Dabholkar:1997zd}
A.~Dabholkar, {\it {Lectures on orientifolds and duality}},
  \href{http://arXiv.org/abs/hep-th/9804208}{{\tt hep-th/9804208}}.

\bibitem{Douglas:1998xa}
M.~R. Douglas, {\it {D-branes and discrete torsion}},
  \href{http://arXiv.org/abs/hep-th/9807235}{{\tt hep-th/9807235}}.

\end{thebibliography}\endgroup
\bibliographystyle{JHEP}

\end{document}